\documentclass[a4paper,11pt]{article}
\pdfoutput=1 

\usepackage{jcappub} 

\usepackage[T1]{fontenc}
\usepackage{soul}
\usepackage{graphicx}
\usepackage{dcolumn}
\usepackage{amssymb,amsmath,bm}
\usepackage{color}
\usepackage[dvipsnames,table]{xcolor}
\usepackage{xfrac}
\usepackage{aas_macros}
\usepackage{mathrsfs}
\usepackage{subcaption}
\usepackage{rotating}
\usepackage{chngcntr}
\usepackage{units}
\usepackage{multirow}
\usepackage{xspace}
\usepackage{xfrac}
\usepackage{mathtools}
\usepackage{comment}
\usepackage{soul}
\usepackage{gensymb}
\usepackage{multirow}
\usepackage{lscape}
\usepackage{longtable}
\newcommand{\degsq}{deg$^2$\xspace}

\newcommand{\Om}{\Omega_{\rm m}}
\newcommand{\Oc}{\Omega_{\rm cdm}}
\newcommand{\Ob}{\Omega_{\rm b}}
\newcommand{\ob}{\omega_{\rm b}}
\newcommand{\oc}{\omega_{\rm c}}
\newcommand{\om}{\omega_{\rm m}}
\newcommand{\As}{A_{\rm s}}
\newcommand{\ns}{n_{\rm s}}

\newcommand{\Ms}{M_{\odot}}
\newcommand{\massunits}{(h^{-1} \Ms)}
\newcommand{\logMc}{\log_{10}(M_{\rm c})}
\newcommand{\logeta}{\log_{10}(\eta)}
\newcommand{\logbeta}{\log_{10}(\beta)}
\newcommand{\logMz}{\log_{10}(M_{z_0, {\rm cen}})}
\newcommand{\logThetaI}{\log_{10}(\theta_{\rm inn})}
\newcommand{\logThetaO}{\log_{10}(\theta_{\rm out})}
\newcommand{\logMi}{\log_{10}(M_{\rm inn})}

\newcommand{\enangle}[1]{\langle #1 \rangle}

\newcommand{\md}{\mathrm{d}}

\newcommand{\AAE}{A_{\rm AE}}

\newcommand{\neff}{n_{\rm eff}}

\newcommand{\thetamin}{\theta_{\rm min}}
\newcommand{\lmax}{\ell_{\rm max}}
\newcommand{\lmin}{\ell_{\rm min}}

\newcommand{\mlim}{m_{\rm lim}}

\newcommand{\nside}{N_{\rm side}}

\newcommand{\Ngal}{\bar{N}_{\rm g}}

\newcommand{\shear}{\boldsymbol{\gamma}}
\newcommand{\ellip}{\boldsymbol{e}}

\newcommand{\Nd}{N_{\rm d}}

\newcommand{\lcdm}{$\Lambda$CDM\xspace}

\newcommand{\txtp}{3$\times$2-point }

\newcommand{\nmt}{\texttt{NaMaster}\xspace}
\newcommand{\cobaya}{\texttt{Cobaya}\xspace}
\newcommand{\healpix}{\texttt{HEALPix}\xspace}
\newcommand{\ccl}{\texttt{CCL}\xspace}

\newcommand{\camb}{\texttt{CAMB}\xspace}
\newcommand{\hfit}{\textsc{HALOFIT}\xspace}
\newcommand{\hmcode}{\textsc{HMCode2020}\xspace}

\newcommand{\baccoemu}{\texttt{Baccoemu}\xspace}
\newcommand{\lensfit}{\textit{lens}fit\xspace}
\newcommand{\euclidemu}{\texttt{EuclidEmulator2}\xspace}

\newcommand{\des}{DES\xspace}

\newcommand{\hsc}{HSC\xspace}
\newcommand{\hscdro}{HSC-DR1\xspace}
\newcommand{\hscdrt}{HSC-DR3\xspace}
\newcommand{\mcal}{\textsc{Metacalibration}\xspace}
\newcommand{\planck}{{\sl Planck}\xspace}

\newcommand{\desyt}{DES-Y3\xspace}
\newcommand{\kidsot}{KiDS-1000\xspace}
\newcommand{\kids}{KiDS\xspace}

\title{Cosmic shear with small scales: \desyt, \kidsot and \hscdro}

\author[a,1]{Carlos Garc\'ia-Garc\'ia,\note{Corresponding author.}}
\author[a]{Matteo Zennaro,}
\author[b]{Giovanni Aric\`o,}
\author[a]{David Alonso,}
\author[c,d]{and Raul E. Angulo}

\affiliation[a]{Department of Physics, University of Oxford, Denys Wilkinson Building, Keble Road, Oxford OX1 3RH, UK}
\affiliation[b]{Institut für Astrophysik (DAP), Universität Zürich, Winterthurerstrasse 190, 8057 Zürich, Switzerland}
\affiliation[c]{Donostia International Physics Center, Manuel Lardizabal Ibilbidea, 4, 20018 Donostia, Gipuzkoa, Spain}
\affiliation[d]{IKERBASQUE, Basque Foundation for Science, 48013, Bilbao, Spain}

\emailAdd{carlos.garcia-garcia@physics.ox.ac.uk}
\emailAdd{matteo.zennaro@physics.ox.ac.uk}
\emailAdd{giovanni.arico@uzh.ch}
\emailAdd{david.alonso@physics.ox.ac.uk}
\emailAdd{reangulo@dipc.org}

\abstract{We present a cosmological analysis of the combination of the \desyt, \kidsot and \hscdro weak lensing samples under a joint harmonic-space pipeline making use of angular scales down to $\lmax=4500$, corresponding to significantly smaller scales ($\delta\theta\sim2.4'$) than those commonly used in cosmological weak lensing studies. We are able to do so by accurately modelling non-linearities and the impact of baryonic effects using \baccoemu. We find $S_8\equiv\sigma_8\sqrt{\Om/0.3}=0.795^{+0.015}_{-0.017}$, in relatively good agreement with CMB constraints from \planck (less than $\sim1.8\sigma$ tension), although we obtain a low value of $\Om=0.212^{+0.017}_{-0.032}$, in tension with \planck at the $\sim3\sigma$ level. We show that this can be recast as an $H_0$ tension if one parametrises the amplitude of fluctuations and matter abundance in terms of variables without hidden dependence on $H_0$. Furthermore, we find that this tension reduces significantly after including a prior on the distance-redshift relationship from BAO data, without worsening the fit. In terms of baryonic effects, we show that failing to model and marginalise over them on scales $\ell\lesssim2000$ does not significantly affect the posterior constraints for \desyt and \kidsot, but has a mild effect on deeper samples, such as \hscdro. This is in agreement with our ability to only mildly constrain the parameters of the Baryon Correction Model with these data.}

\begin{document}
\maketitle
\flushbottom

\section{Introduction}
  Cosmology is going through one of its most exciting moments, as the analysis of Stage-III large scale structure (LSS) surveys winds down \cite{2105.13543, 2105.13544, 1704.05858, 2304.00701, 2304.00702, 1507.00738, 2007.15633,2011.03407,2010.09717}, and we prepare for the unprecedented volume and quality of data from State-IV facilities such as DESI\footnote{\url{https://www.desi.lbl.gov}} \cite{1611.00036}, the Vera Rubin Obseratory LSST\footnote{\url{https://www.lsst.org}} \cite{0805.2366, 1809.01669}, and Euclid\footnote{\url{https://www.euclid-ec.org/.}} \cite{1110.3193}. This is happening at a moment marked by the seemingly ever-increasing cosmological tensions between early- and late-time measurements of the late-time expansion rate, $H_0$, and of the amplitude of density fluctuations at late times, parametrised by $S_8\equiv\sigma_8\sqrt{\Om/0.3}$, where $\Om$ is the fractional energy density of non-relativistic matter, and $\sigma_8$ is the standard deviation of the matter overdensity on spheres of $8\,{\rm Mpc}\,h^{-1}$ radius (with $h\equiv H_0/(100\,\,{\rm km}\,{\rm s}^{-1}\,{\rm Mpc}^{-1})$). On the one hand, the Hubble parameter measured locally using type-Ia supernovae ($h=0.7304\pm0.0104$) \cite{2112.04510} is $5.4\sigma$ higher than that obtained by the \planck CMB satellite ($h = 0.6736\pm0.0054$) \cite{1807.06209}. On the other hand, LSS measurements of $S_8$ present different levels of tension most of them obtaining values that are systematically below that of \planck ($S_8 = 0.832\pm0.013$) (see e.g. \cite{2203.06142} for a review). 

  Focusing on measurements from cosmic shear data alone, the analysis of the 3-year data from the Dark Energy Survey (\desyt hereafter) yields $S_8=0.772^{+0.018}_{-0.017}$ \citep{2105.13543, 2105.13544} and $S_8 = 0.793^{+0.038}_{-0.025}$ \citep{2203.07128}, in real- and harmonic space, respectively (in tension with \planck at $\sim1.5-2.5\sigma$).  The latest analysis of the Kilo-Degree Survey data (\kidsot hereafter) yielded $S_8=0.760^{+0.016}_{-0.038}$ using harmonic-space bandpowers, and $S_8 = 0.759^{+0.024}_{ -0.021}$ using COSEBIs \cite{2007.15633}, corresponding to a higher level of tension ($\sim3\sigma$). Finally, the analysis of the Hyper Suprime-Cam 3-year data yielded $S_8=0.776^{+0.032}_{-0.033}$ \cite{2304.00701} and $S_8=0.769^{+0.031}_{-0.034}$ \cite{2304.00702} in harmonic and real space, respectively, lower than \planck at the $\sim2\sigma$ level. Similar studies have been carried out using other LSS observables. These include \txtp analyses, combining cosmic shear and galaxy clustering \citep{2007.15632,2105.13549}, the CMB lensing power spectrum in combination with BAO data \cite{2206.07773,2304.05203}, CMB lensing tomography, combining galaxy auto-correlations and cross-correlations with CMB lensing maps \citep{2105.03421,2111.09898,2306.17748,2309.05659,2402.05761}, ``5$\times$2-point'' analyses, combining all three tracers and their cross-correlations \citep{2105.12108,2206.10824}, galaxy clustering studies exploiting the non-linear regime \citep{1909.11006,2002.04035}, and the analysis of cluster abundances \citep{1502.01597,2401.02075,2402.08458}. The level of tension in $S_8$ found by these studies varies across them. Intermediate-redshift probes ($z\gtrsim1.5$), such as the CMB lensing power spectrum, or high-redshift CMB lensing tomography \cite{2206.07773,2304.05203,2309.05659,2402.05761}, have generally found constraints that largely agree with \planck. In turn, many lower-redshift probes ($z\lesssim0.5$) have often found lower values of $S_8$ at different levels of tension \cite{1502.01597,1909.11006,2007.15632,2105.13549,2111.09898, 2303.06928} (with the distinct exception of the latest cluster abundance studies by the SPT and eROSITA collaborations \cite{2401.02075,2402.08458}). Understanding the origin of these potential discrepancies, and quantifying the combined level of tension evidenced by current probes, is therefore of paramount importance. Next-generation data will be powerful enough to show whether they are a simple statistical anomaly, caused by observational systematics in current data or, potentially, a signature of new physics beyond the \lcdm model (see some examples in e.g. \cite{2203.06142}). For the time being, however, we can make progress by combining as much of the currently available data as possible, to try to address this question.

  In this paper, we will focus on the case of cosmic shear alone. The clear advantage of galaxy weak lensing over other probes is that it is largely an unbiased tracer of the matter fluctuations, and is thus less sensitive than other observables to the detailed modelling of complex astrophysical processes. Nevertheless, a number of observational and astrophysical systematics must be taken into account in the analysis of cosmic shear data. On the observational side, reliable constraints require a careful characterisation of the redshift distribution of the samples under analysis, as well as marginalising over the associated uncertainties. This can be a complicated task, particularly for deep samples, in the presence of inaccurate photometric redshifts (photo-$z$s) \cite{0801.3822, 1909.09632, astro-ph/0606098, 0805.1409, 1003.0687, 1303.0292, 2012.08569, 1708.01537, 2105.13542}. On the astrophysical side, the two main systematics are the impact of intrinsic alignments (IAs), and baryonic effects. IAs refer to the statistical alignment between intrinsic galaxy shapes, caused not by gravitational lensing, but by local processes (gravitational or otherwise). The most popular IA models, in order of complexity, are the Non-linear Alignment model (NLA) \cite{astro-ph/0406275, 1708.01538}, and its generalisation, the Tidal Alignment and Tidal Torquing model (TATT) \cite{1708.09247}. It has been shown that, due to the lack of constraining power over these models with current data, the choice of TATT over NLA can impact the posterior distributions by $\sim 0.5\sigma$\cite{2303.05537,2305.17173}. The term ``baryonic effects'' refers to all processes modifying the distribution of dark matter owing to the presence of gas and stars. These include gas virialisation, its ejection in outflows driven by Active Galactic Nuclei (AGN) feedback, its impact on the dark matter distribution and, to a lesser extent, the distribution of stars. Baryonic effects may cause a suppression in the matter power spectrum of up to $\sim30\%$ on the scales ($k\gtrsim 0.3 \,{\rm Mpc}^{-1}$) that cosmic shear data is sensitive to. Baryonic effects have been signposted in the literature \cite{2206.11794, 2305.09827, 2303.05537} as a potential driver of the $S_8$ tension, especially in the context of weak lensing.

  Traditionally, a seemingly conservative way to mitigate the impact of baryonic effects has been to remove all scales from the analysis that are significantly affected by the aforementioned baryonic suppression. This may mean rejecting a large fraction of the data (e.g. $\approx$ 45\% in \desyt \citep{2105.13543}). Alternatively, a model for baryonic effects may be built and marginalised over, ideally covering all possible physically meaningful models or, at least, those allowed by current hydrodynamical N-body simulations. In this sense, several approaches have been proposed in the literature, ranging from purely simulation-driven methods (e.g. marginalising over the principal components of a suite of simulations \cite{1405.7423, 1707.02332}), to physics-based analytical parametrisations, such as the popular \hmcode \cite{1505.07833, 1602.02154, 2009.01858}. Any of these approaches must be tested and calibrated against hydrodynamical simulations, to ensure that they are sufficiently accurate over the range of scales used in the analysis. Here, we will make use of the baryonic effects emulator implemented in \baccoemu \cite{2004.06245, 2104.14568, 2011.15018, 2101.12187, 2207.06437}. The model is based on the baryonification approach, originally proposed by \cite{1510.06034}, and based on displacing particles in a gravity-only simulation according to a physical model of baryonic effects. Trained on a large suite of baryonified simulations, with a model depending on 7 different baryonic effects parameters, \baccoemu has been shown to be accurate at the $\sim1-2\%$ level when tested against 74 cosmological hydro simulations. 

  This approach was initially applied in \cite{2303.05537} (A23 hereafter) to the analysis of the \desyt data in real space, showing that the $S_8$ tension could be reduced through a combination of a more accurate non-linear power spectrum parametrisation, a simpler IA model (NLA instead of TATT), and a careful treatment of baryonic effects, while employing a significantly larger range of scales. In this paper we will extend the analysis of A23, combining the three currently available cosmic shear datasets: \desyt, \kidsot, and \hscdro, reanalysed under a common harmonic-space pipeline reaching significantly smaller scales than previous similar studies ($\lmax = 4500$ in our fiducial analysis). We pay particular attention to the modelling of the non-linear regime, both the non-linear matter power spectrum, and baryonic effects. In particular, we use \baccoemu to compute both. It is important to emphasise that, as shown in A23, using a model more accurate than halofit has a non-negligible impact even on relatively large scales, i.e. even when applying scale cuts. Moreover, we carefully account for several other small-scale systematics and validate our analysis against synthetic data vectors with known cosmology. 

  The paper is structured as follows: in Section~\ref{s:methods} we describe our theoretical model and likelihood. In Section~\ref{s:data} we introduce the different datasets used, and the methods employed to analyse them. Section~\ref{s:results} presents the main findings of this work, while Section~\ref{s:robustness} discusses, in detail, the robustness of these results to various analysis choices and, in general, the sensitivity of cosmic shear data to baryonic effects. We conclude in Section~\ref{s:conclusion} with a summary of the main results and a discussion of potential future avenues for further investigation.

\section{Theory model and parameter inference}\label{s:methods}
  \subsection{Cosmic shear}\label{ss:shear}
    The cross-correlation between the $E$-mode cosmic shear signal of sources in two different redshift bins, $i$ and $j$, can be related to the 3D matter power spectrum, $P_{\rm mm}(k, z)$ by the following expression in the Limber approximation \cite{1953ApJ...117..134L},
    \begin{equation}\label{eq:limber}
      C^{ij}_\ell=G_\ell^2\int\frac{d\chi}{\chi^2}q_i(\chi)q_j(\chi)\,P_{\rm mm}\left(k=\frac{\ell+1/2}{\chi},z(\chi)\right)\,
    \end{equation}
    where $\ell$ is the angular multipole, $k$ the modulus of the wave vector, $z$ is the redshift and $\chi$ the comoving radial distance.
    The contribution from weak lensing to the radial kernel $q_i$ is the lensing kernel of the $i$-th bin. Assuming the validity of General Relativity, a flat Universe, and in natural units (i.e. $c = 1$), we can write the lensing kernel as
    \begin{equation}\label{eq:kernel}
      q_{i,L}(\chi) = \frac{3}{2} H_0^2 \Om \frac{\chi}{a(\chi)} \int_{z(\chi)}^\infty \md z' p_i(z') \frac{\chi(z') - \chi}{\chi(z')}.
    \end{equation}
    where $p_i(z)$ is the sample's redshift distribution, and $a=1/(1+z)$ is the scale factor. The prefactor
    \begin{equation}
      G_\ell\equiv\sqrt{\frac{(\ell+2)!}{(\ell-2)!}}\frac{1}{(\ell+1/2)^2}
    \end{equation}
    accounts for the difference between the 3D Laplacian of the gravitational potential and the angular Hessian of the associated lensing potential \citep{1702.05301}. 
    
    In the NLA intrinsic alignment model, in which galaxy shapes align with the local tidal forces at linear order, IAs can be included by simply adding a contribution to the radial kernel $q_i$ of the form \cite{astro-ph/0406275, 1708.01538}:
    \begin{equation}
      q_{{\rm IA},i}(\chi)=-A_{\rm IA}(z)p_i(z)\frac{dz}{d\chi}\,
    \end{equation}
    where the IA amplitude is parametrised as
    \begin{equation}\label{eq:ia}
      A_{\rm IA}(z) = A_{{\rm IA}, 0} \left(\frac{1+z}{1+z_0}\right)^{\eta_{\rm IA}} \frac{0.0139\Om}{D(z)}.
    \end{equation}
    Here, $D(z)$ is the linear growth factor normalised to $D(0) = 1$, and we choose the pivot redshift $z_0 = 0.62$ (as in e.g. \cite{1708.01530, 1708.01538}). We keep $A_{{\rm IA}, 0}$ and $\eta_{\rm IA}$ as free model parameters.

    All theoretical calculations were carried out using the Core Cosmology Library (\ccl)~\cite{1812.05995}, with the linear power spectrum generated by \baccoemu \cite{2104.14568} or \camb~\cite{astro-ph/9911177}, when needed. By default, the non-linear matter power spectrum was computed using the extended version of \baccoemu from \cite{2303.05537}\footnote{The extended version of \baccoemu has broader boundaries than the public version that encompass the priors used here, with the exception of $\Oc + \Ob$. In this case the range, $\Oc + \Ob \in [0.15 0.47]$, which is enough to encompass more than $99\%$ of the samples in all fiducial chains and almost all cases shown in this work.}, although we also quantified it
    using \hfit \cite{1208.2701} in some cases. For instance, when using \baccoemu we resort to \hfit when the parameters lie outside the range of the emulator. However, this only happens in very few cases (less than $1\%$ in all fiducial cases) and does not impact our results. It is also important to note that, although \baccoemu gives predictions of the matter power spectrum down to $k = 10\,h{\rm Mpc}^{-1}$, we use it up to $k = 5\,h{\rm Mpc}^{-1}$ (which are small enough scales for our data), to match the scale cut of the baryon emulator, and we use the \ccl default extrapolation scheme (quadratic spline in logarithmic space) to model the smaller scales. We show in Appendix~\ref{ap:extrap}, that this extrapolation does not impact our results. Finally, the model used to describe baryonic effects is outlined in Section \ref{ss:baryons}.

  \subsection{Baryonic effects}\label{ss:baryons}
    The distribution of baryonic matter (gas and stars) on small, halo scales is strongly affected by complex astrophysical processes, including gas accretion, AGN and supernova feedback. These effects, particularly the ejection of gas outside of the virial radius, can lead to a significant suppression in the small-scale power spectrum ($k\sim1\, h{\rm Mpc}^{-1}$), while the formation of galaxies can lead to an enhancement on much smaller scales ($k\gtrsim10\,h{\rm Mpc}^{-1}$). Moreover, the effect of AGN feedback can extend to even much larger scales than those traditionally considered safe in scale-cut-based analyses. In the most extreme scenarios allowed by cosmological hydrodynamical simulations, scales as large as $k \sim 0.3 \,h{\rm Mpc}^{-1}$ are already significantly affected by baryonic processes \cite{1104.1174}. Finally, these modifications also alter the distribution of dark matter as it adapts to the new gravitational potential. To account for these physical effects we will explore various models.

    \paragraph{Baryonification emulator.} As our fiducial model, we employ the baryonification algorithm implemented in \cite{1911.08471,2011.15018}. Baryonification (or the baryon correction model, BCM), first introduced by ~\cite{1510.06034} (ST15 hereafter), is a method to modify in post-processing the 3D matter field given by a gravity-only $N$-body simulation, to take into account several baryonic processes. Inspired by the halo model, the density profiles of the simulated haloes are perturbed by employing analytical stellar and gas density profiles, and the particles are displaced accordingly. The density profiles are described using a few free parameters with a clear physical interpretation, which can be constrained using either hydrodynamical simulations or astronomical observations. Baryonification can reproduce at per-cent level the power spectra and bispectra of several different hydrodynamical simulations \citep{1911.08471,2011.15018}.

    In this work, we employ a neural network emulator of the matter power spectrum suppression predicted by baryonification, publicly available in \baccoemu \citep{2011.15018}. This emulator is accurate at 1-2$\%$ level down to $k = 5\,h$Mpc$^{-1}$, and was tested fitting the power spectra of a library of 74 hydrodynamical simulations. It features 7 free baryonic parameters: $M_{\rm c}$ and $\beta$ describe the fraction of gas mass still residing in the halo (i.e. not expelled by feedback processes), with $M_{\rm c}$ the halo mass at which half of the gas mass is lost, and $\beta$ the slope of the halo mass dependence. $\theta_{\rm inn}$, $M_{\rm inn}$, and $\theta_{\rm out}$ parametrise the scale dependence of the virialised gas density profile. $\eta$ characterises the distance to which gas is ejected. $M_{\rm 1,z0,cen}$ is the characteristic halo mass scale for central galaxies at $z=0$ (see \cite{2009.14225} for details). Note that, implemented this way, this parametrisation does not consider the redshift dependence of any of its parameters. This is not necessarily a valid assumption, particularly given the range of redshifts covered by some of our data \cite{1911.08471}. However, given the moderate sensitivity to baryonic effects we find, we do not expect this to have a large effect on our results.

    When the parameters are outside the boundaries of \baccoemu, we need to resort to extrapolation. In this case, we follow A23 and assign the baryon boost corresponding to the closest cosmology within the range of the emulator, while keeping fixed the (cold) baryon fraction ($\Ob / (\Ob + \Oc)$). This is a reasonable approximation since the dominant dependence of baryonic effects on cosmology is through the cosmic baryon fraction \cite{1810.08629, 1906.00968, 1911.08471, 2011.15018}. Finally, the baryon emulator's smallest scale is $k = 5\,h{\rm Mpc}^{-1}$. As with the matter power spectrum, in order to access the smaller scales, when needed, we extrapolate with \ccl (quadratic spline in logarithmic space). Again, we show in Appendix~\ref{ap:extrap}, that this extrapolation does not impact our analysis. 
    
    \paragraph{Other parametrisations.} To interpret our results, we will also employ two simpler parametrisations of the baryonic suppression. First, we will consider the simple analytical parametrisation provided by ST15 in the context of the BCM. The model in this case depends on three parameters: $M_c$ and $\eta_b$ (analogous to the $M_c$ and $\eta$ parameters described above), and $k_s$, the characteristic scale of the stellar component. We fix $k_s = 55 h\,{\rm Mpc}^{-1}$, since our data is sensitive to much larger scales that are weakly impacted by the stellar mass distribution. It is important to note that, in contrast to the \baccoemu parametrisation, when exploring the ST15 fitting function, we will impose top-hat priors on $M_c$ and $\eta_b$ of the form $M_c>10^{12}\,h^{-1}M_\odot$, and $\eta_b<1$, approximately covering the range of parameters for which this parametrisation was validated by ST15.

    We will also explore the simple parametrisation of \cite{2206.11794,2305.09827} (AE hereafter). This model parametrises the scale dependence of a generic suppression in the matter power spectrum in terms of the difference between the linear and non-linear matter power spectra. In detail, the suppressed non-linear power spectrum is $P_k = P_k^{\rm L} + \AAE (P_k^{\rm NL} - P_k^{\rm L})$, where $P_k^{\rm L}$ and $P_k^{\rm NL}$ are the gravity-only linear and non-linear power spectra, respectively, and $\AAE$ is a free amplitude parameter (with $\AAE=1$ corresponding to the case with no suppression).

  \subsection{Likelihood}
    We assume a Gaussian likelihood for these measured power spectra, justified by the Central Limit Theorem on the scales used in this analysis~\cite{0801.0554}. We describe the measurements of the angular power spectra and their covariance matrix in Section~\ref{s:data}.

    Our theoretical model follows closely the analysis choices of the DESY3 analysis \cite{2105.13543, 2105.13544, 2203.07128}, in particular the reanalysis of A23 including small scales. Table~\ref{tab:priors} collects the set of free parameters and priors. In all cases, we vary 6 cosmological parameters, $\{\Om,\, \As,\, h,\, \Ob,\, \ns,\, \sum m_\nu\}$. In addition, we assume an independent set of IA parameters $\{A_{\rm IA},\, \eta_{\rm IA}\}$ for each galaxy survey. Thus, when combining \desyt, \kidsot and \hscdro, we have a total of 6 IA free parameters. We also marginalise over the uncertainty in the calibration of the redshift distributions by introducing a free parameter per tomographic bin, $\Delta z^i$ that shifts their redshift distribution $p_i(z)$ so that $p_i(z) \rightarrow p_i(z + \Delta z^i)$. This has been shown to accommodate the uncertainty in the $p(z)$ calibration of weak lensing samples \cite{1809.09148, 1708.01532, 2301.11978}. We use the same priors used in the official analysis of each survey. As justified in \cite{2210.13434}, we ignore potential statistical correlations between the redshift uncertainties of different surveys due to their use of a common calibration sample. Similarly, we marginalise over the multiplicative shape measurement biases of each sample, using the priors defined in the official analyses. Finally, we introduce additional parameters to describe baryonic effects. In the fiducial case of the \baccoemu implementation of BCM, we add the 7 extra free parameters described in Section~\ref{ss:baryons}, $\{\logMc,\, \logeta,\, \logbeta,\, \logMz,\, \logThetaI,\, \logThetaO,\, \logMi\}$\footnote{Note that here we are abusing the notation: the masses have units of $\massunits$ and, therefore, we need to divide by them to make the argument of the logarithm dimensionless. As such,  $\logMc \equiv \log_{10}(M_{\rm c} / \massunits)$, $\logMi \equiv \log_{10}(M_{\rm inn} / \massunits)$ and $\logMz \equiv \log_{10}(M_{z_0, {\rm cen}}/ \massunits)$.}. 
    In principle, the BCM parameters may depend on redshift. In \baccoemu, an explicit redshift dependence is considered only for the stellar-to-halo mass relation ($\logMz$, for more details see \cite{1911.08471}). We note that, however, both observations and hydrodynamical simulations suggest such redshift dependence to be below the sensitivity of our data \cite{2111.10080, 1810.08629, 1911.08471}.       
    Therefore, in our analysis, we use a fixed set of BCM parameters for all redshift bins, in line with the \desyt real space analysis of A23. In the case of the analytic BCM parametrisation of ST15, we also vary two global parameters $\{\logMc, \logeta \}$ (within a smaller range than for \baccoemu, as described in Section~\ref{ss:baryons}) and keep $k_s = 55 h {\rm Mpc}^{-1}$ fixed. Finally, for the AE parametrisation we vary a single parameter, $\AAE$. 

    We sample the posterior distribution with \cobaya \cite{2005.05290, 2019ascl.soft10019T}, using the Metropolis-Hasting Markov Chain Monte Carlo (MCMC) method  \cite{metropolismc,hastingsmc}. We address the convergence with the Gelman-Rubin parameter \cite{gelmanrubin}, requiring $R-1<0.03$ in the diagonalised parameter space\footnote{We have checked that our choice of $R-1<0.03$ instead of the more usual $R-1<0.01$ virtually recovers the same posteriors while significantly reducing the CPU consumption.}. We assess the goodness of fit by reporting and comparing the best-fit $\chi^2 \equiv -2\log(\mathcal{L}_{\rm WL})$, where $\mathcal{L}_{\rm WL}$ is the Gaussian likelihood of the weak lensing data, and the corresponding probability-to-exceed (PTE) or $p$-value, which, for a $\chi^2$ distribution, is given by $p = 1 - F(\chi^2 | \nu)$. Here, $F$ is the cumulative probability distribution and $\nu$ the number of degrees of freedom, which we approximate as $\nu = N_{\rm d} - 2$, where $N_{\rm d}$ is the number of data points and 2 corresponds to the parameters that weak lensing constraints best ($S_8$ and $\Om$). Given the size of the data vectors, a variation of $\nu$ of order $O(1)$ will not impact our goodness of fit assessment. We obtain the best-fit parameters as the step in the MCMC chains with minimum $\chi^2$, to avoid running costly minimisers. Although this introduces some noise, it is a good approximation given the typical size of our chains ($O(5 \times 10^5)$ samples). For comparison, we also report the $\chi^2$ corresponding to the Maximum a Posteriori (MAP), which takes into account the effect of priors, in Table~\ref{tab:results}. As with the best-fit $\chi^2$, we estimate it from the chain directly. Although we discuss the results in terms of the best-fit $\chi^2$, the conclusions would remain the same provided we had used the MAP. Finally, we consider a good fit to the data anything with $0.05 \leq p \leq 0.95$, corresponding to a $2\sigma$ region for a Gaussian distribution.

\begin{table}
    \centering
    \begin{tabular}{|l|ll|ll|}
           \hline
           &\multicolumn{2}{c|}{\bf Cosmology}&  \multicolumn{2}{c|}{\bf Baryons (\baccoemu)}\\
           &$\Om$&  $U(0.1, 0.7)$&  $\logMc$& $U(9, 15)$\\
           &$\As 10^{9}$&  $U(0.5, 5)$&  $\logeta$& $U(-0.7, 0.7)$\\
           &$h$&  $U(0.55, 0.90)$&  $\logbeta$& $U(-1.0, 0.7)$\\
           &$\Ob$&  $U(0.03, 0.07)$&  $\logMz$& $U(9, 13)$\\
           &$\ns$&  $U(0.87, 1.07)$&  $\logThetaI$& $U(-2, -0.53)$\\
           &$\sum m_\nu$ [eV]&  $U(0.0559, 0.400)$&  $\logThetaO$& $U(-0.48, 0)$\\
           &$\tau$&  0.08&  $\logMi$& $U(9, 13.5)$\\
           \hline
           &\multicolumn{2}{c|}{\bf Amon-Efsthatiou} & \multicolumn{2}{c|}{\bf Schneider \& Teyssier 2015}\\
    \multirow{3}{3cm}{Other baryonic effects models}
           & $A_{\rm AE}$& $U(0, 2)$& $\logMc$  & $U(12, 15)$\\
           & & & $\logeta$& $U(-0.7, 0)$\\
           & & & $k_s$& 55 $h {\rm Mpc}^{-1}$\\
           \hline
           \hline
           &\multicolumn{2}{c|}{\bf Photo-$z$ shift}&  \multicolumn{2}{c|}{\bf Shear calibration}\\
    \multirow{4}{*}{\desyt}
           &$\Delta z^0$&  $N(0, 0.018)$&  $m^0$& $N(-0.0063, 0.0091)$\\
           & $\Delta z^1$& $N(0, 0.015)$& $m^1$&$N(-0.0198, 0.0078)$\\
           & $\Delta z^2$& $N(0, 0.011)$& $m^2$&$N(-0.0241, 0.0076)$\\
           & $\Delta z^3$& $N(0, 0.017)$& $m^3$&$N(-0.0369, 0.0076)$\\
           \hline
    \multirow{5}{*}{\kidsot}
           & $\Delta z^0$& $N(0, 0.0106)$& $m^0$&$N(0, 0.019)$\\
           & $\Delta z^1$& $N(-0.002, 0.0113)$& $m^1$&$N(0, 0.020)$\\
           & $\Delta z^2$& $N(-0.013, 0.0118)$& $m^2$&$N(0, 0.017)$\\
           & $\Delta z^3$& $N(-0.011, 0.0087)$& $m^3$&$N(0, 0.012)$\\
           & $\Delta z^4$& $N(0.006, 0.0097)$& $m^4$&$N(0, 0.010)$\\
           \hline
    \multirow{4}{*}{\hscdro}
           & $\Delta z^0$& $N(0, 0.0285)$& $m^0$&$N(0, 0.01)$\\
           & $\Delta z^1$& $N(0, 0.0135)$& $m^1$&$N(0, 0.01)$\\
           & $\Delta z^2$& $N(0, 0.0383)$& $m^2$&$N(0, 0.01)$\\
           & $\Delta z^3$& $N(0, 0.0376)$& $m^3$&$N(0, 0.01)$\\
           \hline
           &\multicolumn{4}{c|}{\bf Intrinsic Alignments}\\
    All surveys
           & $A_{\rm IA}$& $U(-5, 5)$& $\eta_{\rm IA}$ &$U(-5, 5)$\\
           \hline
           
    \end{tabular}
    \caption{Free parameters and prior distributions. $U(a, b)$ stands for an uniform distribution with boundaries $a$ and $b$ and $N(\mu, \sigma)$ for a Normal distribution centred at the mean $\mu$ and with variance $\sigma$. For the photo-$z$ shifts we are neglecting any correlation between redshift bins. We consider an independent set of IA parameters for each survey.}
    \label{tab:priors}
\end{table}

\section{Data}\label{s:data}
  \subsection{Stage-3 weak lensing surveys}
\begin{figure}
      \begin{minipage}{0.49\linewidth}
        \centering
        \includegraphics[width=\linewidth]{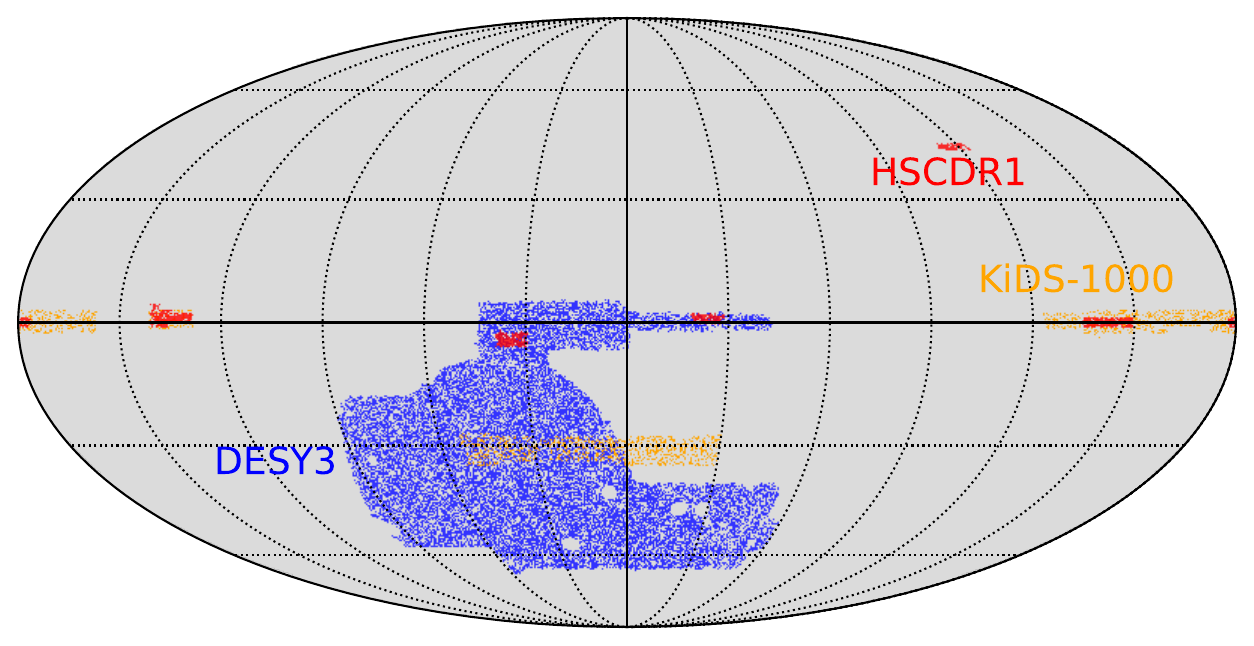}
        \includegraphics[width=\linewidth]{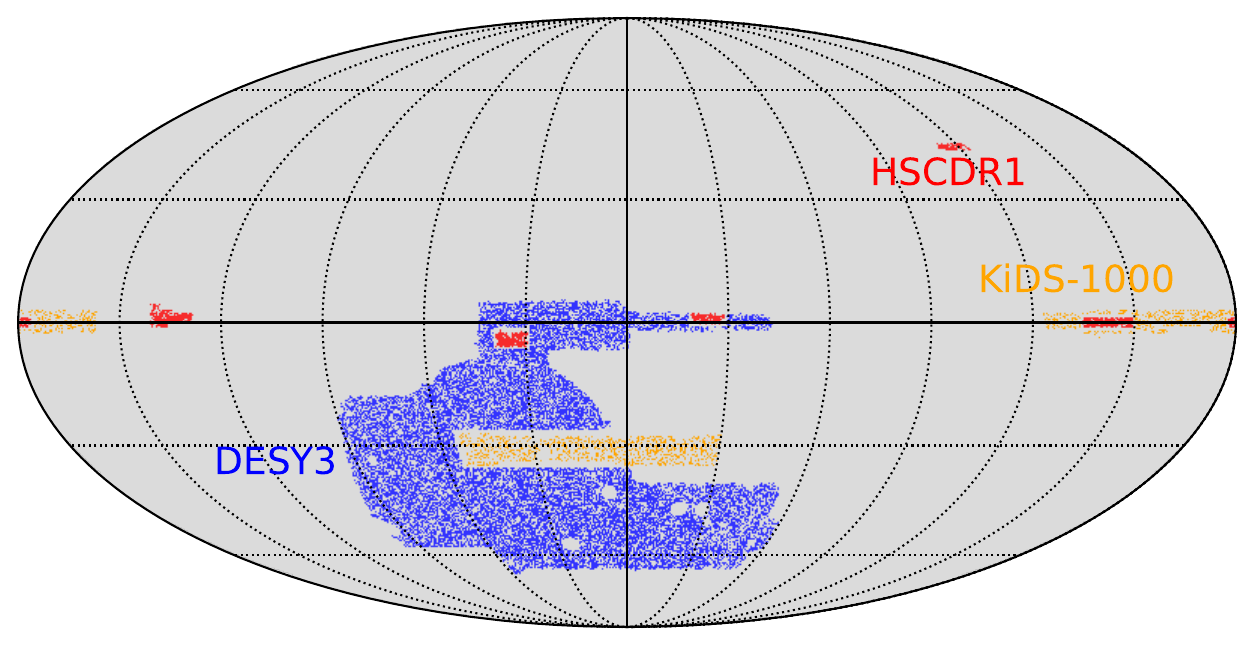}
      \end{minipage}
      \begin{minipage}{0.49\linewidth}
        \centering
        \includegraphics[width=\linewidth]{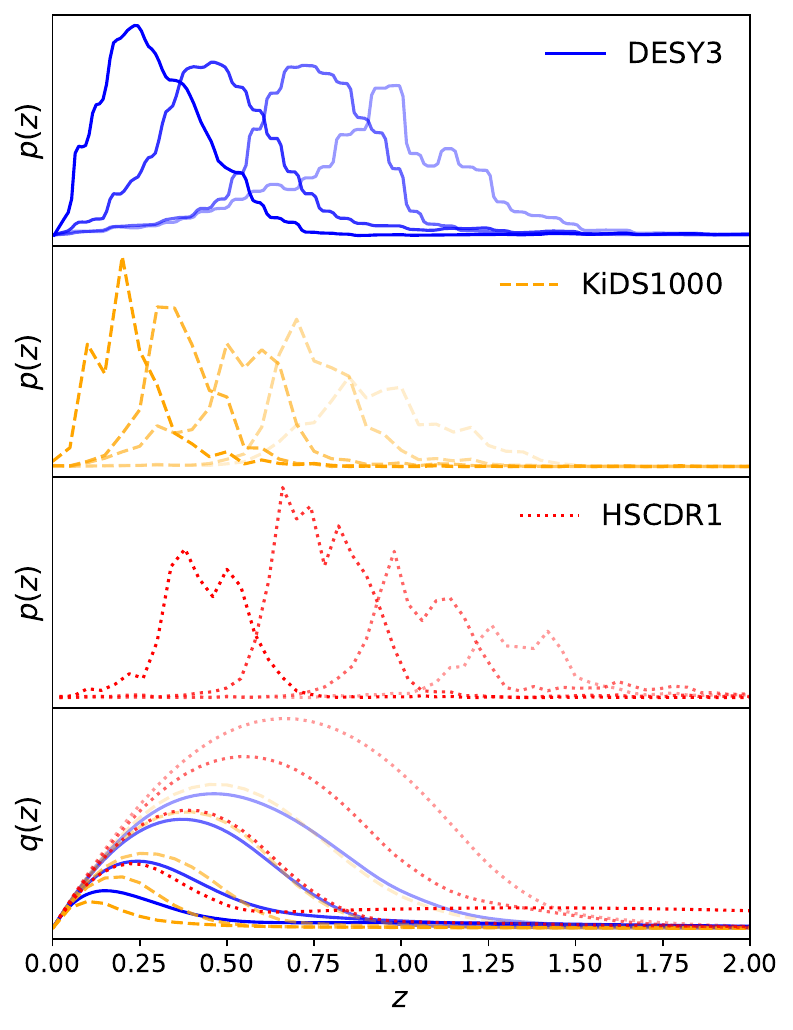}
      \end{minipage}
      \caption{\ul{Top left:} \desyt, \kidsot and \hscdro footprints. \ul{Bottom left:} Same but with the overlapping areas removed. This is done when combining them to avoid modelling the covariance between the different surveys. We leave about 1 arcmin of separation between the different footprints. \ul{Right:} Redshift distributions for the different weak lensing samples considered in this work. In the bottom panel, we show the associated lensing kernels (Eq.~\ref{eq:kernel}).}
      \label{fig:footprint.pz}
    \end{figure}

    We analyse data from three publicly available weak lensing surveys: \desyt, \kidsot, and \hscdro. Starting from the public catalogues, we process all data following the same steps (or as similar as possible), to ensure a self-consistent analysis under the same assumptions. The left panel in Fig.~\ref{fig:footprint.pz} shows the redshift distributions of the three different surveys (split into 4 redshift bins for \desyt and \hscdro, and 5 bins for \kidsot). The bottom sub-panel shows the associated lensing kernels. It is interesting to note that the lensing kernels of the \desyt bins resemble those of bins 2-5 in the \kidsot sample. Thus, both surveys are sensitive to largely the same range of redshifts. In turn, HSC is markedly deeper, and will thus allow us to explore the impact of baryonic effects at higher redshifts. The left panels show the angular footprint of the three surveys and the regions we cut out around them when combining their data. The rest of this section provides a brief description of each sample (further details can be found in the original papers), as well as any specific analysis choices made.

    \paragraph{\desyt.} The Dark Energy Survey is a photometric survey that has observed about 5000\degsq in 6 years of observations \cite{astro-ph/0510346, 1801.03181, 2101.05765} in 5 filter bands ($grizY$). The observations were taken from the Cerro Tololo Inter-American Observatory (CTIO) with the 4m Blanco Telescope, using the 570-Mpix Dark Energy Camera (DECam\cite{1504.02900}). In this work, we use the publicly available weak lensing cosmology catalogues from 3 years of observations\footnote{\url{https://des.ncsa.illinois.edu/releases/y3a2/Y3key-catalogs}} \cite{2011.03408}. This catalogue contains 100,204,026 galaxies over an effective area of 4143\degsq and a weighted number density of $\neff = 5.59$ gal/arcmin$^2$. We closely follow \cite{2011.03408} in the production of the cosmic shear maps. In particular, we divide the sample into four broad tomographic bins in the range $z_{\rm ph}\in[0, 1.5]$, with roughly the same number of objects. We also use the official redshift distributions for each of these bins \cite{2012.08566}. In each bin, we remove the mean ellipticity and correct for the multiplicative bias. We estimate the later es prescribed for \mcal; i.e. by measuring the response tensor $R$ and making the approximation $(1+m) = (R_{11} + R_{22})/2$, as described in \cite{2011.03408}. We verified that, using the power spectrum pipeline described in Section \ref{ss:pseudo-cl}, we are able to reproduce the official harmonic-space data vector from \cite{2203.07128} at high accuracy (see Fig.~\ref{fig:cl.des_official}).

    \paragraph{\kidsot.} The Kilo-Degree Survey (\kids) is an optical survey that has mapped 1350\degsq in four bands, $ugri$, using the VLT Survey Telescope (VST), in the ESO Paranal Observatory. We use data from the Gold Sample, made available with the public Data Release 4 (DR4)\footnote{\url{https://kids.strw.leidenuniv.nl/DR4/KiDS-1000_shearcatalogue.php}} \cite{1902.11265}, which comprises a selection of galaxies with reliable shapes and redshift distributions \cite{1902.11265, 2007.01845}. The Gold Sample includes galaxy images and forced photometry from the VIKINGS survey in five additional bands, covering 1006 \degsq, before masking (777.4 \degsq after masking). Following the official analysis of \cite{2007.15633}, we split the sample into 5 tomographic bins, with each galaxy being assigned by its best-fitting photometric redshift. We use the official redshift distributions provided with the DR4, obtained with the self-organising map (SOMPZ) method \cite{1909.09632, 2007.15635}. In each bin, we remove any residual mean ellipticity \cite{2007.01845}, and correct for the multiplicative biases from Table 1 of \cite{2007.15633}, which were obtained using image simulations. The maps are built as in Section \ref{ss:pseudo-cl}, assigning the \lensfit weights to each galaxy.

    \paragraph{\hscdro.} The Hyper Suprime-Cam Subaru Strategic Program (HSC SSP) is a wide-field imaging survey that will cover 1000\degsq in five filter bands ($grizy$). We use the publicly available data from the first Data Release (\hscdro)\footnote{\url{https://hsc-release.mtk.nao.ac.jp/doc/}} of the Wide layer, covering $\sim 108$ \degsq and with a limiting magnitude $\mlim = 26.4$ and median $i$-band seeing of $\sim 0.6''$. We follow \cite{1705.06745} and select the cosmic shear sample imposing a magnitude limit $\mlim = 24.5$. We then divide the sample into four tomographic bins with edges $[0.3, 0.6, 0.9, 1.2, 1.5]$ assigning each galaxy using the \texttt{Ephor\_AB} method photo-$z$. We use the official redshift distributions from \cite{1809.09148}, calibrated with the high-quality photometric galaxies from the COSMOS 30-band photometric catalogue \cite{1604.02350} using the direct calibration (DIR) method \cite{0801.3822} with weights obtained with SOM. Further details can also be found in \cite{2010.09717,2210.13434}. From Fig.~\ref{fig:footprint.pz}, it is clear that \hscdro is deeper than the other two weak lensing surveys. Furthermore, it is interesting to note how the kernel of the first bin does not go to 0 until $z\sim 3$. This is a consequence of the high redshift ($z\sim 3.5$) bump on the tail of the redshift distribution for this particular bin and makes the lensing kernel receive non-negligible contributions from high-$z$.

  \subsection{Pseudo-$C_\ell$ pipeline}\label{ss:pseudo-cl}
    The procedure used to process the shear catalogues and estimate their power spectra follows closely the methodology described in \cite{2010.09717}. 
    
    We start by generating shear maps $\shear_p$ by averaging ellipticities of all galaxies within each pixel $p$:
    \begin{equation}\label{eq:sh.signal}
      \shear_{p} = \frac{\sum_{i\in p} w_i \ellip_i}{\sum_{i\in p} w_i}\,,
    \end{equation}
    where $w_i$ is the weight of the $i$-th galaxy. Second, we build the mask $w_p$ as the weighted number count in each pixel
    \begin{equation}\label{eq:sh.mask}
      w_p = \sum_{i \in p} w_i\,.
    \end{equation}
    From these maps, we estimate the cosmic shear angular power spectrum, accounting for the mode coupling caused by the presence of the mask, with the pseudo-$C_\ell$ method, as implemented in \nmt \cite{1809.09603}. The details can be found in \cite{2010.09717}. The bias to the shear auto-correlations caused by shape noise is estimated analytically as described in \cite{2010.09717} and subtracted from the data vector.
    
    We consider three different contributions to the power spectrum covariance matrix \cite{1807.04266}: the ``Gaussian'' covariance, corresponding to the fully disconnected component of the matter overdensity trispectrum, the Super-Sample covariance (SSC), sourced by the coupling of observed modes with modes larger than the survey footprint via the connected trispectrum in the squeezed limit, and the connected non-Gaussian covariance (cNG), caused by all remaining contributions to the connected trispectrum. The largest contribution to the power spectrum covariance, particularly in the presence of significant shape noise, comes from the Gaussian component \cite{0801.0554,1707.04488}. We calculate this component using analytical methods (see e.g. \cite{1906.11765,astro-ph/0307515}), specifically the improved Narrow Kernel Approximation (NKA) of \cite{2010.09717}. In doing so, we assume all fields to be spin-0 quantities (i.e. we treat the $E$- and $B$-mode components as uncorrelated scalar fields at the map level), which has been shown to provide a better estimate of the Gaussian covariance than naively accounting for the spin-2 nature of cosmic shear in the NKA \cite{1906.11765}. In the Gaussian covariance estimator we make use of the mode-coupled angular power spectra directly measured from the data (see Eq. 2.36 of \cite{2010.09717}), containing both signal and noise. This allows us to avoid any possible modelling inaccuracies of the signal and noise on small scales. The next leading contribution is the SSC, which we estimate using the halo model as described in \cite{1601.05779}. More in detail, we compute the variance of the projected linear density field, $\sigma_B^2$, using the angular power spectrum of the masks involved.  Finally, we neglect the cNG component for \desyt and \kidsot since its effect is negligible for all the scales and noise levels of the data used in this work \cite{1807.04266}. However, as a safety measure, given the smaller sky area covered by \hscdro, we also include the cNG term for it.

    All maps are constructed using a \healpix\footnote{\url{http://healpix.sourceforge.net}} \cite{gorski_healpix_2005} pixelisation with $\nside = 4096$, corresponding to a resolution of $\delta\theta \sim 0.85'$. The smallest scale that is reliably accessible in this case is $\lmax = 2\nside = 8192$, corresponding to $\sim 1.3'$. In our analysis, we will use a fiducial scale cut of $\lmax = 4500$, corresponding to scales $\delta\theta \sim 2.4'$, in order to approximately match the scales over which the real-space shear correlation functions measured by \desyt were found to be robust against contamination from e.g. PSF systematics \cite{2011.03408}. We will also provide results for $\lmax = 1000$, closer to the typical conservative scale cuts used in previous analyses (e.g. \cite{2007.15633, 2111.07203, 2203.07128}), and $\lmax = 2048$, to study the impact of baryonic effects on other previous, less conservative analyses (e.g. \cite{2105.12108, 1809.09148}). On large scales, we use the following cuts: although the official harmonic space analysis of \desyt does not set a large scale cut \cite{2203.07128}, we set $\lmin = 20$ in order to remove the first bandpower, for which our analytical covariances are usually unreliable. We use $\lmin = 100$ for \kidsot, following  \cite{2007.15632,2007.15633}, and $\lmin = 300$ for \hscdro, as in the official analysis \cite{1809.09148}. We bin the measured power spectra into bandpowers using the following scheme for \desyt and \kidsot: bins with constant linear width $\Delta \ell = 30$ in the range of $0 \leq \ell \leq 240$, and logarithmic down to $\ell = 3\nside-1 = 12,287$ with $\Delta \log_{10}(\ell) = 0.055$. For \hscdro, we use the data from \cite{2010.09717}, which is binned as shown in Table 2 therein.

  \subsection{Angular power spectrum measurements}
    The measured angular power spectra are displayed in Appendix~\ref{ap:cl} for the different cases studied in this work. For \desyt and \kidsot, we apply the pseudo-$C_\ell$ method directly to the full cosmic shear maps, given their wide coverage. In the case of \hscdro, we use the data from \cite{2010.09717} which was computed independently in each patch, using the flat-sky approximation (given their small footprint), and then co-added afterwards. When combining the different surveys we take a ``layer cake'' approach, following \cite{2208.07179, 2305.17173}, removing the patches of the deeper survey from the fainter one. E.g. when combining \desyt and \kidsot, we remove the \kidsot patches from the \desyt footprint. Similarly, when combining \desyt, \kidsot and \hscdro, we remove the \hscdro patches from \desyt and \kidsot. When removing a given patch, we also cut out a narrow strip approximately $\sim 1-2'$-wide from the most exterior pixels in the footprint of the fainter survey to further reduce the spatial correlation between both samples (which we further validated by ensuring that the cross-survey correlations are compatible with zero). The resulting set of combined footprints is displayed in the bottom left panel of Fig.~\ref{fig:footprint.pz}. Note that, unlike in \cite{2305.17173}, we remove these patches at the map level by masking the overlapping areas and, therefore, the \mcal response tensors and the mean shear are not recomputed. We verified that this approximation does not affect the final cosmological constraints. Once this is done we assume that each survey is uncorrelated with the others (i.e. the covariance matrix of the combination of power spectra from different surveys is block-diagonal). This approximation was tested in \cite{2305.17173}, for the case of the \desyt and \kidsot combination, with lognormal simulations. Note, however, that, as discussed in \cite{2210.13434}, the use of the same training sample in the calibration of the redshift distributions introduces correlations between different, otherwise independent, data sets. Fortunately, these correlations were shown to have a negligible impact on the posterior distributions and we, therefore, neglect them in this analysis. Finally, since the cosmic shear signal is dominated by $E$-modes, we treat any positive detection of $B$-modes as a signature of possible systematics. We find no evidence of $B$-modes at more than $2\sigma$ ($p \geq 0.05$) in all cross-correlations up to the smallest scale explored in this work ($\lmax=8192$) except for $C_\ell^{1,3}$ of \desyt ($p= 0.03$), $C_\ell^{2,3}$ of \kidsot ($p=0.03$) and $C_\ell^{0,2}$ of \hscdro ($p = 0.02$). This small number is compatible with the look-elsewhere effect and, in fact, a Kolmogorov-Smirnov test on the full set of $\chi^2$ from all possible spectra with respect to a $\chi^2$ distribution returns $p=0.99,\, 0.46,\, 0.58$ for \desyt, \kidsot and \hscdro, respectively. Therefore, we conclude that we have no significant evidence of $B$-modes in any of the three samples.
    
\begin{figure}
        \centering
        \includegraphics[width=\textwidth]{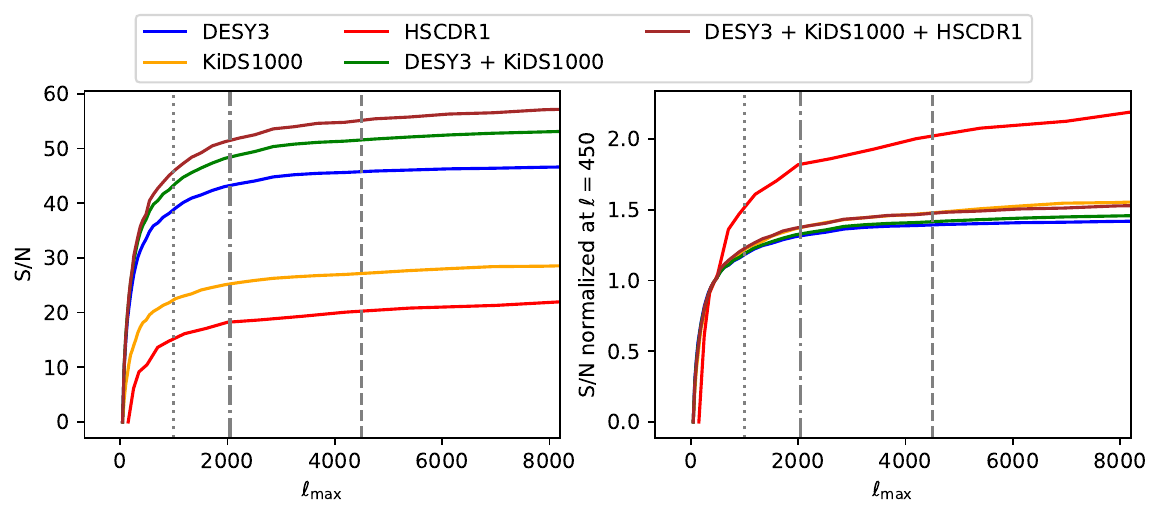}
        \caption{Signal-to-noise ratio for each studied case. In the \ul{left} panel, we see that the SNR is dominated by \desyt, with a small increase when combined with \kidsot. In the \ul{right} panel, we show the SNR normalised to $\ell = 450$ and see that the SNR of HSC, as deeper survey, increases more than in the other shallower surveys at smaller scales. The vertical gray lines show different scale cuts used in the analysis (note that $\lmax = 8192$ was also used).}
        \label{fig:snr}
    \end{figure}

    Before carrying out the cosmological analysis, it is worth comparing the constraining power of the different surveys on different scales. Fig.~\ref{fig:snr} shows the cumulative signal-to-noise ratio (SNR) as a function of scale cut for each survey and combination of them. We clearly see that, given its significantly larger area,  \desyt dominates the SNR and will drive the overall constraining power. However, it is interesting to see that, while the SNR for \desyt and \kidsot saturates around $\ell=2000$, HSC, being a deeper survey, gains significant constraining power going to smaller scales. This is easier to see in the right panel of the same figure, where the SNR curves are normalised at $\lmax=450$. It is also important to highlight that our survey-overlap-removal strategy to avoid having to compute the covariance between the three data sets does not significantly impact the SNR (and hence the corresponding cosmological constraints), for both \desyt and \kidsot. Finally, by combining all three surveys we gain a $\sim 20\%$ increase on the SNR (at $\lmax=4500$) with respect to \desyt alone.
    
\section{Comparison with official results}\label{s:official}
  In this Section we assess the compatibility of our results with the official analyses of \desyt and \kidsot. The comparison with the \hscdro results was carried out in \cite{2210.13434,2010.09717}. Note that, although we have tried to match the analysis choices of the official analyses where possible, we have not been able to match every choice (e.g.: real vs. harmonic space, choice of samplers, sampled parameters, scale cuts, matter power spectrum modelling, etc.). 

  \begin{figure}
    \centering
    \includegraphics[width=\textwidth]{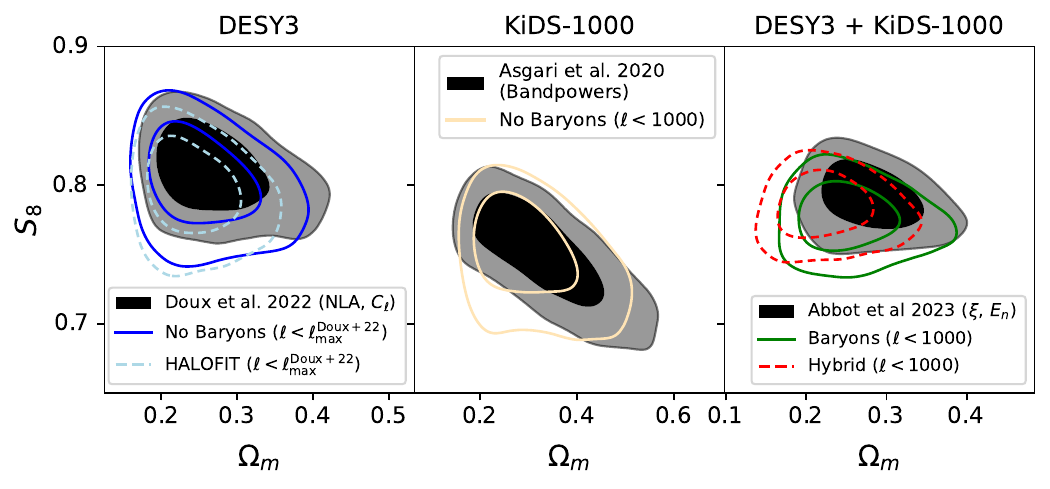}
    \caption{Cosmological parameters posterior distributions (68\% and 95\% C.L. regions) for \desyt (left), \kidsot (center) and their combination (right). Black contours show the results from the official analyses, while the results found with our pipeline are shown as coloured lines. The choice of statistic used in the official analyses is shown in the legend ($C_\ell$, correlation functions $\xi$, and COSEBIs $E_n$).
    \ul{Left panel:} in the case of \desyt we find good agreement when we use the  \baccoemu matter power spectrum  (or \hfit for a model closer to the official analysis) without accounting for baryonic effects, once we apply scale cuts similar to those in \cite{2203.07128}. The shift apparent for \hfit is likely due to the different effective scales probed: the different binning makes the scale cuts slightly different. This shift is then partially corrected by using the more accurate non-linear matter power spectrum of \baccoemu, which is known to shift $S_8$ to higher values \cite{2303.05537,2305.17173}.
    \ul{Centre:} our pipeline returns results that are compatible with the \kidsot analysis of \cite{2007.15633}. We compare the results with $\lmax=1000$ (note that the constraints with $\lmax=2048$ are virtually the same) in comparison with the official  $\lmax = 1500$. We find reasonable agreement despite the different sampled parameters and priors, IA modelling, binning and scale cuts. \ul{Right panel:} we compare our results for the combination of \desyt and \kidsot with the official analysis from \cite{2305.17173}, using real-space correlations and COSEBIs. The \texttt{Hybrid} constraints are obtained using \hmcode and sampling the parameters of the Hybrid analysis of \cite{2305.17173}, with their priors. The parameter shifts are compatible with the known discrepancies between real and harmonic-space analyses \cite{1906.06041, 2011.06469}.} 
    \label{fig:post.official}
  \end{figure}

  \paragraph{\desyt.} This is, to our knowledge, the first independent reanalysis of the publicly available \desyt at the catalogue level. We find negligible differences between the power spectra obtained with our pipeline and the official bandpowers of \cite{2203.07128} (see Fig.~\ref{fig:cl.des_official}). The two main differences between our measurements and the official ones are: first, we only account for the Gaussian and SSC contributions to the covariance matrix, whereas \cite{2203.07128} also includes the connected non-Gaussian covariance. As shown in \cite{1807.04266}, this is a good approximation for cosmic shear data, and should not impact the final results. Furthermore, we remove the large scales ($\ell < 20$), where the NKA has been shown to lose accuracy. Secondly, in order to go to smaller scales, we use a higher resolution map ($\nside = 4096$) to reanalyse the data. This allows us to go to scales as small as $\ell = 2\nside = 8192$. As a consequence, we also use a different $\ell$-binning, as shown in Fig.~\ref{fig:cl.des} and discussed in Section~\ref{ss:pseudo-cl}, more appropriate to the range of scales covered. In order to assess the impact of these differences at the level of the cosmological parameter constraints we compare our results applying similar scale cuts as those from the official analysis \cite{2203.07128}. The difference in binning and resolution means that the scale cuts cannot be exactly the same. The result can be seen in Fig.~\ref{fig:post.official}, where we find a good agreement with the official analysis. Interestingly, we find a better agreement using the non-linear matter power spectrum from \baccoemu ($0.2\sigma$ shift) than \hfit ($0.4\sigma$). This may be a consequence of the different effective scales probed, or of the different binning schemes (which may have different sensitivity to the detailed scale dependence of the $C_\ell$). The shift between \baccoemu and \hfit is compatible with the findings of \cite{2303.05537, 2305.17173}: a more precise model of the non-linear power spectrum leads to an upwards shift in $S_8$.

  \paragraph{\kidsot.} As with \desyt, we reanalyse \kidsot at the catalogue level. A similar reanalysis was carried out in \cite{2105.12108}. The main change with respect to that work is our inclusion of the SSC contribution to the covariance matrix. It was shown in \cite{2109.04458}, that the pseudo-$C_\ell$ estimator used here, as implemented in \nmt, is able to recover the official bandpower measurements of \cite{2007.15633} at high accuracy. As shown in the central panel of Fig.~\ref{fig:post.official} we obtain cosmological constraints that are in reasonable agreement with the official analysis of \cite{2007.15633}, despite sampling in different parameters, using a different IA model, and different binning and scale cuts. The main difference comes from a looser $S_8$, which is caused by using a more general IA model, including the time dependence of the IA amplitude (see Eq.~\ref{eq:ia}). This effect has been shown in previous works \cite{1906.09262,2105.12108,2305.17173}. Since the official analysis used $\lmax=1500$, we also check that we recover compatible results with $\lmax=2048$ (not shown in Fig.~\ref{fig:post.official} for clarity but in equally good agreement).

  \paragraph{\desyt + \kidsot.} Finally, we compare our reanalysis of the combined \desyt and \kidsot data sets with the official result of \cite{2305.17173}. As shown in Fig.~\ref{fig:post.official}, using the fiducial modelling of this paper (i.e. \baccoemu non-linear matter power spectrum and BCM baryonic feedback) we obtain constraints that are in good agreement with the official ones, albeit with a mild shift on $S_8$ of about $0.6\sigma$. Differences are expected due to several different analysis choices. First and most notably, we use different summary statistics: we reanalyze both \desyt and \kidsot in harmonic space, whereas \cite{2305.17173} keeps the fiducial summary statistics of each survey (real space for \desyt and COSEBIs for \kidsot). We follow the same procedure to avoid accounting for the cross-correlation between \desyt and \kidsot data sets by removing the overlapping areas of \kidsot from \desyt. This difference in the chosen summary statistics means that we cannot exactly match the scale cuts. Instead, we use $\lmax=1000$ for this comparison as an approximate scale cut, similar to those used in \cite{2305.17173}. The second major difference is the modelling of the matter power spectrum and baryon effects. Whereas \cite{2305.17173} used \hmcode for both the non-linear matter power spectrum and baryonic feedback, we use \baccoemu. \hmcode is in agreement with several emulators, including \baccoemu and EuclidEmulator \cite{1809.04695, 2010.11288}, at $<2.5\%$ down to $k < 10 h {\rm Mpc}^{-1}$ \citep{2009.01858}, whereas the agreement between \baccoemu and EuclidEmulator is at the $1\%$ level \citep{2010.11288}. In this work, we employ the version of \baccoemu first used in A23, which features a much larger cosmological volume thanks to an additional suite of simulations.  Therefore, one would not expect large differences in the parameter posterior distributions due to the modelling of the non-linear power spectrum. On the other hand, \hmcode models baryonic effects with a single parameter $T_{\rm AGN}$, calibrated against the BAHAMAS suite of simulations \cite{2009.01858}. Within \hmcode, changes in $T_{\rm AGN}$ are mapped onto changes on several parameters of the halo model sensitive to baryonic feedback. Instead, \baccoemu models baryonic effects with 7 free parameters, linked to physical properties of the gas as explained in Section~\ref{ss:baryons}. This allows for additional freedom in the model that may not be captured by the single parameter $T_{\rm AGN}$. Apart from the different parameters for the baryonic effects model, both analyses diverge on the choice of cosmological parameters to sample over. In \cite{2305.17173}, the sampled parameters are $\{S_8, h, \oc, \ob, \ns, m_\nu\}$, whereas in our analysis we sample over $\{\Om,\, \As,\, h,\, \Ob,\, \ns,\, \sum m_\nu\}$. The fact that we also marginalise over the multiplicative biases of \kidsot at the likelihood level is not expected to significantly impact our results, given their tight priors. As discussed in e.g. \cite{2305.17173}, the different choices of sampled cosmological parameters can produce noticeable changes in the posterior distribution. Ultimately, it is worth noting that the parameter shifts we observe are in agreement with the known discrepancies between real and harmonic space analyses (e.g. see \cite{1906.06041,2011.06469}). In order to quantify this further, we show in Fig.~\ref{fig:post.official} the results obtained using \hmcode, while sampling the parameters, with their priors, of the Hybrid analysis of \cite{2305.17173}: the constraints are in agreement with the official ones albeit a $0.3\sigma$ shift on $S_8$ and a $0.8\sigma$ shift on $\Om$. The extension to lower $\Om$ respect to our fiducial result, which lies between both results, is due to the effective change of priors when sampling our set of parameters, or those from \cite{2305.17173}. This means that the data is still not precise enough to be insensitive to the choice of priors. In conclusion, the differences seen between our analysis and the official ones are consistent with the different modeling, sampled parameters and, mainly, the different statistics used. 

\section{Fiducial results}\label{s:results}
  We now present the results of our fiducial analysis, which makes use of our angular power spectrum measurements up to $\lmax=4500$. This choice is motivated by attempting to match the smallest angle used in the real space analysis of A23, $\thetamin=2.5'$, using the approximate relation $\lmax=\pi/\thetamin$. As we will discuss later in this section (and as shown in Appendix~\ref{ap:real_vs_fourier}), this is not necessarily a good approximation: a real-space cut at scale $\theta$, contains information about harmonic scales beyond $\ell=\pi/\theta$, and vice-versa. Nevertheless, although our measurements are reliable up to $\lmax=8192$, we choose a more conservative scale cut to avoid, in as much as possible, real-space scales over which PSF systematics have been reported in \desyt \cite{2011.03408}. We will study the robustness of our results to this choice in Section~\ref{ss:lmax}.

  \begin{figure}
    \centering
    \includegraphics[width=0.495\linewidth]{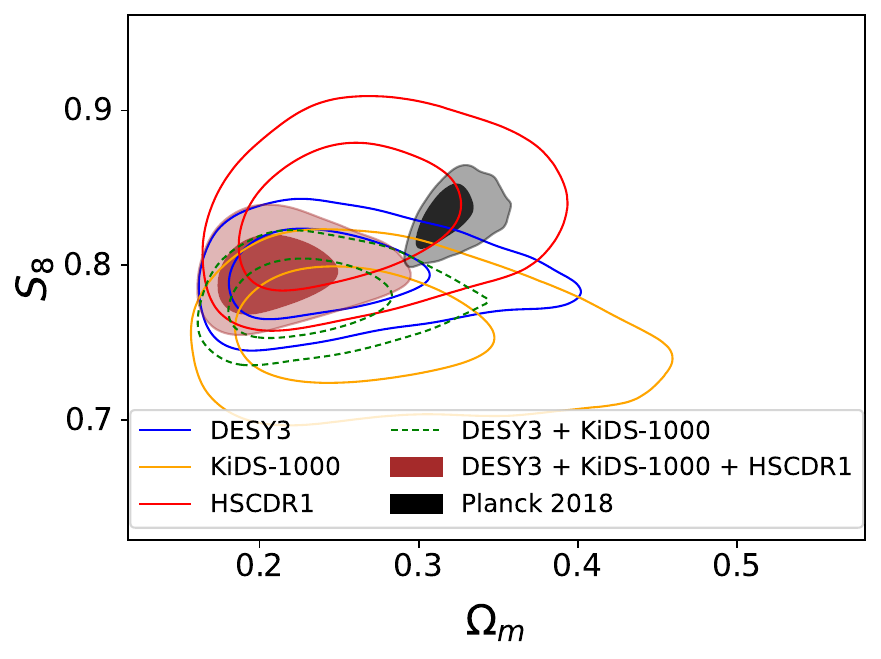}
    \includegraphics[width=0.495\linewidth]{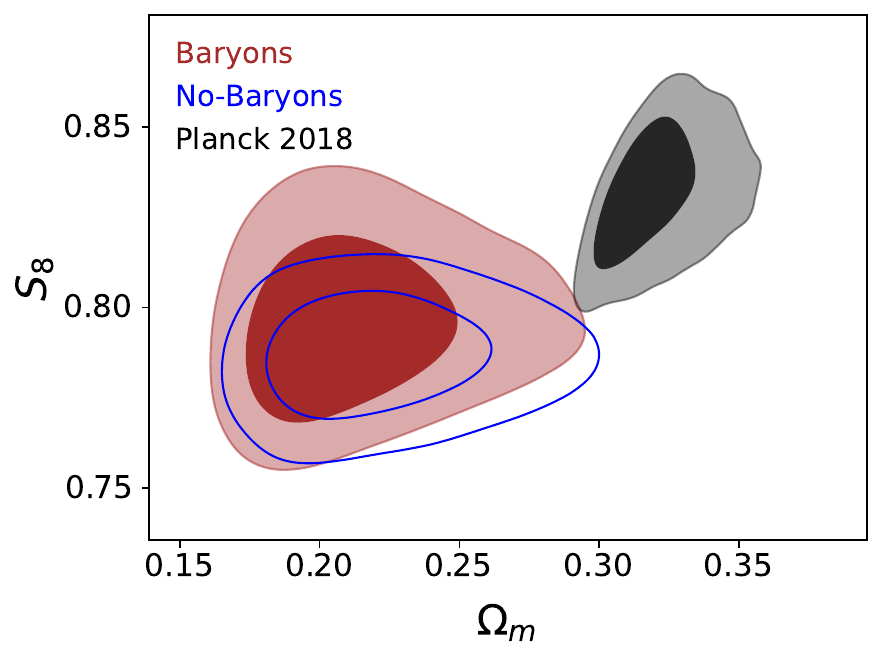}
    \caption{Cosmological constraints in the $(S_8,\Om)$ plane (68\% and 95\% C.L. regions) obtained with our fiducial scale cut $\lmax=4500$. \ul{Left panel:} constraints from \desyt, \kidsot, \hscdro, \desyt+\kidsot and \desyt+\kidsot+\hscdro, using \baccoemu matter power spectrum and modelling baryonic effects. The final constraints on $S_8$ are very similar to those found by \desyt, since it dominates the overall SNR. \ul{Right panel:} constraints from the combination of all data sets with and without modelling baryonic effects. In both panels, the results from \planck ($TT$, $TE$, $EE$+lowE+lensing, \cite{1807.06209}) are shown in black. The inclusion of baryons broadens the posterior distribution of $S_8$, allowing it to take larger values, and thus reducing the level of tension with \planck to $1.8\sigma$. However, we recover a value of $\Om$ that is in tension with \planck at the $\sim3\sigma$ level. This trend is not associated with any particular survey, and only arises from the combination of all of them (especially after \hscdro has been added to \desyt and \kidsot).}
    \label{fig:deskidshsc.post.comb}
  \end{figure}

  Fig.~\ref{fig:deskidshsc.post.comb} shows the posterior distribution in the $S_8$-$\Om$ plane, the parameters weak lensing is most sensitive to. We find that the constraints found from different surveys are in reasonable agreement with one another. Their combination leads to an improvement in the uncertainty on $S_8$ by a factor 0.1, 0.4, and 0.5, for \desyt, \kidsot, and \hscdro, respectively. In turn, the improvement factor for $\Om$ is 0.4 for \desyt, and 0.5 for both \kidsot and \hscdro. As expected, given our discussion of the SNR (see Fig.~\ref{fig:snr}), the constraints are mainly driven by \desyt, particularly in the case of $S_8$. On the other hand, we see that the constraint on $\Om$ gets significantly tighter when combining our datasets. Overall, we find that \lcdm is an excellent fit to our data, with $\chi^2 = 312$ ($\Nd = 300$, $p = 0.28$) for \desyt, $\chi^2 = 403$ ($\Nd = 420$, $p = 0.69$) for \kidsot, $\chi^2 = 105$ ($\Nd = 100$, $p = 0.31$) for \hscdro and $\chi^2 = 841$ ($\Nd = 820$, $p = 0.28$) for the combination of all probes.

  \paragraph{Constraints on $S_8$.} The right panel of Fig.~\ref{fig:deskidshsc.post.comb} shows the posterior constraints on $S_8$ and $\Om$ plane obtained in our fiducial case (burgundy) and when completely ignoring baryonic effects (blue), compared to the CMB results from \planck, with varying neutrino mass \cite{1807.06209} (black). In our fiducial case, we infer $S_8 = 0.795^{+0.015}_{-0.017}$, lower than the \planck value but in statistical agreement ($\Delta S_8 = 1.8\sigma$\footnote{We define the parameter shifts and tensions based on their mean posterior value and assuming the distributions to be Gaussian and uncorrelated. Hence we calculate $\Delta X [\sigma] = (\langle X_{A} \rangle - \langle X_{B} \rangle) / \sqrt{\sigma_{X,A}^2 + \sigma_{X,B}^2}$, for a given parameter $X$ in the cases $A$ and $B$.}). When ignoring baryonic effects, we obtain instead $0.787\pm 0.011$, lower than \planck at $2.6\sigma$. Marginalising over baryonic effects with the BCM model as implemented in \baccoemu produces a $0.4\sigma$ upwards shift in the posterior mean of $S_8$, accompanied by a $\sim30\%$ broadening of the uncertainties. Both conspire towards improving the agreement with \planck. As discussed in \cite{2303.05537}, the $S_8$ tension is also alleviated by the use of a more accurate model of the non-linear matter power spectrum than \hfit.

  \paragraph{Constraint on $\Om$.} As shown in the left panel of Fig. \ref{fig:deskidshsc.post.comb}, the three cosmic shear datasets used here recover constraints on $\Om$ that, while not individually in tension with \planck, lie consistently below the CMB measurements. Combining all three surveys thus dramatically reduces the upper bound on $\Om$, and we infer $\Om = 0.212^{+0.017}_{-0.035}$, lower than the \planck value by $3.5\sigma$ (in this case, slightly enhanced by the inclusion of baryonic effects). Importantly, this is not driven by a single survey, but emerges from the combination of all of them. We have verified that this tension is driven by the data, and not caused by measurement systematics (e.g. pixelisation effects), residual stochastic noise in the power spectra, or projection effects due to the lack of constraining power over $\Ob$. In particular, using a BBN prior on this parameter does not alleviate the tension. It is worth noting that, interestingly, we find that the uncertainty on $\Om$ depends on the location of its best fit value in the $(S_8,\Om)$ plane. As shown in Appendix \ref{ap:mock}, analysing a mock dataset centred at our best-fit parameters leads to constraints on $\Om$ that are $\sim30\%$ tighter than those found for data centred at the \planck cosmology (using the same covariance as our data in both cases). 

  \begin{figure}
    \centering
    \includegraphics[width=\linewidth]{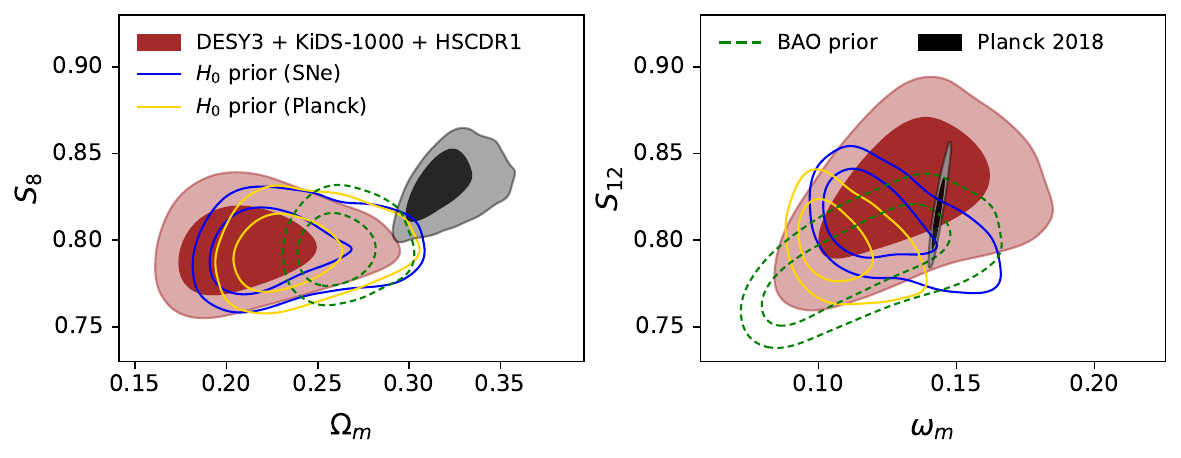}
    \caption{Posterior distributions in the $(S_8,\Om)$ (left) and $(S_{12},\om)$ (right) planes obtained with the fiducial scale cut $\lmax=4500$, and including different external priors (see main text for details).}
    \label{fig:deskidshsc.post.S12}
  \end{figure}

  In order to further investigate this potential $\Om$ tension, we study our constraints in different parameter spaces. In particular, following the rationale in \cite{2002.07829}, we avoid the use of parameters whose definition depends on the value of the Hubble constant $h$, using $\om\equiv\Om h^2$, and $S_{12} = \sigma_{12} (\om/0.14)^{0.4}$, where $\sigma_{12}$ is the standard deviation of the linear matter overdensity on spheres of radius $R=12\,{\rm Mpc}$ (note the ``non-$h$'' units). Although cosmic shear data can obtain tight constraints on the overall amplitude of matter fluctuations, they are weakly sensitive to the current expansion rate $h$. Thus, as discussed in \cite{2002.07829}, the marginalisation over the unconstrained $h$ mixes the value of $\sigma_8$ corresponding to a wide range of physical radii (since $\sigma_8$ is defined in terms of spheres of constant radius in units of ${\rm Mpc}\,h^{-1}$). Within our prior on $h$, a scale of $8\,{\rm Mpc}\,h^{-1}$ corresponds to radii from $\sim9$ to $\sim15\,{\rm Mpc}$. In the case of \planck, which is able to obtain tight constraints on $h$ ($h=0.6736\pm0.0054$), the radii are much better defined $R_8 = 11.88 \pm 0.095$ Mpc. The results are shown in Fig.~\ref{fig:deskidshsc.post.S12}, with the left and right panels showing the constraints in the $(S_8,\Om)$ and $(S_{12},\om)$ planes, respectively. We find that, in the $(S_{12},\om)$ plane, our constraints are in good agreement with \planck. Overlaid on our fiducial constraints and those from \planck we show the constraints obtained imposing a prior on $h$ using the \planck measurement (yellow), and the local measurement from SH0ES \cite{2112.04510} ($h=0.7304 \pm 0.0104$, blue). Focusing on the right panel, we can see that our constraints in the $(S_{12},\om)$ plane improve dramatically in the presence of an $h$ prior. Interestingly, as was pointed out in \cite{2002.07829, 2209.12997}, using the \planck prior on $h$ we recover again the tension with \planck, whereas using the SH0ES prior restores the concordance with \planck. A similar result was also observed in \cite{2306.17748}. Thus, as discussed in \cite{2002.07829, 2209.12997}, tensions in the $(S_8,\Om)$ plane can seemingly be recast in terms of the $H_0$ tension. It is worth noting, however, that the fit to the data is still good in both cases with a $\chi^2 = 842.1$ ($\Nd = 820$, $p = 0.27$) with \planck's prior, and $\chi^2 = 841.6$ ($\Nd = 820$, $p = 0.28$), with SH0ES' prior. These values are only mildly higher than the fiducial result ($\chi^2 = 841.1$).

  To further examine the dependence of our results on external priors constraining the background evolution (or, in particular, the distance-redshift relation), we repeated our analysis including the Baryon Acoustic Oscillation measurements of the Sloan Digital Sky Survey (SDSS) DR16  \cite{1607.03155, 2007.08993}, which are sensitive to $\Om$ and $h$ in a different combination, and thus intersect the $(S_8,\Om,h)$ and $(S_{12},\om,h)$ hyperplanes in a different direction to the $h$-priors. The results are shown as green dashed lines in Figure~\ref{fig:deskidshsc.post.S12}. The BAO prior shifts the value of $\Om$ towards \planck, reducing the level of tension to $2.8\sigma$. This is done without worsening the fit to the weak lensing data. The (non-reduced) $\chi^2$ value for the weak lensing data vector grows only by $\Delta\chi^2=0.3$, with a negligible decrease in the associated $p$-value of $\Delta p=0.002$. The BAO prior also retains the good agreement with \planck in the $(S_{12},\om)$ plane (left panel of Fig. \ref{fig:deskidshsc.post.S12}). This highlights that the tension we see in $\Om$, although quantified at $3.5\sigma$, may be caused by a prior volume effect.  To better quantify the relevance of this tension, taking into account the full multi-dimensional posterior distribution and not only the $\Om$ 1D-projection, we compute the ``suspiciousness'' metric, defined in \cite{2007.08496,2102.11511}. The result, $\log(S) = -3.9$, corresponds to a $2-3\sigma$ tension between the weak lensing and \planck data sets depending on the method used to estimate the dimensionality of the model (see Appendix~\ref{ap:suspiciousness} for details). As before, this points towards a better agreement between weak lensing and \planck data than is apparent from the $\Om$ plane.

  \begin{figure}
    \centering
    \includegraphics[width=0.495\textwidth]{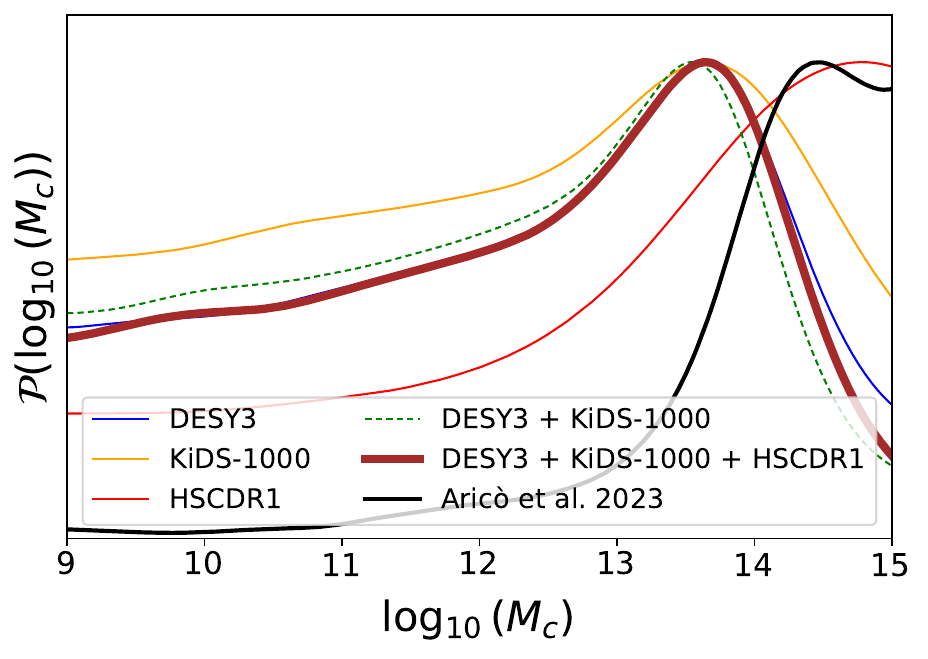}
    \includegraphics[width=0.495\textwidth]{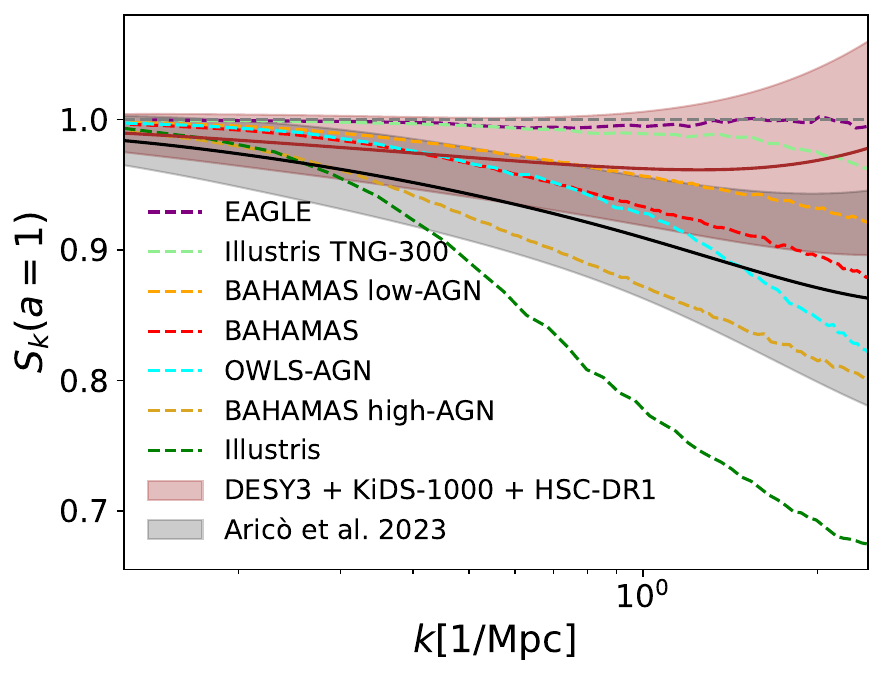}
    \caption{\ul{Left}. Constraints on the $\logMc$ BCM parameters from different combinations of data sets. \ul{Right}. Reconstruction of the baryonic boost ($1\sigma$ level). We compare our results (brown) to those obtained in \cite{2303.05537} (black) using our same modelling but with \desyt real space correlation functions and results from some of the main hydrosimulations (dashed lines) as fitted by \cite{2011.15018}. The lack of strong preference for the modelling of baryonic effects translates here into a $S_k$ compatible with 1 (see Tab.~\ref{tab:diff}). It is worth mentioning here that our results are mostly driven by \desyt only.} 
    \label{fig:boost.baccoemu}
  \end{figure}

  \paragraph{Constraints on baryonic effects.} We exploit \baccoemu to provide independent constraints on the strength of baryonic effects using our weak lensing datasets. In Fig.~\ref{fig:boost.baccoemu} we show both the posterior distribution of $\logMc$, the only BCM parameter that our data can constrain, and the constraint on the baryonic boost, $S_k$, defined as the ratio between matter power spectrum including baryonic physics and the gravity-only one. We find $\logMc \sim 14$ when combining all probes\footnote{Since $\logMc$ is widely prior dominated, the mean and error are not very informative. Instead, we report the approximate value of their posterior distribution maximum.}. As expected, \desyt is the most constraining survey and drives the final constraints. Importantly, the three surveys yield compatible results, with \kidsot and \desyt peaking on roughly the same $\logMc$. Interestingly, \hscdro prefers a slightly higher value of $\logMc$, and therefore more baryonic feedback, than \kidsot and \desyt. \hscdro is a deeper survey and is thus more sensitive to higher redshifts and smaller scales, which explains the different sensitivity to baryonic effects. Alternatively, it could also be that higher redshift data prefer a different value of $\logMc$, as seen e.g. in \cite{1911.08471}. The BCM parametrisation used here will not be able to fully capture this evolution, but the impact on the final constraints should be negligible, given the mild sensitivity to baryonic effects and the overall dominance of \desyt on the final constraints. Finally, the \hscdro posterior is in good agreement with the \desyt real-space analysis of A23 ($\logMc = 14.38^{+0.60}_{-0.56}$, see also \cite{2206.08591}), whereas our fiducial $\logMc$, driven by \kidsot and \desyt is lower but still statistically compatible. For our fiducial scale cuts, our constraining power is limited and we do not see a clear departure from $S_k = 1$ (i.e. no baryon suppression), in contrast to A23. At $k=2.5$ Mpc$^{-1}$ we find $S_k = 0.980^{+0.097}_{-0.078}$. This lack of clear support for a larger baryon suppression is in agreement with the negligible change in the best fit $\chi^2$ with and without baryonic effects: $\Delta \chi^2 = -0.3$, out of a $\chi^2 \sim 841$ ($\Nd = 820$, $p=0.3$). This is also evident in Figs.~\ref{fig:cl.des}, \ref{fig:cl.kids}, and \ref{fig:cl.hsc}, where we show the best-fit angular power spectra obtained with and without baryonic effects, along with the data for each survey.
  
  In the right panel of Fig.~\ref{fig:boost.baccoemu}, we compare the baryonic boost we infer from data with the one predicted by several cosmological hydrodynamical simulations. Given our constraining power, we find good agreement with most simulations, except those predicting extreme levels of baryonic suppression. Simulations that predict weaker baryonic effects, such as EAGLE \cite{1407.7040, 1501.01311} or Illustris TNG-300 \cite{1707.03395,1707.03396, 1707.03406,1707.03401,1707.03397} all lie within our $68\%$ bounds. Simulations with stronger AGN feedback, such as BAHAMAS \cite{1603.02702, 1712.02411}, and OWLS \cite{0909.5196, 1104.1174}, lie between the $68\%$ and $95\%$ bounds of our constraints on scales $k\gtrsim1 \, {\rm Mpc}^{-1}$. Finally, simulations with extreme levels of baryonic suppression, such as Illustris \cite{1405.2921}, which predict a $30\%$ suppression at $k\sim 2\,{\rm Mpc}^{-1}$, are more clearly ruled out by our data.

  As mentioned above, the baryonic suppression we find here is weaker than that obtained in A23 using \desyt correlation functions, raising questions about the compatibility of both analyses. In order to assess the compatibility, and make sure both pipelines are in agreement, we check that A23 best fit is a good fit to our \desyt harmonic space data, and that our fiducial best fit for \desyt is a good fit to A23 real space data, using each pipeline respectively. Specifically, using the best fit model obtained in A23, we find that the fit to the harmonic space data with our pipeline gives $\chi^2 = 323$ ($\Nd = 300$, $p = 0.15$) and, therefore, is a good fit. Similarly, the A23 pipeline, using correlation functions, obtains $\chi^2 = 419$ ($\Nd = 400$, $p = 0.22$) with our \desyt best fit parameters, also a good fit. It is a good sign, however, that in both cases, each statistic still prefers its respective best-fitting model: $\chi^2 = 312$ ($\Nd = 300$, $p=0.28$) with our pipeline and harmonic space, and $\chi^2 = 414$ ($\Nd = 400$, $p = 0.27$) with A23 pipeline and in real space. 
  Second, we have checked that both pipelines produce angular power spectra that deviate less than 1$\sigma$; i.e. always fully compatible within the data uncertainties, with the largest deviation coming from small scales, where the matter power spectrum is extrapolated in different ways and the errors of the measurements are quite large. The difference must then come from the difference between angular power spectra and correlation functions. In fact, as warned at the beginning of this section, we see in Fig.~\ref{fig:xi.int} that, indeed, our scale cut $\lmax=4500$ corresponds to a larger scale than the one the correlation function $\xi_-$ is sensitive to at $\theta=2.5'$.  However, we see that, given the size of the error bars of $\xi_-(2.5')$, the difference in scales might not be the only reason. We leave for future work a systematic study of the differences between real- and harmonic-space measurements that might cause the different sensitivity to baryonic effects.  We thus conclude that this is likely due to the different sensitivity of real- and harmonic-space measurements to different scales, and is similar in nature to the cosmological parameter shifts observed discussed in Section~\ref{s:official} 

  Finally, we report here the value of our best fit parameters with $\chi^2 = 841$ ($\Nd = 820$, $p = 0.28$): 
    \begin{itemize}
        \item {\bf Cosmological parameters:} 
          $\As = 4.93\times 10^{-9}$, $\Om = 0.186$, $\Ob = 0.0514$, $h = 0.851$, $\ns = 1.058$, $\sum m_\nu = 0.061\,{\rm eV}$, with derived parameters $\sigma_8 = 1.029$ and $\sigma_{12} = 0.880$.
        \item {\bf Baryon parameters:} 
        $\logMc = 13.35$, $\logeta = 0.0446$, $\logbeta = -0.0388$, $\logMz = 10.19$, $\logThetaI = -1.450$, $\logThetaO = 0.319$, $\logMi = 12.80$ .
        \item {\bf Multiplicative biases:} $m^{\rm \desyt} = \{ -0.0014, -0.0125, -0.0280, -0.0458 \}$, $m^{\rm \kidsot} = \{-0.0290, 0.0047, 0.0178, -0.0148, 0.0028\}$, $m^{\rm \hscdro} = \{-0.0151, 0.0084, 0.0018, 0.0103\}$
        \item {\bf Redshift shifts:} $\Delta z^{\rm \desyt} = \{-0.0063, -0.0100, -0.0077, 0.0066\}$, $\Delta z^{\rm \kidsot} = \{-0.0231, 0.0250, -0.0174, -0.0288, -0.00422\}$  and $\Delta z^{\rm \hscdro} = \{0.00982, 0.00880,$ $0.0944, 0.0127\}$ 
        \item {\bf Intrinsic Alignments:} $\{A_{\rm IA}, \eta_{\rm IA}\}$ = $\{-0.709, 4.480\}$ for \desyt, $\{1.052, 2.733\}$ for \kidsot and $\{0.469, -0.987\}$ for \hscdro.
    \end{itemize}

  \paragraph{Summary.} In summary, we provide the first cosmological constraints from the combination of \desyt, \kidsot and \hscdro, down to $\lmax=4500$, with a careful modelling of baryonic effects. This results in $S_8 = 0.795^{+0.015}_{-0.017}$ , $\Om = 0.212^{+0.017}_{-0.035}$, and $\logMc \sim 14$. The resulting $S_8$ constraints are compatible with \planck within $1.8\sigma$, but we find an apparent tension in $\Om$ at the $3.5\sigma$ level. This tension is alleviated by introducing priors on the background expansion (e.g. from BAO data) without worsening the fit to the weak lensing data. Furthermore, we show that the tension disappears in the $(S_{12},\om)$ space, and may instead be recast in terms of an $H_0$ tension. Through a joint analysis with \planck, we measure the suspiciousness of our results to be $\log(S) = -3.9$, corresponding to a statistical tension at the level of $2$-$3\sigma$. Finally, we do not find strong evidence of baryonic suppression on the scales probed when marginalising over both cosmology and baryonic effects. Nevertheless, the inclusion of baryonic effects in the model is vital to provide robustness to the final cosmological constraints, given the currently large uncertainties on their impact. These must be fully propagated to avoid introducing biases on the cosmological parameters, or under-estimating their uncertainties. 

\section{Robustness to modelling and analysis choices}\label{s:robustness}
  In this Section we study the impact of the most relevant aspects of our analysis to our final results. In particular, we will investigate how our results depend on the scale cut $\lmax$, how the cosmological parameter constraints change by fixing or imposing strong priors on the BCM parameters, how the constraints on the BCM parameters from weak lensing change if we fix the cosmology, and how our results depend on the choice of parametrisation to describe baryonic effects.
  \begin{figure}
    \centering
    \includegraphics[width=\textwidth]{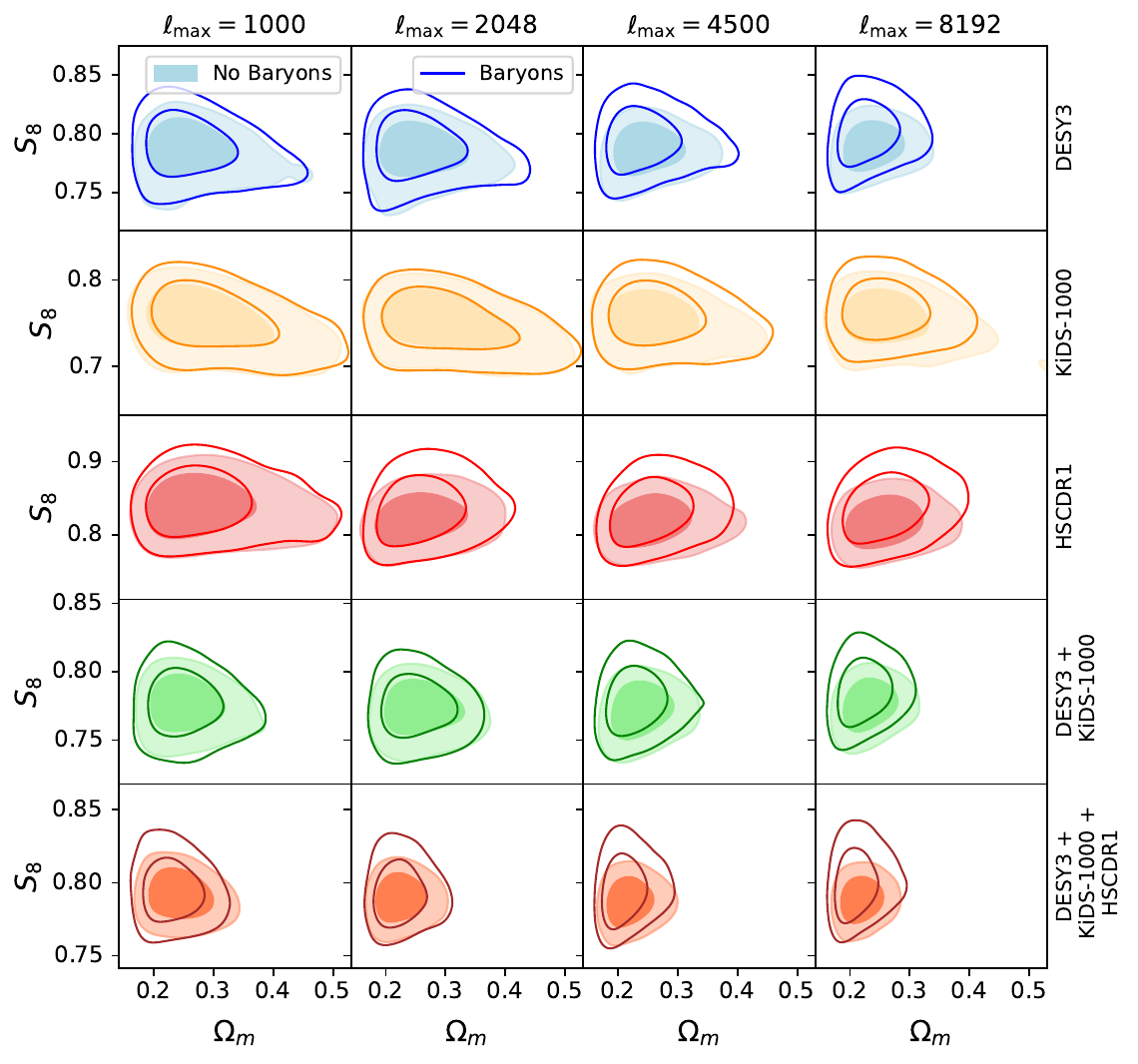}
    \caption{Posterior distribution for \desyt (blue), \kidsot (orange), \hscdro (red), the combination of \desyt and \kidsot (green) and the combination of all surveys (brown). The solid lines show the constraints found when marginalising over baryonic effects, while the filled contours show the result of ignoring baryonic effects altogether. Going from left to right, results are shown for different scale cuts: $\lmax=1000$, $\lmax=2048$, $\lmax=4500$ (fiducial), and $\lmax=8192$. We see little effect up to, and including, $\lmax=2048$ for \desyt and \kidsot, but not for \hscdro, given its higher number density, which makes it more sensitive to small scales, coupled with its stricter large-scale cut ($\lmin=300$), which removes the anchor to the amplitude of the fluctuations. For \desyt, \kidsot, and their combination, modelling baryonic effects becomes important at $\lmax=4500$, whereas for the combination of all surveys, baryonic effects become relevant already for $\lmax=1000$.}
    \label{fig:baryons.nobaryons}
  \end{figure}

  \subsection{Scale cuts}\label{ss:lmax}
\begin{figure}
      \centering
      \includegraphics[width=\textwidth]{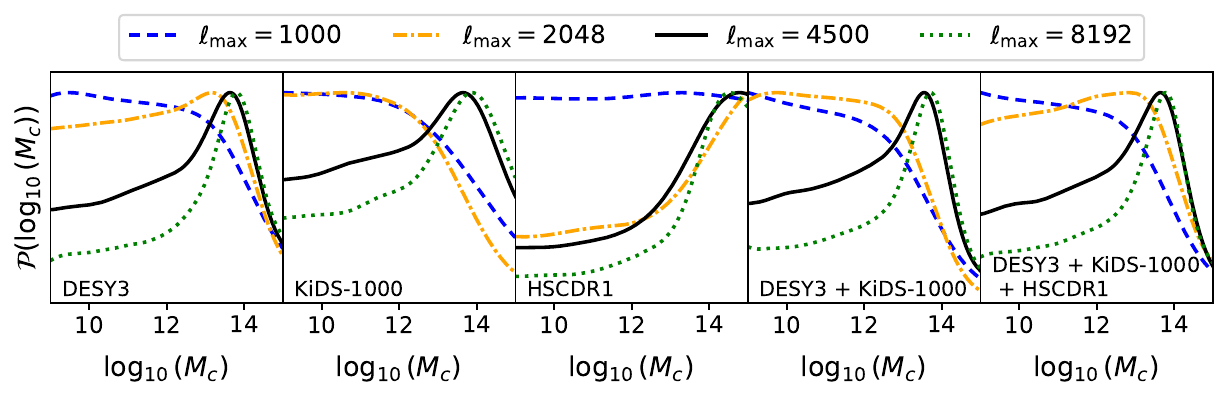}
      \caption{1D posterior distributions of the BCM parameter $\logMc$ from different data combinations and scale cuts. We are only able to start placing meaningful constraints on $\logMc$ at our fiducial scale cut $\lmax=4500$.}
      \label{fig:deskidshsc.post.Mc}
    \end{figure}

    Fig.~\ref{fig:baryons.nobaryons} shows the posterior distribution of $S_8$ and $\Om$ as a function of $\lmax$, with and without baryonic effects modelling for \desyt, \kidsot, \hscdro, and their combination. Similar results for $\logMc$, the only BCM parameter we are mildly sensitive to, are shown in Fig. \ref{fig:deskidshsc.post.Mc}.
    
    When modelling baryonic effects, $S_8$ constraints do not improve significantly by going to small scales. In turn, the constraints on $\Om$ and $\logMc$ improve significantly: the uncertainty in $\Om$ is reduced by $\sim30\%$, and $\logMc$ goes from being unconstrained to $\logMc \sim 14$. Specifically, this preference is only evident when $\lmax\gtrsim 4500$ (see Fig.~\ref{fig:deskidshsc.post.Mc}). Including smaller scales pushes $\Om$ towards lower values, leading to the $\Om$ tension with \planck discussed in Section \ref{s:results}. At small scales, $\Om$ changes the tilt of the power spectra in a way that is no longer degenerated with $\sigma_8$, increasing the constraining power over it. \desyt shows a similar behaviour for $\Om$ and $\logMc$. In \kidsot, we find that the main improvement in the $\Om$ constraints occurs when including scales up to $\lmax=4500$, whereas in \hscdro, this happens already at $\lmax=2048$, and gains are marginal when going to smaller scales. Looking at baryonic effects, \desyt and \kidsot start constraining $\logMc$ with $\lmax\gtrsim 4500$. On the other hand, we find \hsc is able to start constraining $\logMc$ already at $\lmax=2048$. The reason for this is twofold. First, \hscdro is a deeper survey and can extract more information by going to smaller, scales as seen in Fig.~\ref{fig:snr}. Second, by removing the largest scales ($\lmin=300$) we lose the anchor of the overall amplitude of the matter fluctuations. These two effects make \hscdro the survey most sensitive to baryonic effects. However, its absolute constraining power is considerably lower than both \desyt and \kidsot and, therefore, its impact on the final constraints is small.

    We can use these results to determine the scale $\lmax$ at which modelling baryonic effects becomes a requirement to obtain robust posterior distributions. As shown in Fig.~\ref{fig:baryons.nobaryons}, \kidsot seems to be the dataset with the smallest sensitivity to baryonic effects, in spite of recovering the lowest value of $S_8$ out of all three surveys when baryonic effects are ignored. Even at $\lmax = 8192$ the bias is negligible ($0.3\sigma$ shift in $S_8$), and the recovered uncertainties are only slightly underestimated (by 6\% in $S_8$). In the case of \desyt, with a higher SNR, the effect is more important, but still small, even at $\lmax=2048$, with just slightly smaller error bars for $S_8$ (by 10\%), but with no significant bias ($0.2\sigma$). The effect is larger at $\lmax=4500$ (17\% on $S_8$), and certainly significant when considering scales up to $\lmax=8192$ ($0.5\sigma$ shift on $S_8$, and $20\%$ smaller errors).  In contrast, \hscdro is significantly more sensitive to the inclusion of baryonic effects in the modelling. For $\lmax\geq 2048$, we find that ignoring baryonic effects leads to a consistent under-estimation of the $S_8$ errors by $\sim22\%$ overconstrained, in agreement with the official analysis of \cite{1809.09148}. However, we observe no major shift at any scale, the maximum being $0.5\sigma$ at $\lmax=8192$.  Looking now at the combination of \desyt and \kidsot, we see that modelling baryons become necessary at $\lmax\gtrsim 4500$. At $\lmax=4500$, the impact is a downwards $0.3\sigma$ shift on $S_8$, becoming $0.5\sigma$ at $\lmax=8192$. Finally, for the combination of the three weak lensing surveys, we see a clear overestimation of the $S_8$ errors even at $\lmax=1000$ and a shift of $0.4\sigma$ at already $\lmax=4500$, growing to $0.6\sigma$ at $\lmax=8192$. In summary, the main impact of neglecting baryonic effects on the scales explored here (up to $\lmax=8192$) is a downward shift in $S_8$ of up to $\sim0.5\sigma$, and an under-estimation of its uncertainties by $\sim30\%$. This is in good agreement with the findings of A23. A quantitative summary of the effect of not including baryonic effects in the analysis can be found in Tab.~\ref{tab:diff}.

  \subsection{Fixing baryonic parameters}\label{ss:fixbaryons}
    \begin{figure}
      \centering
      \begin{minipage}{0.52\linewidth}
        \centering
        \includegraphics[width=\linewidth]{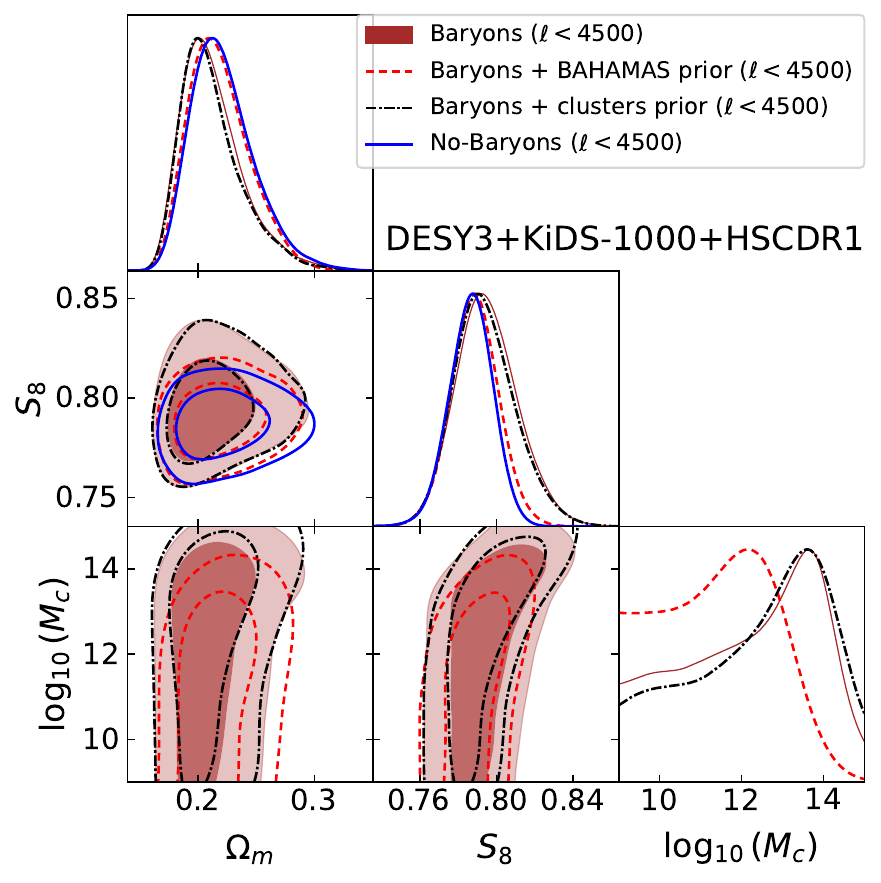}
      \end{minipage}
      \begin{minipage}{0.45\linewidth}
        \centering
        \includegraphics[width=\textwidth]{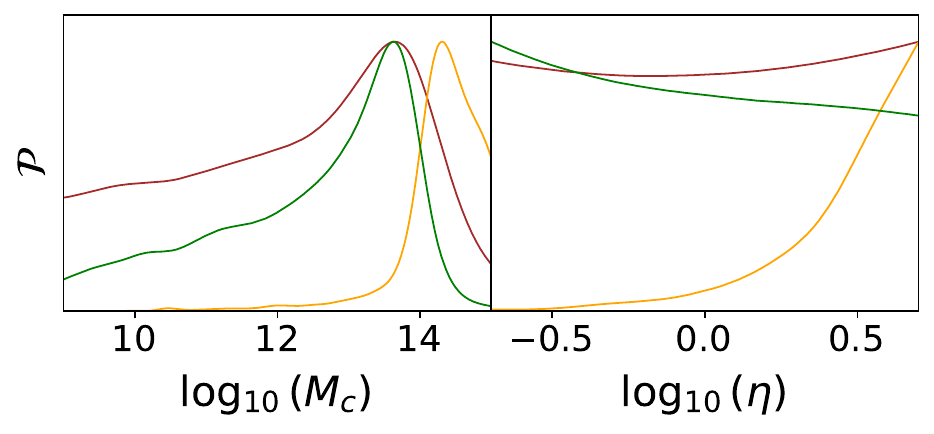}
        \includegraphics[width=\textwidth]{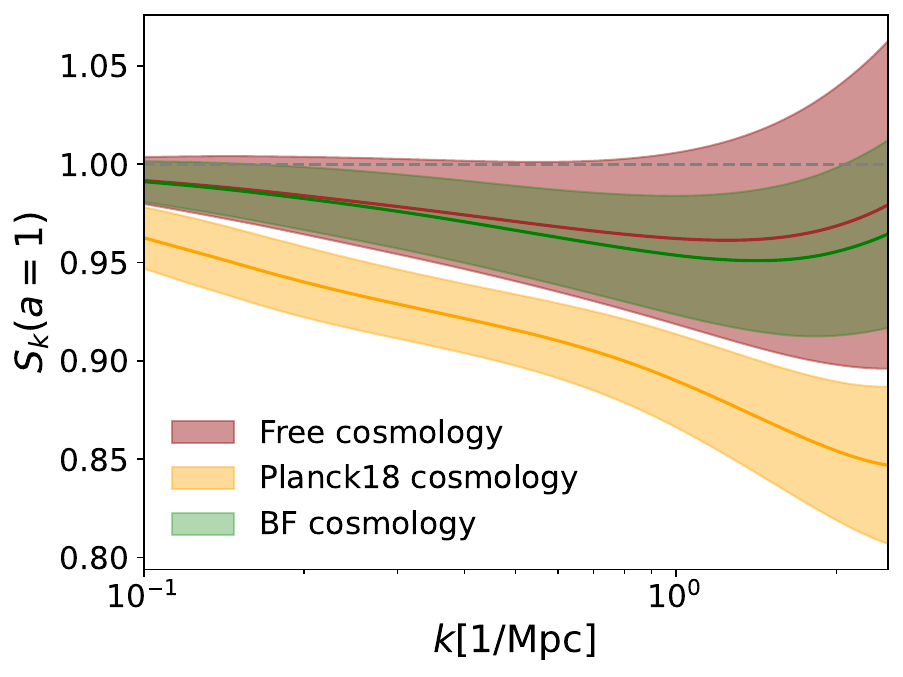}
      \end{minipage}
      \caption{\ul{Left panel:} constraints on $S_8$, $\Om$, and $\logMc$ (68\% and 95\% C.L.) in our fiducial run, freeing up all the BCM parameters free (burgundy contours). We over-plot the result of fixing all BCM parameters except $\logMc$ to the best-fit values from BAHAMAS (red dashed), as well as the result of placing a prior on $\logbeta$ and $\logMz$ from cluster measurements in \cite{2309.02920} (black). For comparison, we also show the result of ignoring baryonic effects (blue). \ul{Right panel:} Constraints on the BCM parameters $\logMc$ and $\logeta$ in our fiducial analysis (burgundy), when fixing the cosmological parameters to the \planck best-fit (orange), and when fixing them to our best fit $\Lambda$CDM model (green). The lower panel shows the resulting $1\sigma$ constraints on the baryonic suppression factor $S_k$. Fixing the cosmology to \planck leads to a preference for a significantly higher $\logMc$, which causes a dramatically stronger baryonic suppression. This is necessary in order to accommodate the lower values of $S_8$ and $\Om$ found in our analysis. Note that, nevertheless, the resulting best-fit model is still a good fit to the data.
      }
      \label{fig:fixing.params}
    \end{figure}

    It is interesting to explore the effect which tight priors on the BCM parameters may have on our cosmological constraints. This can be achieved by combining, in the same analysis, direct probes of the gas, such as thermal and kinematic Sunyaev-Zel'dovich cross-correlations \cite{2009.05558,2109.04458,2101.08373} or X-rays maps \cite{2309.11129}. Alternatively, the impact of baryonic effects may be constrained from direct measurements of the bound gas mass fraction in clusters \cite{1911.08494,2110.02228,2309.02920}. These additional measurements can put tight effective priors on the baryonic effect parameters, beyond the constraints achievable by weak lensing alone (for which baryonic feedback is a secondary effect). Placing priors on the BCM parameters, we are also able to explore the robustness of our cosmological constraints. Here we explore two different scenarios:
    \begin{itemize}
        \item First, we fix the BCM parameters that we are not able to constrain (i.e. all except $\logMc$), reducing the model dimensionality by 6. In this case, which we will label the ``BAHAMAS prior'', we fix these parameters to the values found by \cite{2011.15018} to provide a good fit to the power spectrum in the BAHAMAS simulations: $\logeta=-0.33$, $\logbeta=-0.28$, $\logMz=10.21$, $\logThetaO = 0.12$, $\logThetaI=-0.62$  and $\logMi=9.95$. Note that setting these parameters while leaving $\logMc$ free still leaves significant freedom for the possible values of the baryonic suppression factor, since its amplitude is largely controlled by this parameter.
        \item In the second case, we use the measurements of $\logbeta$ and $\logMz$ (but not $\logMc$, as our data is sensitive to it) obtained using cluster observations from \cite{2309.02920}. We thus impose Gaussian priors of the form $\logbeta = -0.29 \pm 0.11$ and $\logMz = 12.00 \pm 0.06$, as measured in \cite{2309.02920}, and leave the other parameters free with the fiducial priors of Table~\ref{tab:priors}. We label this scenario ``clusters prior'' in what follows.
    \end{itemize}
    The results of this study can be found in Fig.~\ref{fig:fixing.params}. First, we note that applying the BAHAMAS prior and fixing 6 of the 7 BCM parameters shrinks the posteriors of the cosmological parameters, almost matching those obtained when baryonic effects are ignored. In particular, the marginalisation over the unconstrained baryonic parameters points towards lower $\Om$ (enhancing slightly the $\Om$ tension with \planck discussed in the previous section), higher $S_8$ (in this case, reducing the tension with \planck), and higher $\logMc$. When fixing the other BCM parameters to the BAHAMAS best fit, we recover a preference for lower values of $\logMc$, with the posterior peaking at $\logMc\sim12$, and decaying sharply above $\logMc\sim13.5$, corresponding to rather weak baryonic suppression. This is probably caused by the fact that by fixing the BCM parameters to BAHAMAS, we might be favouring a too strong baryon suppression, not compatible with our data. As a consequence, $\logMc$ moves towards lower values to reduce the overall impact of baryonic effects. In any case, the results are not incompatible with our fiducial analysis, given our inability to place a lower bound on $\logMc$ with our harmonic-space scale cuts.

    On the other hand, when we impose Gaussian priors for $\logbeta$ and $\logMz$ from the clusters analysis of \cite{2309.02920}, we recover almost the same constraints as when marginalising over all the BCM parameters with flat priors. This is due to the fact that, as in our fiducial case, these priors are not informative in the most relevant parameters: $\logMc$ and $\logeta$. In fact, once again we cannot constrain the latter, which, after $\logMc$, has the largest impact on the baryonic suppression at large scales (since it parametrises the distance out to which gas is ejected by the central AGN). It is worth noting that our prior on this parameter is rather conservative $\logeta \in [-0.7,0.7]$, allowing values far beyond those preferred by hydrodynamical simulations, $\logeta = -0.32 \pm 0.22$ (see e.g. \cite{2011.15018,2309.02920}). 
    To verify if such a large prior might induce a volume effect on our results, we repeated our analysis imposing a tighter prior $-0.7 < \logeta < 0$. We find negligible changes in the posteriors distributions, of all parameters (including $\logMc$, $S_8$, and $\Om$).

  \subsection{Fixing cosmological parameters}\label{ss:fixcosmo}
    We now explore the impact of fixing the cosmological parameters on the constraints of the BCM model. In particular, we consider two cosmologies:
    \begin{itemize}
        \item {\bf BF cosmology:} our best-fit $\Lambda$CDM parameters for $\lmax=8192$ and ignoring the impact of baryonic effects. In this case:
        \begin{equation}
          \{\As,\Om,\Ob,h,\ns,\sum m_\nu\}=\{4.93\times 10^{-9}, 0.208, 0.0395, 0.711, 1.014, 0.228\,{\rm eV}\}.
        \end{equation}
        \item {\bf Planck18:} the best-fit $\Lambda$CDM cosmology found by \planck \cite{1807.06209}:
        \begin{equation}
          \{\As,\Om,\Ob,h,\ns,\sum m_\nu\} = \{2.10 \times 10^{-9}, 0.3153, 0.0493, 0.6736, 0.9649, 0.06\,{\rm eV}\}.
        \end{equation}
    \end{itemize}
    The results of these two cases can be seen in the right panels of Fig.~\ref{fig:fixing.params}.

    As could be expected, cosmology has a strong impact on the posterior distributions of the BCM parameters. Using our BF cosmology leads to a constraint on $\logMc$ that peaks at the same value as our fiducial analysis, but has significantly suppressed tails at both high and low masses.
    The resulting constraints on $S_k$ are, consequently, tighter. In turn, assuming the Planck18 cosmology (which, as we described in Section \ref{s:results} is in tension with our best fit at the $\sim2$-3$\sigma$ level) has a more dramatic effect. First, the posterior distribution of $\logMc$ shows a clear preference for values above $\logMc=14$, corresponding to a significantly stronger level of baryonic feedback and power spectrum suppression. Interestingly, in this case we are also able to discard low values of $\logeta$, with a clear preference for positive values that are much larger than those favoured by hydrodynamical simulations $\logeta = -0.32 \pm 0.22$ (see e.g. \cite{2011.15018,2309.02920}). These dramatic changes are needed to compensate for the higher value of $S_8$ and $\Om$ preferred by \planck ($1.8\sigma$ and $3.5\sigma$ higher than our fiducial analysis). As shown in the bottom-right panel of Fig.~\ref{fig:fixing.params}, the resulting baryonic suppression is significantly stronger than that obtained in our fiducial analysis over all the scales considered. Interestingly, the value of $\logMc$ inferred assuming the Planck18 cosmology is in good agreement with the value found in the real-space analysis A23, although a much lower value of $\logeta$ was preferred in that analysis. This suggests that the $S_8$ tension can be fully solved even when keeping the gas fractions in haloes consistent with cosmic shear and galaxy clusters observations (e.g. analysing a collection of galaxy clusters, \cite{2309.02920} infer $\logMc=13.82\pm0.36$), by ejecting the gas much further away than normally found in hydrodynamical simulations. Nevertheless, we caution the reader against over-interpreting the predictions from the BCM model in such extreme scenarios.

  \subsection{Comparison with alternative baryonic models}\label{ss:baryonsmodels}
    
    Finally, we explore the impact on our results of using alternative parametrisations for baryonic effects. In particular, we explore the simpler BCM fitting function of Schneider \& Teyssier (ST15) \cite{1510.06034}, and the effective power spectrum suppression model of Amon-Efstathiou (AE) \cite{2206.11794, 2305.09827}. The choice of ST15 is meant to test whether we need the much more general and complex model of \baccoemu, or the simpler model of ST15, with just two free parameters, is enough to describe these data. On the other hand, using the AE model will allow us to test whether we can recover the lack of evidence for baryonic suppression found in our fiducial analysis.

\begin{figure}
      \centering
      \includegraphics[width=0.6\linewidth]{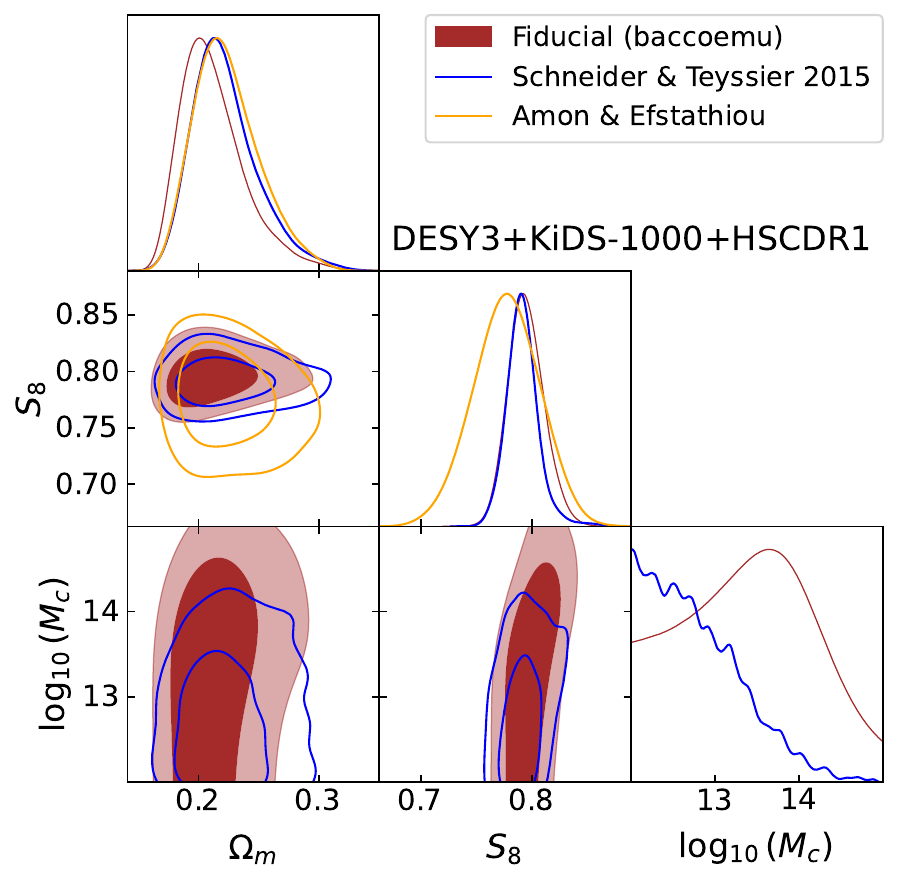}
      \includegraphics[width=0.35\linewidth]{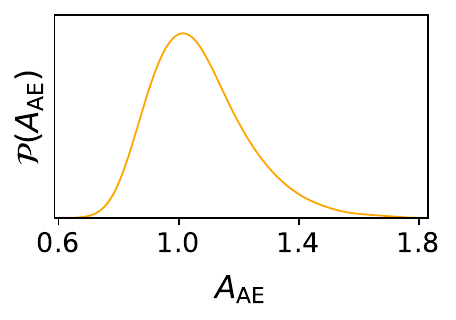}
      \caption{Parameter constraints for the fiducial case (burgundy) and the simpler parametrisations of ST15 \cite{1510.06034} (blue) and AE \cite{2206.11794,2305.09827} (orange). We have restricted the range of $\logMc$ to match the allowed regime of the ST15 model (i.e. $\logMc > 12$) \cite{1510.06034}.}
      \label{fig:baryons.models}
    \end{figure}

    Fig.~\ref{fig:baryons.models} shows the posteriors of cosmological and baryonic parameters obtained with these models compared to our fiducial results. First, we see that the mean of the posterior distributions of the cosmological parameters is not significantly affected by the choice of model for baryonic effects, with shifts of $0.5\sigma$ at most in $S_8$ for the AE model. Moreover, we see that the lack of strong evidence for baryonic suppression in our data is also confirmed by these two new models: ST15 is not able to place tight constraints on $\logMc$, and the results are compatible with our fiducial measurement (which implies a weaker baryon suppression, compatible with $S_k = 1$ within errors). We remind the reader that ST15 is restricted to the range of halo masses $\logMc > 12$ and $\logeta < 0$ to match the range of parameters for which this parametrisation was validated \cite{1510.06034}. We find similar results in the AE case, with $\AAE = 1.07^{+0.11}_{-0.19}$, perfectly compatible with no suppression ($\AAE=1$). Interestingly, in agreement with our results in Section \ref{ss:fixcosmo}, and with the results of \cite{2206.11794,2305.09827}, the value of $\AAE$ parameter changes significantly when assuming the Planck18 cosmology: $\AAE = 0.860\pm 0.032$, showing a $4\sigma$ preference for a suppression of the non-linear scales. Finally, it is worth noting that both models provide also a good fit to the data: $\chi^2_{\rm AE} = 841.6$ ($\Nd = 820$, $p = 0.28$) and $\chi^2_{\rm ST15} = 842.3$ ($\Nd = 820$, $p = 0.28$), very close to our fiducial $\chi^2 = 841.1$ ($\Nd = 820$, $p = 0.28$).

    Coming back to the cosmological constraints, we find a good agreement with the mean values of our fiducial analysis. In fact, ST15 recovers very similar constraints, with just small shifts on $S_8$ ($0.15\sigma$) and $\Om$ ($0.25\sigma$) and about 8\% tighter constraints on $S_8$. In the case of AE, however, we obtain noticeably larger error bars on $S_8$ (not so on $\Om$). 
    Although one could naively expect that a model with fewer free parameters should yield tighter constraints, these depend instead on the space of possible power spectra suppression allowed by a given model. The BCM, although flexible enough to fit a large collection of state-of-the-art hydrodynamical simulations, is limited by construction by our knowledge of gas and stellar distribution in halos. On the other hand, the phenomenological model of AE can predict arbitrarily strong $S_k$, as we cannot impose physical priors on it. 

    Overall, the stability of our results under change in baryonic models shows the robustness of our fiducial constraints.
  
\section{Conclusion}\label{s:conclusion}
  We have presented the first joint analysis of the three major, publicly available, cosmic shear datasets: \desyt, \kidsot and \hscdro. We reanalyse these datasets, at the catalogue level, in a consistent manner, using a harmonic-space pipeline, and homogenising the analysis choices across all surveys. Furthermore, by carefully modelling non-linearities and baryonic effects with \baccoemu, we are able to exploit, for the first time, scales down to $\lmax = 4500$ (and, more boldly, $\lmax=8192$).

  We find that the constraints derived from each survey independently, are in broad agreement with each other, both in terms of cosmological and baryonic effects parameters. Combining them with our fiducial analysis choices ($\lmax=4500$ and baryonic effects described by \baccoemu with 7 free parameters), we infer $S_8=0.795^{+0.015}_{-0.017}$ and $\Om=0.212^{+0.017}_{-0.032}$. Compared with the \planck constraints, we find a compatible value for $S_8$ ($1.8\sigma$ lower than \planck), but a value of $\Om$ that is lower and in tension at the $3.5\sigma$ level. This is due to a combination of two effects: first, each survey individually recovers constraints on $\Om$ that are lower than (but in reasonable agreement with) \planck. Secondly, including information from smaller scales leads to additional constraining power on $\Om$ (more so than $S_8$), reducing its upper limit. We have studied different potential effects that could cause this shift: pixelisation effects (see Section \ref{ap:pixel}), residual stochastic noise, prior and volume effects (e.g. from unconstrained parameters, such as $\omega_b$ or $\logeta$). As shown in Section \ref{ap:mock}, when applied to synthetic data with a similarly low $\Om$, our pipeline recovers unbiased cosmological constraints. Ultimately, the effect could be caused by unknown systematics (e.g. PSF effects below scales $\theta\sim3'$, which our harmonic space scale cuts could still have some sensitivity to), which may be unveiled by future data releases from these collaborations.

  In order to further investigate this tension, we study our constraints in the $(S_{12},\om)$ plane, minimising the dependence of our constraints on the poor ability of cosmic shear data to constrain $h$. In this space, we observe no tension. The tension reappears after applying a prior on $h$ from \planck \cite{1807.06209} although, interestingly, it disappears again if using a prior from local measurements of $H_0$ \cite{2112.04510}. This is in agreement with the rationale of \cite{2002.07829, 2209.12997}, where it was shown that the current $S_8$ tension may be recast in terms of the $H_0$ tension under this reparametrisation. We also observe that our $\Om$ tension is alleviated by $0.7\sigma$ when including an external prior from BAO measurements, without worsening our fit to the weak lensing data, which might signal to a prior volume effect on $\Om$. Finally, to fully estimate the level of tension accounting for the full dimensionality of the parameter scale, we computed the suspiciousness statistic \cite{2007.08496}, obtaining $\log(S) = -3.9$, which corresponds to a  $2-3\sigma$ tension (see Section~\ref{ap:suspiciousness} for details). This highlights that the statistical tension between current weak lensing data and \planck (in this work in terms of $\Om$, not $S_8$), although worth investigating, might not be intrinsic to the LSS, but instead either a different incarnation of the $h$ tension, or a moderate statistical anomaly.

  Our constraints on the BCM parameters are in broad agreement with those of \cite{2303.05537}, although we find looser constraints on the dominant parameter $\logMc$. Unlike \cite{2206.08591, 2303.05537}, we recover an associated baryonic suppression factor compatible with $S_k =1$ within $1\sigma$. As in the case of cosmological parameters, this is a consequence of employing a harmonic-space pipeline, as opposed to the real-space analysis of these works. One reason is that a correlation function measured down to a scale $\thetamin$ will contain information about multipoles larger than $\lmax=\pi/\thetamin$ -- see Appendix \ref{ap:real_vs_fourier}. However, given the error of the real space measurements, we could expect a smaller discrepancy than what we find. A more careful study of the differences between real- and harmonic-space analysis is needed to fully understand the reason behind the different sensitivity levels to baryonic effects. 
  This lack of sensitivity to baryonic effects is confirmed with mock analyses employing hydrodynamical simulations (see Appendix~\ref{ap:mock}), and we verified that our harmonic-space theory prediction pipeline agrees with the predictions used in \cite{2303.05537}. More importantly, we find that modelling baryonic effects is critical to fully propagate all sources of uncertainty to the final cosmological constraints. The inclusion of baryons allows for larger values of $S_8$, thus ameliorating the tension with CMB data, and improving the compatibility of different weak lensing datasets. In turn, the posterior constraints on BCM parameters depend critically on the cosmological model assumed. Fixing the cosmology to the \planck best fit, we recover strong evidence for baryonic suppression, while it remains compatible with $S_k=1$ for our fiducial best-fit model. We find these qualitative results to be robust to the choice of baryonic effects parametrisation.

  We provide results for four different scale cuts: $\lmax=1000, 2048, 4500$, and $8192$, and address the impact of baryonic effects at these different scales. We find that for \desyt- and \kidsot-like data, the impact of baryons can be largely ignored up to scales $\lmax\simeq 2000$, although deeper surveys, such as \hscdro (and future Stage-IV experiments), are sensitive to baryonic effects over a broader range of scales. This is due to the smaller impact of shot noise in the measurement uncertainties, and to the different sensitivity to different redshift ranges. Furthermore, we find that, when including smaller scales (beyond $\lmax\sim2000$), most of the information is used to reduce the uncertainties on $\Om$, while the constraints on $S_8$ largely saturate. With the advent of Stage-4 surveys and the expected significant gain on SNR at smaller scales, we should be able to improve our $S_8$ constraints by going to higher $\ell$. 

  We leave a careful assessment of the impact of scale cuts, and the correspondence between real- and harmonic-space cuts for future work, aiming to find an approximate equivalence between both. It would be interesting to assess if the $\Om$-tension observed with our harmonic space pipeline is also reproduced by a joint analysis of \desyt, \kidsot, and \hscdro in real space. It will also be interesting to study methods to incorporate the theoretical error, given the limited accuracy of most models for astrophysical systematics and non-linear effects (e.g. emulators, \hfit, \hmcode, IA models), into the analysis in a self-consistent manner. Propagating these theoretical uncertainties will be vital to ensure that any potential future tensions in cosmological parameters can be properly quantified. Finally, a more thorough study of the impact of the model used to describe IAs would be desirable. Current datasets show no evidence of strong IA contamination ($A_{\rm IA}$ is compatible with 0 within $2\sigma$ for all the datasets explored here), and hence models more sophisticated than the basic NLA are currently not warranted by these data. Nevertheless, future datasets will be more sensitive to the impact of IAs, and a thorough characterisation of the accuracy requirements for IA models will be of paramount importance \cite{2311.16812}.

  In light of the $\Om$-tension obtained in our work, it would be particularly interesting the extension of the current weak lensing only analysis to incorporate galaxy clustering measurements in a 3x2pt study. This would increase the constraining power on $\Om$ and remove the possible prior volume effect that might be causing it. Whereas this would be straightforward if we restrict the galaxy clustering to large scales, if we want to push to smaller scales, we need to calibrate the effect of baryons in the non-linear galaxy bias expansion. This is left for future work. In the coming months, the data from six years of observations of \des and three years from \hsc \cite{2304.00701} will be publicly available. In particular, the increase in constraining power from \hscdrt may allow us to cast light on the impact of baryonic effects on scales smaller than less dense surveys, as well as their evolution at higher redshifts. A joint analysis of these datasets, together with the legacy data release from \kids, will be of key importance to determine our ability to reliably use cosmic shear data on small scales in the context of near-future Stage-IV experiments, such as LSST or Euclid.

\acknowledgments
We would like to thank Cyrille Doux for sharing the \desyt Fourier analysis chains and data vectors. We would also like to thank Elisa Chisari, Pedro Ferreira, Benjamin Joachimi, Will Handley, Danielle Leonard and Martin Rey for useful discussions. We thank the referee for their comments and suggestions. Finally, we would also like to thank all collaborations for taking the time and resources to make not only the data publicly available but also the chains from their analyses.

CGG and DA are supported by the Beecroft Trust. MZ is supported by STFC. REA acknowledges the support of the E.R.C. grant 716151 (BACCO) and of the Project of Excellence Prometeo/2020/085 from the Conselleria d’Innovació, Universitats, Ciéncia i Societat Digital de la Generalitat Valenciana, and of the project PID2021-128338NB-I00 from the Spanish Ministry of Science. 

We made extensive use of computational resources at the University of Oxford Department of Physics, funded by the John Fell Oxford University Press Research Fund.

For the purpose of Open Access, the authors have applied a CC BY public copyright licence to any Author Accepted Manuscript version arising from this submission.

\ul{Software}:  We made extensive use of the {\tt numpy} \citep{oliphant2006guide, van2011numpy}, {\tt scipy} \citep{2020SciPy-NMeth}, {\tt astropy} \citep{1307.6212, 1801.02634}, {\tt healpy} \citep{Zonca2019}, {\tt matplotlib} \citep{Hunter:2007} and {\tt GetDist} \citep{Lewis:2019xzd} python packages. We also made extensive use of the \healpix \cite{gorski_healpix_2005} package. The codes used to produce these results can be found in \url{https://gitlab.com/carlosggarcia/shear_baryons}.

This paper makes use of software developed for the Large Synoptic Survey Telescope and Euclid. We thank the LSST Project for making their code available as free software at \url{http://dm.lsst.org}. 

\ul{Data:} 

  \ul{DES}. This project used public archival data from the Dark Energy Survey (DES). Funding for the DES Projects has been provided by the U.S. Department of Energy, the U.S. National Science Foundation, the Ministry of Science and Education of Spain, the Science and Technology Facilities Council of the United Kingdom, the Higher Education Funding Council for England, the National Center for Supercomputing Applications at the University of Illinois at Urbana-Champaign, the Kavli Institute of Cosmological Physics at the University of Chicago, the Center for Cosmology and Astro-Particle Physics at the Ohio State University, the Mitchell Institute for Fundamental Physics and Astronomy at Texas A\&M University, Financiadora de Estudos e Projetos, Funda{\c c}{\~a}o Carlos Chagas Filho de Amparo {\`a} Pesquisa do Estado do Rio de Janeiro, Conselho Nacional de Desenvolvimento Cient{\'i}fico e Tecnol{\'o}gico and the Minist{\'e}rio da Ci{\^e}ncia, Tecnologia e Inova{\c c}{\~a}o, the Deutsche Forschungsgemeinschaft, and the Collaborating Institutions in the Dark Energy Survey.
        
  The Collaborating Institutions are Argonne National Laboratory, the University of California at Santa Cruz, the University of Cambridge, Centro de Investigaciones Energ{\'e}ticas, Medioambientales y Tecnol{\'o}gicas-Madrid, the University of Chicago, University College London, the DES-Brazil Consortium, the University of Edinburgh, the Eidgen{\"o}ssische Technische Hochschule (ETH) Z{\"u}rich,  Fermi National Accelerator Laboratory, the University of Illinois at Urbana-Champaign, the Institut de Ci{\`e}ncies de l'Espai (IEEC/CSIC), the Institut de F{\'i}sica d'Altes Energies, Lawrence Berkeley National Laboratory, the Ludwig-Maximilians Universit{\"a}t M{\"u}nchen and the associated Excellence Cluster Universe, the University of Michigan, the National Optical Astronomy Observatory, the University of Nottingham, The Ohio State University, the OzDES Membership Consortium, the University of Pennsylvania, the University of Portsmouth, SLAC National Accelerator Laboratory, Stanford University, the University of Sussex, and Texas A\&M University.
    
  Based in part on observations at Cerro Tololo Inter-American Observatory, National Optical Astronomy Observatory, which is operated by the Association of Universities for Research in Astronomy (AURA) under a cooperative agreement with the National Science Foundation.
    
  \ul{HSC}. The Hyper Suprime-Cam (HSC) collaboration includes the astronomical communities of Japan and Taiwan, and Princeton University. The HSC instrumentation and software were developed by the National Astronomical Observatory of Japan (NAOJ), the Kavli Institute for the Physics and Mathematics of the Universe (Kavli IPMU), the University of Tokyo, the High Energy Accelerator Research Organisation (KEK), the Academia Sinica Institute for Astronomy and Astrophysics in Taiwan (ASIAA), and Princeton University. Funding was contributed by the FIRST program from Japanese Cabinet Office, the Ministry of Education, Culture, Sports, Science and Technology (MEXT), the Japan Society for the Promotion of Science (JSPS), Japan Science and Technology Agency (JST), the Toray Science Foundation, NAOJ, Kavli IPMU, KEK, ASIAA, and Princeton University. 
    
  The Pan-STARRS1 Surveys (PS1) have been made possible through contributions of the Institute for Astronomy, the University of Hawaii, the Pan-STARRS Project Office, the Max-Planck Society and its participating institutes, the Max Planck Institute for Astronomy, Heidelberg and the Max Planck Institute for Extraterrestrial Physics, Garching, The Johns Hopkins University, Durham University, the University of Edinburgh, Queen’s University Belfast, the Harvard-Smithsonian Center for Astrophysics, the Las Cumbres Observatory Global Telescope Network Incorporated, the National Central University of Taiwan, the Space Telescope Science Institute, the National Aeronautics and Space Administration under Grant No. NNX08AR22G issued through the Planetary Science Division of the NASA Science Mission Directorate, the National Science Foundation under Grant No. AST-1238877, the University of Maryland, and Eotvos Lorand University (ELTE) and the Los Alamos National Laboratory.
    
  Based in part on data collected at the Subaru Telescope and retrieved from the HSC data archive system, which is operated by Subaru Telescope and Astronomy Data Center at National Astronomical Observatory of Japan.

    \ul{KiDS}. Based on observations made with ESO Telescopes at the La Silla Paranal Observatory under programme IDs 177.A-3016, 177.A-3017, 177.A-3018 and 179.A-2004, and on data products produced by the KiDS consortium. The KiDS production team acknowledges support from: Deutsche Forschungsgemeinschaft, ERC, NOVA and NWO-M grants; Target; the University of Padova, and the University Federico II (Naples).
    
    We use the gold sample of weak lensing and photometric redshift measurements from the fourth data release of the Kilo-Degree Survey (Kuijken et al. 2019, Wright et al. 2020, Hildebrandt et al. 2021, Giblin et al 2021), hereafter referred to as KiDS-1000. Cosmological parameter constraints from KiDS-1000 have been presented in Asgari et al 2021 (cosmic shear), Heymans et al 2021 (3x2pt) and Tröster et al 2021 (beyond $\Lambda$CDM), with the methodology presented in Joachimi et al 2021.

\clearpage
\appendix

\section{Pixelisation effects}\label{ap:pixel}
  \begin{figure}
    \centering
    \includegraphics[width=0.9\textwidth]{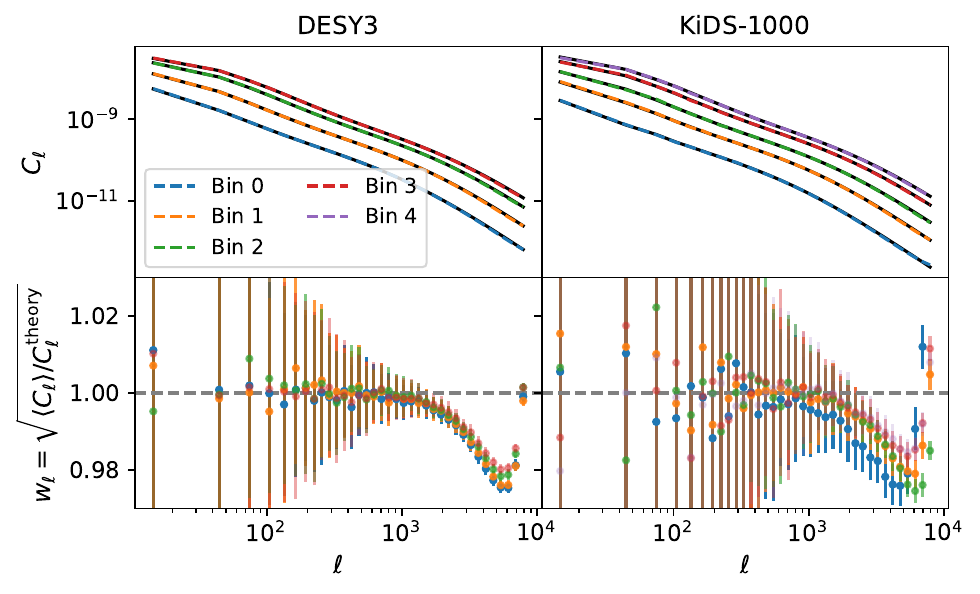}
    \caption{Effective pixel window function measured from 100 Gaussian simulations for \desyt and \kidsot. In the top panel, the black line is the input $C_\ell^{\rm theory}$ and the colour dashed lines are the mean of the 100 realisations. In the bottom panel, the resulting window function. The effect, at most 2\%, does not affect the parameter constraints found in our analysis.}
    \label{fig:pixwin}
  \end{figure}
  A key novelty of this work is our use of small scale cosmic shear data in harmonic space. One concern on such scales is the impact of the pixelisation scale inherent to the maps from which the angular power spectra were computed. As discussed in \cite{2010.09717}, two limiting regimes exist, depending on the average number of sources per pixel in the survey, $\Ngal$. In the limit $\Ngal \lesssim 1$, the cosmic shear field is effectively sampled at the galaxy positions, and the effect of pixelisation is negligible. In turn, for $\Ngal\gg1$, the field is effectively averaged within each pixel, and the effect of pixelisation can be corrected by accounting for the corresponding theoretical pixel window function. In this work, we use high-resolution maps with $\nside = 4096$, and the average number of galaxies per pixel in the case of \desyt is $\Ngal\sim 1.1$ for all tomographic bins and, therefore, we are in the ``sampling'' regime. Similarly, for \kidsot, $\Ngal\lesssim1.3$ in all bins, and $\Ngal<1.47$ for HSC. Thus, we affect the impact of pixelisation on the measured power spectra to be small. The effects of pixelisation on \hscdro were found to be negligible in \cite{2010.09717} and thus we do not study it further. Nevertheless, given the key role played by small-scale lensing in our analysis, we compute the effective pixel window function for each redshift bin in the \desyt and \kidsot samples as follows: we generate 100 Gaussian simulated maps of the cosmic shear field, without shape noise, at high resolution ($\nside=8192$). We then produce synthetic shear catalogs by sampling this field at the positions of the \desyt and \kidsot sources. We then compute the power spectrum bandpowers $\hat{C}_q$\footnote{$q$ here labels each of the $\ell$ bins.} from these catalogs using our pipelines (which produces maps of the cosmic shear field from the catalogs at $\nside = 4096$). The resulting estimated bandpowers, averaged over simulations, $\langle \hat{C}_q\rangle$, are then compared against the theoretical bandpowers $C^{\rm th}_q$, estimated by convolving the input power spectrum with the associated bandpower window functions. The effective pixel window function is then estimated as $w_q = \sqrt{\langle\hat{C}_q\rangle / C_q^{\rm th}}$. The results are shown in Fig. \ref{fig:pixwin}. The effective pixel window function deviates from 1 by $\sim2\%$ at most in all cases on the smallest scales. The measured $w_q$ is virtually the same across different tomographic bins in \desyt, and shows some mild variation in the case of \kidsot. This is as expected, since the number densities of the different \desyt bins are similar, while they vary from $0.62\,{\rm arcmin}^{-2}$ to $1.85\,{\rm arcmin}^{-2}$ in the case of \kidsot.
  
  We repeat the cosmological analysis of our measured power spectra, correcting them for this effective pixel window function at $\ell>2000$. The resulting cosmological constraints are almost identical to those found in our fiducial analysis, and hence our results are robust to pixelisation effects. This will likely not be the case for future weak lensing surveys, given their higher statistical power, and developing estimators that are able to accurately account for the effects of pixelisation, or to avoid the use of pixels altogether, will be of paramount importance \cite{2312.12285,WolzPrep,TessorePrep}.

\section{Validation against mock data}\label{ap:mock}
\begin{figure}
    \centering
    \includegraphics[width=0.495\linewidth]{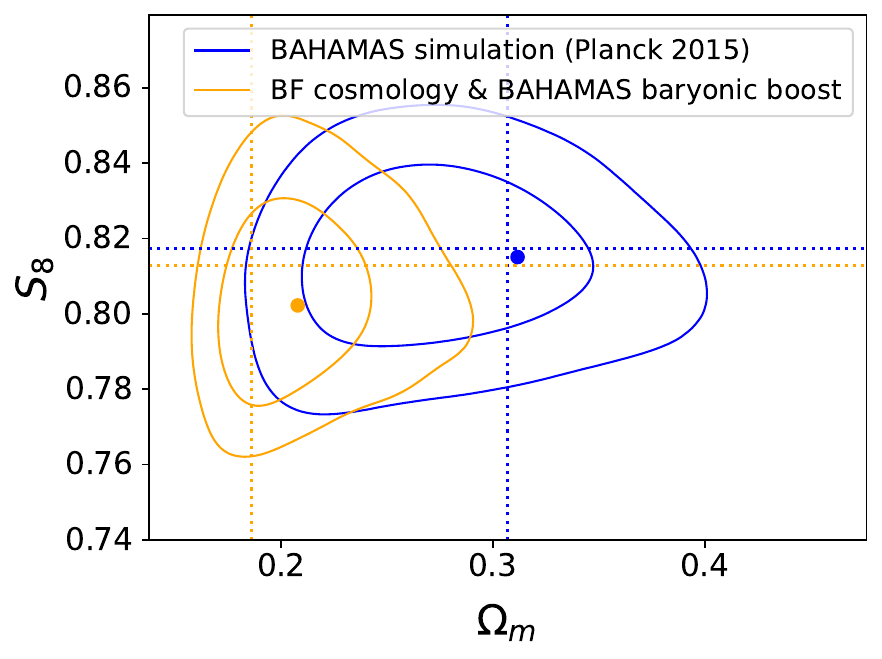}
    \includegraphics[width=0.495\linewidth]{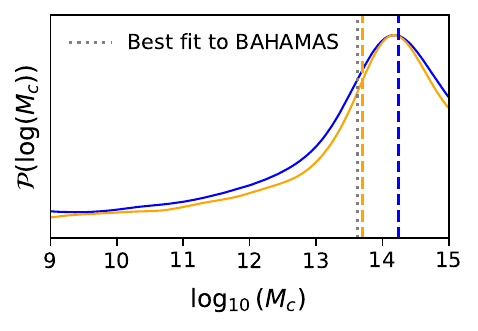}
    \includegraphics[width=0.495\linewidth]{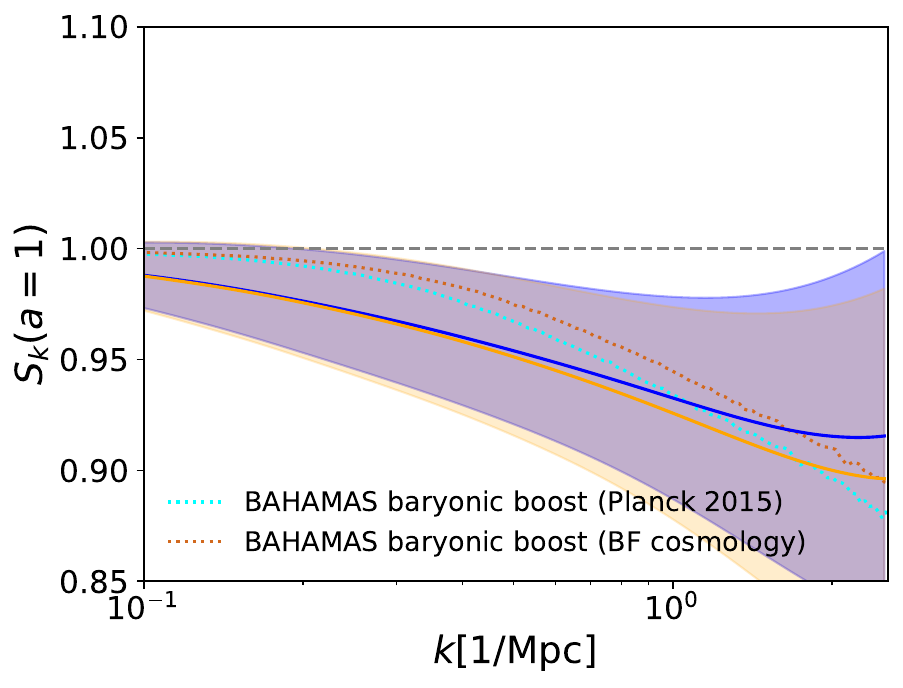}
    \caption{Results found using in mock data. Constraints in the $(S_{8},\Om)$ plane, showing the 68\% and 95\% C.L. regions (top left), on $\logMc$ (top right) and on the baryonic suppression factor (bottom), with $1\sigma$ errors. Note that the difference in the baryon suppression factors is only due to the units in 1/Mpc. We find good agreement with the true input parameters and suppression factors (dotted lines).} 
    \label{fig:mock}
  \end{figure}

  In order to validate our theory prediction pipeline we generate a mock data vector for the combined \desyt, \kidsot and \hscdro data sets. The mock data are generated using the redshift distributions, covariance and bandpower window functions from the real data. We account for the shifts in the redshift distribution, and the multiplicative biases, for which we use the mean of their priors. We do not include intrinsic alignments although we marginalise over them. We carry out two tests. First, we use the non-linear matter power spectrum and the baryon suppression measured from one BAHAMAS hydrodynamical simulation, specifically the fiducial Planck2015-$T_{\rm AGN}=10^{7.8} \, {\rm K}$. To avoid sample variance, we model the gravity-only power spectrum at $k<1\,{\rm Mpc}^{-1}$ with \euclidemu, and apply the corresponding baryon boost measured from the hydrodynamical simulations. At $k>1\,{\rm Mpc}^{-1}$, we use directly the power spectrum from BAHAMAS. Although this leads to a discontinuity in the power spectrum, it is within the cosmic variance of the simulation and has a negligible impact on the angular power spectrum once projected. The cosmological model is set to that of the simulation: $\As\,10^9=1.9909$, $\Oc=0.2572$, $\Ob=0.0482$, $\ns=0.9701$, $h=0.6933$ and $\sum m_\nu=0.06$, and the associated baryon suppression can be seen in Fig.~\ref{fig:mock}. This test allows us to verify that our pipeline can recover a state-of-the-art simulated power spectrum down to small scales (both the gravity-only non-linearities, and the baryonic suppression factor). Second, to understand the effect of cosmology on the posterior distributions, we generate the non-linear matter power spectrum of our fiducial best-fit cosmology when modelling baryonic effects with \baccoemu  ($\As\, 10^9=4.9320$, $\Oc=0.13397$, $\Ob=0.05138$, $\ns=1.0582$, $h=0.8507$ and $\sum m_\nu=0.0612$), and a baryonic suppression factor estimated from the BAHAMAS simulation, as before. This allows us to test the impact of having a best-fit cosmology with low $\Om$ on the shape of the final posterior distribution. Note that, in this case, the baryon suppression and the cosmology are not necessarily in physical agreement, since the measured $S_k$ is obtained from a simulation with the best-fit \planck 2015 cosmology \cite{1502.01589}. The result of these tests are shown Fig.~\ref{fig:mock}. We are able to recover the input cosmology and baryonic suppression in both cases, demonstrating the flexibility of our model to reproduce realistic observables. Remarkably, we observe the same lack of constraining power beyond the $S_8$, $\Omega_m$ and $\logMc$ parameters as with the real data. In fact, despite having an input baryon boost that suppresses the matter power spectrum by about 10\% at $k\sim 2\,{\rm Mpc}^{-1}$, the recovered $S_k$ is compatible with $S_k = 1$ within less than $2\sigma$. This is also true even if we repeat the analysis using scales up to $\lmax=8192$. Interestingly, we also find that the final cosmological parameter uncertainties do depend on the underlying best-fit cosmology, particularly in the case of $\Om$. The $\Om$ uncertainty shrinks by $\sim50\%$ in the low-$\Om$ cosmology with respect to the \planck 2015 case. However, we find the constraints on $\logMc$ and the baryon boost to be insensitive to the input cosmology. 

\section{Small-scale extrapolation}\label{ap:extrap}
\begin{figure}
    \centering
    \includegraphics[width=0.495\linewidth]{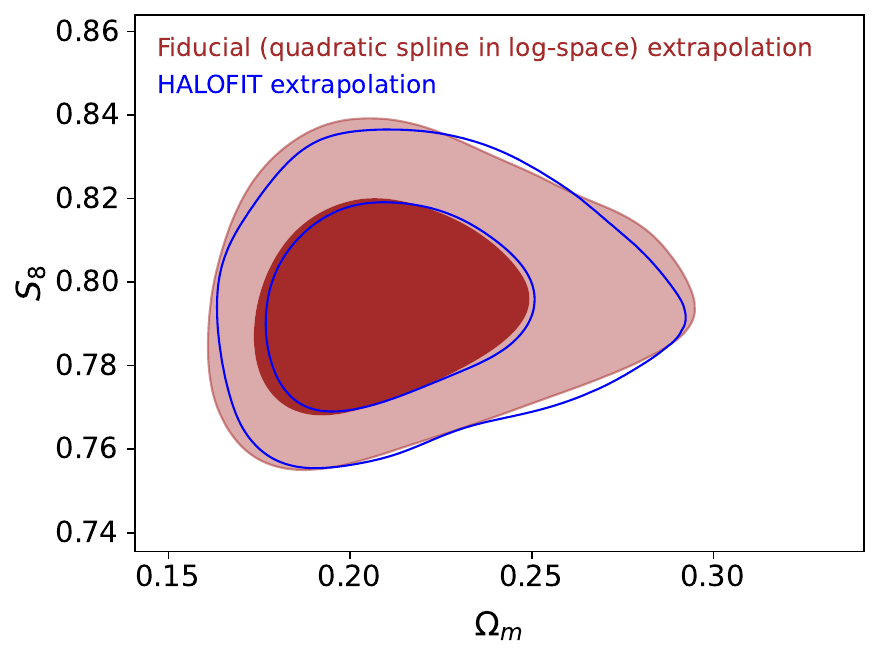}
    \includegraphics[width=0.495\linewidth]{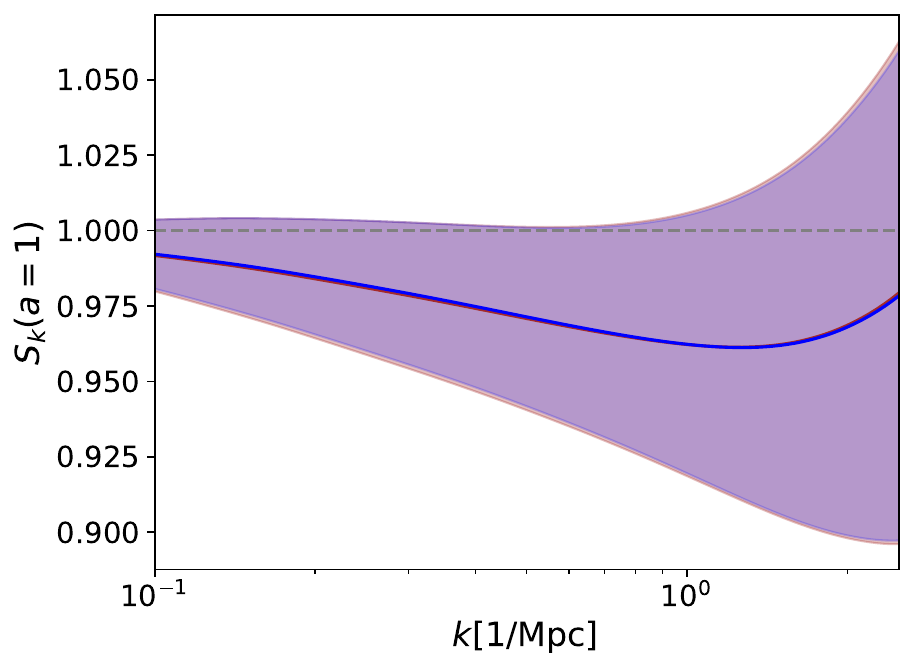}
    \caption{Effect of the power spectrum extrapolation at $k>5h\,{\rm Mpc}^{-1}$. $S_{8}$-$\Om$ (left) and baryon boost (right) posterior distributions (68\% and 95\% C.L. regions) with the fiducial scale cut $\lmax=4500$. We see that the effect of the extrapolation at $k>5h\,{\rm Mpc}^{-1}$ is negligible in our fiducial constraints.}
    \label{fig:extrap}
\end{figure}

In this Section we explore the effect of the power spectrum extrapolation at scales beyond the reach of \baccoemu's baryon emulator ($k > 5h\,{\rm Mpc}^{-1}$). As mentioned in the text, although the non-linear boost emulator can reach accurately $k = 10h\,{\rm Mpc}^{-1}$, we only model the scales up to $k = 5h\,{\rm Mpc}^{-1}$ with \baccoemu to match those from the baryon emulator.  We explore two alternatives: a quadratic spline in logarithmic space for both the matter power spectrum and the baryon boost (our fiducial way) and using \hfit to estimate the matter power spectrum at $k>5h\,{\rm Mpc}^{-1}$ and the quadratic spline for the baryonic boost. We have checked the results in both mock and real data. We report in Fig.~\ref{fig:extrap} the results on real data for the combination of \desyt, \kidsot and \hscdro, our most constraining case. We see that the impact of the extrapolation is negligible for both the cosmological and baryonic effect constraints.

\section{Real space vs Fourier space}\label{ap:real_vs_fourier}

\begin{figure}
    \centering
    \includegraphics[width=0.6\linewidth]{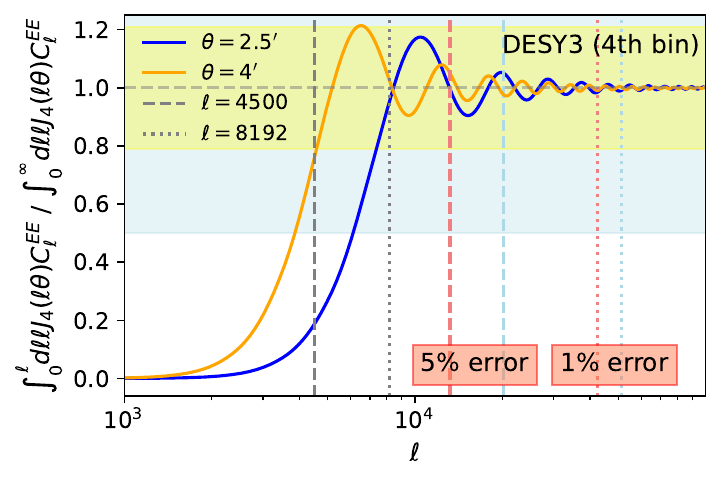}
    \caption{Relative error in the correlation function $\xi_{-}(\theta)$ as a function of the maximum $\ell$ used in the calculation for $\theta = 2.5',\, 4'$, in the 4th redshift bin of \desyt. The colour band are the $\xi_-$ $1\sigma$ error measurement: yellow for $\theta = 4'$ and light blue for $\theta = 2.5'$.  Multipoles significantly larger than the naive estimate (e.g. $\lmax=4500$ for $\theta=2.5'$) are needed to obtain a moderately reliable estimate of $\xi_-(\theta)$, although given the large error bar (light blue band), one might not need to that high $\ell$.}
    \label{fig:xi.int}
\end{figure}

In this Section we explore the equivalence of real and harmonic space scales.  The mathematical relation between real space correlation functions and harmonic space angular power spectra, in the flat-sky approximation, is given by 
\begin{equation}
    2\pi \xi_\pm(\theta) = \int_0^\infty \md\ell J_{0/4}(\theta \ell) C_\ell^{EE}\,,
\end{equation}
with $J_{0/4}$ the Bessel functions of the first kind. The approximate relation $\thetamin=\pi / \lmax$, in which $\lmax$ is the Nyquist frequency for a grid with spacing $\thetamin$, is not a good estimate of the largest multipole that must be used in order to reconstruct the correlation function at high accuracy. This is shown in Figure~\ref{fig:xi.int}, which shows the relative precision of the correlation function calculation at $\theta=2.5'$ and $\theta=4'$ achieved as a function of $\lmax$, in the 4th redshift bin of \desyt. Multipoles up to several times $\sim10^4$ are required to recover the correlation function with moderate ($\sim5\%$) precision, in contrast with the naive estimate $\lmax=4500$ for $\theta=2.5'$. However, given the actual error bars of these measurements (from $50\%$ to $20\%$ for $\xi_-$ for $\theta \in [2.5',\, 4.5']$), one does not actually need to go to that high $\ell$. In order to carefully address this question, we would need to study the interplay of the $\xi_-$ accuracy and the data error. Furthermore, we might need to explore other possible differences between harmonic and real space.

\section{Suspiciousness}\label{ap:suspiciousness}
  In order to address the statistical tension of the weak lensing and \planck datasets, we use the suspiciousness metric of \cite{2102.11511, 2007.15632}:
  \begin{equation}
    \log(S) = \enangle{\log(L_{AB})}_{P_{AB}} - \enangle{\log(L_A)}_{P_{A}} - \enangle{\log(L_B)}_{P_{B}}\,,
  \end{equation}
  where $L_{AB}$ is the likelihood from the combination of datasets $A$ and $B$, $L_{A(B)}$ is the likelihood for $A(B)$, and $\enangle{.}_{P_X}$ is the average over the posterior distribution of $X$, which we compute as the average over all the MCMC samples. 

  For multivariate Gaussian distributions, this is simply~\cite{2007.08496}
  \begin{equation}\label{eq:logS_Gauss}
    \log(S) = \frac{d}{2} - \frac{\chi^2}{2} \text{ with } \chi^2 = (\mu_A - \mu_B) ({\rm Cov_A + \rm Cov_B})^{-1} (\mu_A - \mu_B)^T\,,
  \end{equation}
  with $\mu_X$ and ${\rm Cov_X}$ the mean and covariance of the posterior distributions constrained with the likelihood $X$ and $d$ the effective number of dimensions.

  Since our posterior distributions are not Gaussian, we use three different approximations to interpret the fiducial results in Section \ref{s:results}:
  \begin{enumerate}
    \item Following \desyt analysis (Eq. E7 of \cite{2207.05766}), we estimate the Bayesian Model Dimensionality, $d_{\rm BMD} = d_A + d_B - d_{AB}$, with $d$ the number of parameters constrained by one of the cases, and obtained as $d = 2 (\enangle{\log(L)^2} - \enangle{\log(L)}^2)$. Using this approach we find $d_{\rm BMD} = -0.7$ which, interpreted as in \cite{2207.05766}, would mean that the datasets are completely in agreement ($p = 1$), as there cannot be tensions if no constrained parameters are shared. Note that this value is roughly compatible with the value measured by \desyt of $d_{\rm BMD} = 1.5 \pm 1.6$.
    \item Using the Gaussian approximation above ($\log(S) = d/2 - \chi^2/2$), we can solve for $\chi^2$ so that $\chi^2 = d - 2 \log(S)$, and estimate the number of constrained parameters $d$, as the number of parameters that we constrain with the Gaussianised \planck likelihood ($d=6$). With this approach, we find $\chi^2 = 14$ ($p=0.03$), corresponding to a $2.1 \sigma$ tension. In a very pessimistic case, setting $d=1$, $\chi^2 = 8.8$ ($p=0.003$), corresponding to a $3.0\sigma$ tension.
    \item If instead of solving for $\chi^2$, we compute it as in Eq.~\ref{eq:logS_Gauss}, $\chi^2 = 20$ and  $p = 0.003$, corresponding to a $2.9 \sigma$ tension. Note, however, that in this case: $\log(S) = -6.8$ instead of the measured $\log(S) = -3.9$.
  \end{enumerate}
  The value that we report in the main text corresponds to a tension with \planck at the $2-3\sigma$ level, based on the last two approximations. We disregard the first approach, as we do not believe these two datasets can be fully uncorrelated as shown, for instance, by \desyt \cite{2207.05766}.

\section{Power spectrum measurements}\label{ap:cl}
\begin{figure}[h]
    \centering
    \includegraphics[width=\textwidth]{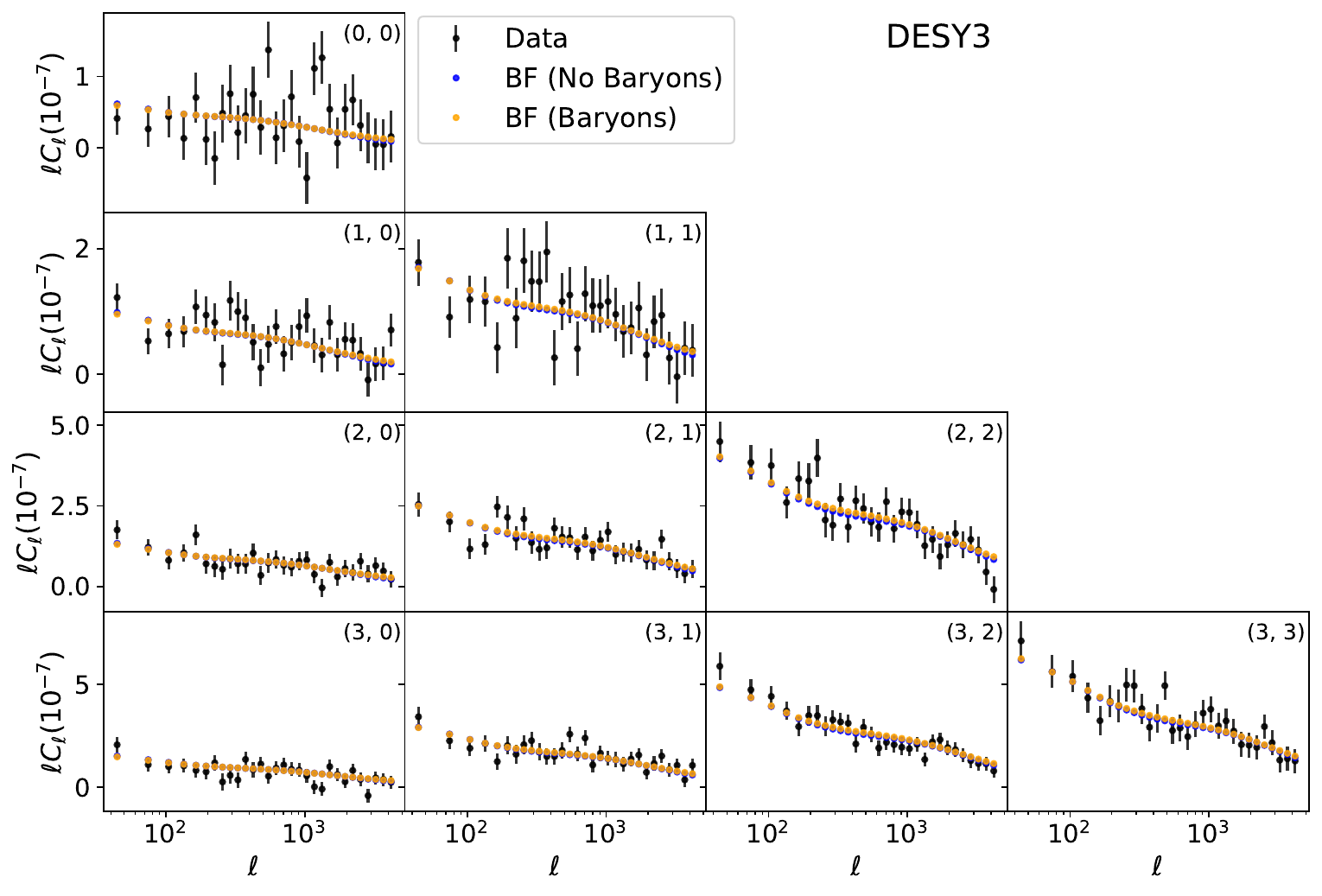}
    \caption{\desyt angular power spectra (black) and the best fit of all three surveys with (orange) and without (blue) modelling baryonic effects with $\lmax=4500$. The effect of baryons in the best fit is quite small, in agreement with our results.}\label{fig:cl.des}
  \end{figure}

\begin{figure}
    \centering
    \includegraphics[width=\textwidth]{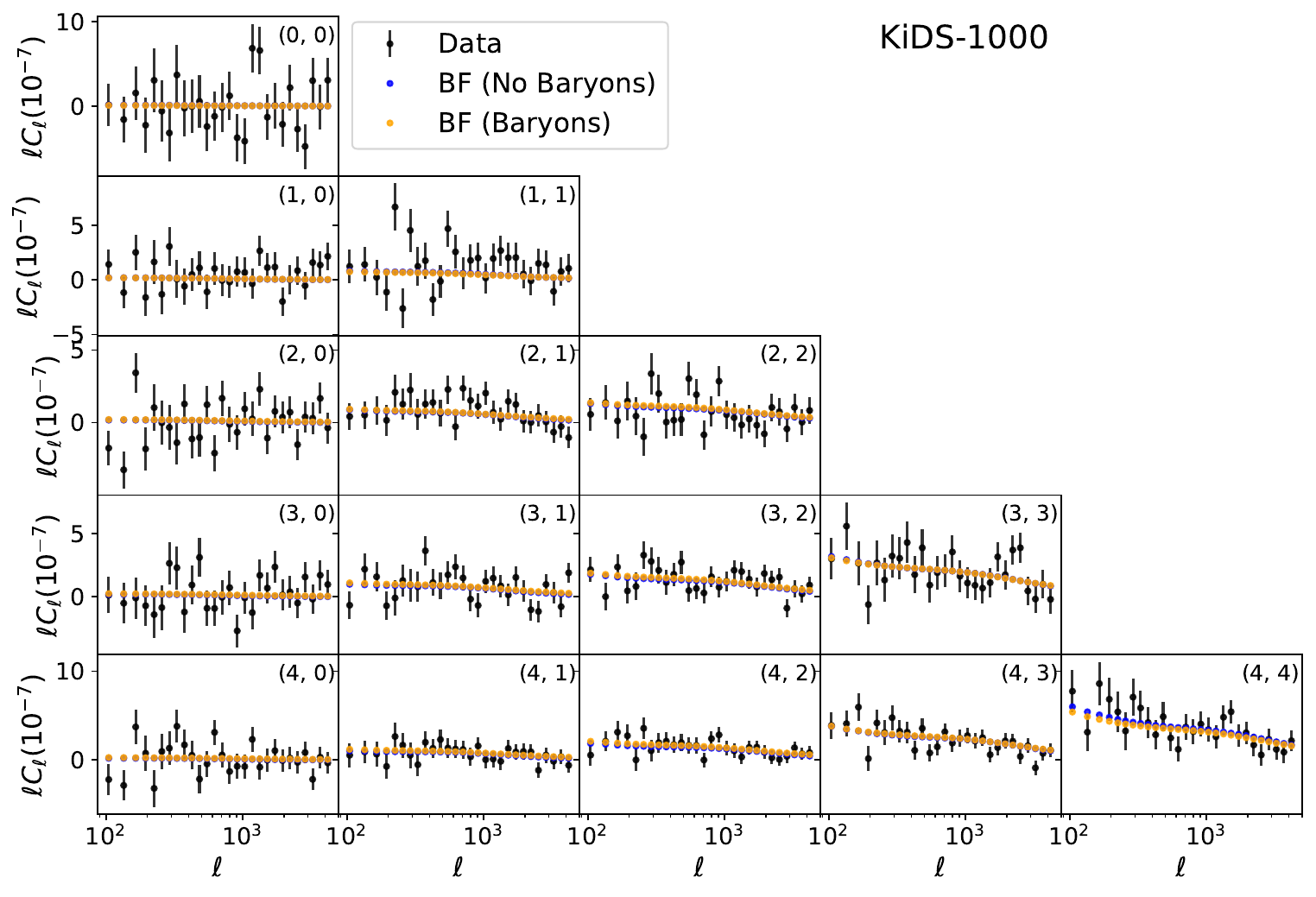}
    \caption{As Fig. \ref{fig:cl.des} for the \kidsot sample.} 
    \label{fig:cl.kids}
  \end{figure}

\begin{figure}
    \centering
    \includegraphics[width=\textwidth]{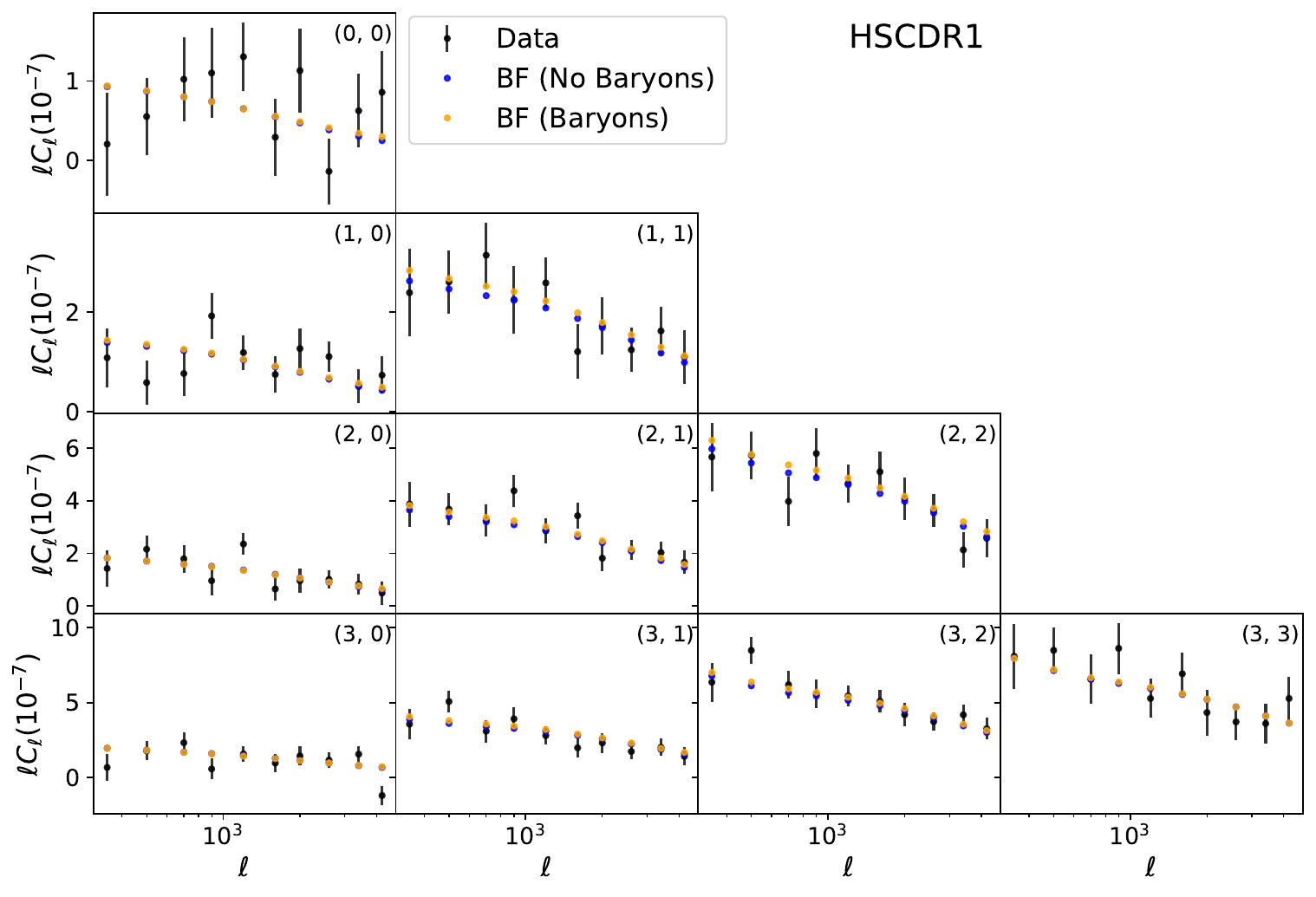}
    \caption{As Fig. \ref{fig:cl.des} for the \hscdro data.} 
    \label{fig:cl.hsc}
  \end{figure}

\begin{figure}
    \centering
    \includegraphics[width=\textwidth]{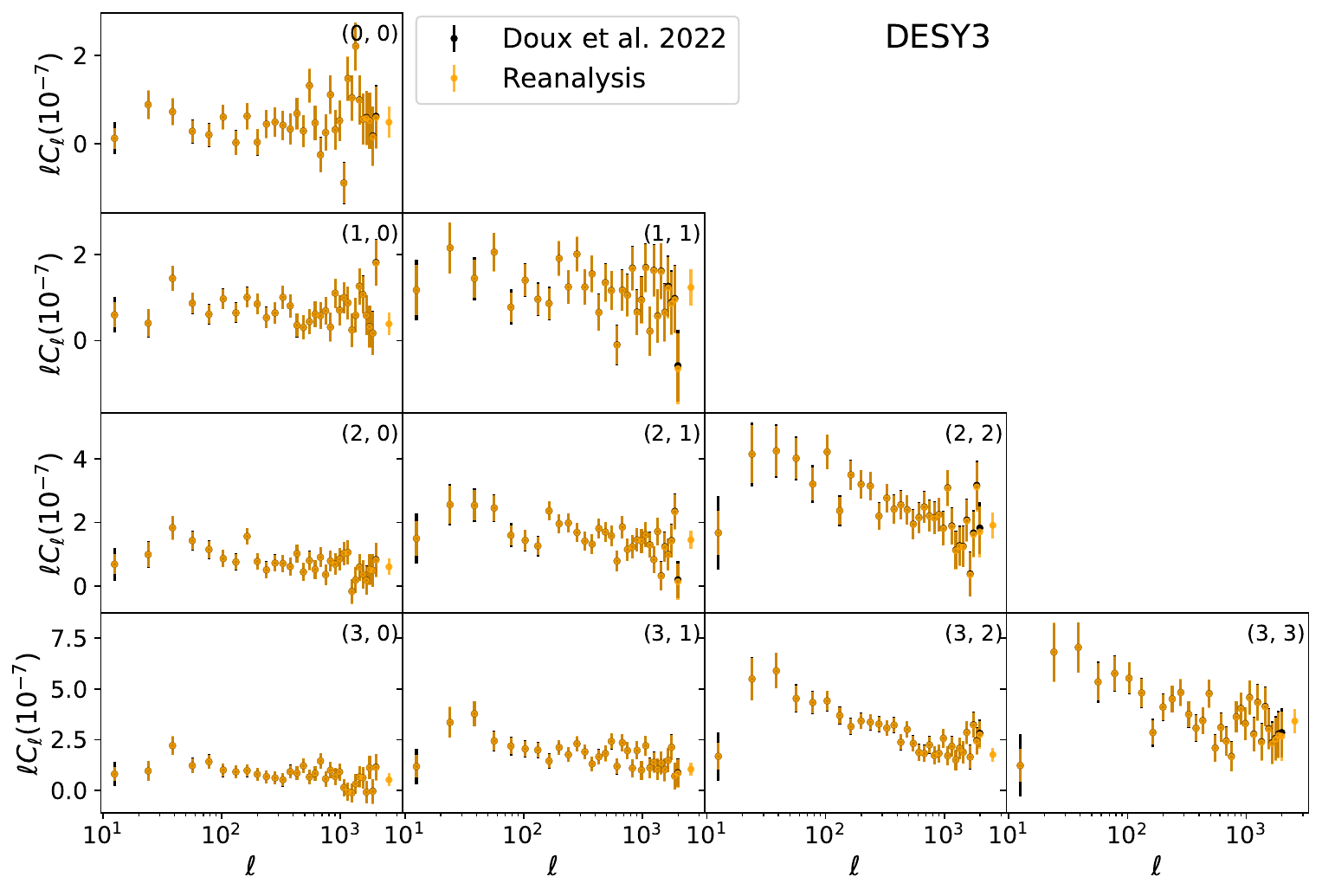}
    \caption{\desyt angular power spectra computed at $\nside=1024$ with the same binning used in the official analysis \cite{2203.07128}. The official results are shown in black, with our measurements shown in orange. The main difference is in the statistical uncertainties of the first bin, which we discard in this analysis.}
    \label{fig:cl.des_official}
\end{figure}

  This section presents the power spectrum measurements that make up the data vector used in this analysis. Figs. \ref{fig:cl.des}, \ref{fig:cl.kids}, and \ref{fig:cl.hsc} show the measurements from \desyt, \kidsot, and \hscdro, respectively, while Fig. \ref{fig:cl.des_official} compares the result of our measurement pipeline with the official results of \cite{2203.07128}.

\section{Summary of results}
  \begin{figure}
    \centering
    \includegraphics[width=\textwidth]{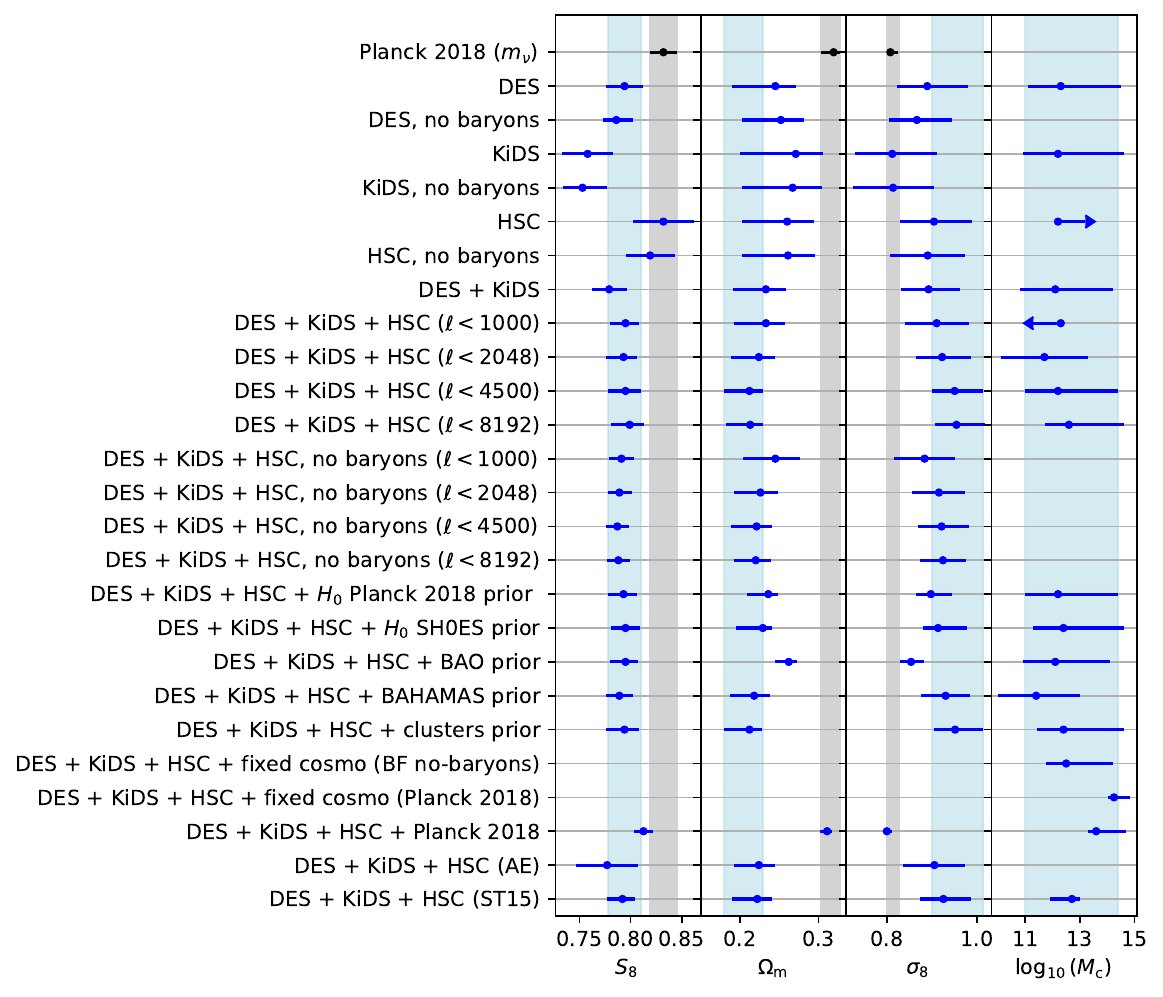}
    \caption{Summary plot showing the constraints for all the cases studied in this work. The lines with missing points are those for which such parameters were kept constant or not introduced in the analysis. If not stated otherwise, the fiducial $\lmax = 4500$ is used.}
    \label{fig:results}
  \end{figure}
  This appendix gathers all the results discussed in this paper. Fig. \ref{fig:results} shows the constraints on $S_8$, $\Om$, and $\logMc$ obtained with different analysis choices, and compares them with Planck 2018 constraints \cite{1807.06209}. These results are shown quantitatively in Table \ref{tab:results}. Table \ref{tab:P18} shows the level of tension with respect to the \planck 2018 best-fit cosmology, while Table \ref{tab:diff} shows the shifts on $S_8$ and $\Om$ (and the change in their posterior uncertainties) when ignoring the impact of baryonic effects. Note that, care must be exercised when interpreting the numerical constraints on $\logMc$ quoted in these tables. As described in the main text, in most cases we are not able to fully constrain $\logMc$, its posterior distribution is highly non-Gaussian, and the quoted intervals depend, at some level, on the priors adopted for this parameter. The results presented in these tables and figures are only a subset of all data and analysis choice combinations explored as part of this work. 

\begin{landscape}
    \footnotesize
    \begin{longtable}{|p{8cm}|c|c|c|c|c|c|c|c|c|}
        \hline
        {\bf Combination} & \multicolumn{2}{c|}{$\chi^2_{\rm WL}$} & $N_{\rm d}$ & \multicolumn{2}{c|}{$p$-value} & $S_8$ & $\Omega_{\rm m}$ & $\sigma_8$ & $\logMc$ \\
        \hline
                          & BF & MAP                              &             & BF & MAP                      &       &                  & & \\
			\hline
			DES & 311.6 & 312.5 & 300 & 0.28 & 0.27 & $0.794\pm 0.018$ & $0.245^{+0.026}_{-0.055}$ & $0.889^{+0.090}_{-0.068}$ & $12.3^{+2.2}_{-1.2}$ \\
			\hline
			DES, no baryons & 312.1 & 313.0 & 300 & 0.28 & 0.26 & $0.786^{+0.016}_{-0.013}$ & $0.252^{+0.029}_{-0.049}$ & $0.866^{+0.077}_{-0.063}$ & -- \\
			\hline
			KiDS & 403.3 & 409.3 & 420 & 0.69 & 0.61 & $0.758\pm 0.025$ & $0.271^{+0.035}_{-0.071}$ & $0.811^{+0.10}_{-0.082}$ & $12.2^{+2.4}_{-1.3}$ \\
			\hline
			KiDS, no baryons & 403.3 & 408.9 & 420 & 0.69 & 0.62 & $0.753^{+0.024}_{-0.019}$ & $0.267^{+0.037}_{-0.065}$ & $0.813\pm 0.090$ & -- \\
			\hline
			HSC & 104.5 & 105.6 & 100 & 0.31 & 0.28 & $0.832\pm 0.030$ & $0.260^{+0.034}_{-0.058}$ & $0.904^{+0.085}_{-0.076}$ & $> 12.2$ \\
			\hline
			HSC, no baryons & 105.3 & 106.5 & 100 & 0.29 & 0.26 & $0.819\pm 0.024$ & $0.261^{+0.034}_{-0.058}$ & $0.890\pm 0.084$ & -- \\
			\hline
			DES + KiDS & 715.1 & 723.9 & 720 & 0.52 & 0.43 & $0.779\pm 0.017$ & $0.233^{+0.025}_{-0.042}$ & $0.892^{+0.071}_{-0.062}$ & $12.1^{+2.1}_{-1.3}$ \\
			\hline
			DES + KiDS + HSC ($\ell < 1000$) & 486.8 & 493.4 & 460 & 0.17 & 0.12 & $0.795^{+0.013}_{-0.015}$ & $0.233^{+0.024}_{-0.041}$ & $0.910\pm 0.071$ & $< 12.3$ \\
			\hline
			DES + KiDS + HSC ($\ell < 2048$) & 670.9 & 679.1 & 640 & 0.18 & 0.13 & $0.793^{+0.013}_{-0.017}$ & $0.224^{+0.021}_{-0.035}$ & $0.922^{+0.064}_{-0.057}$ & $11.7\pm 1.6$ \\
			\hline
			DES + KiDS + HSC ($\ell < 4500$) & 841.1 & 848.8 & 820 & 0.28 & 0.22 & $0.795^{+0.015}_{-0.017}$ & $0.212^{+0.017}_{-0.032}$ & $0.950^{+0.064}_{-0.050}$ & $12.2^{+2.2}_{-1.2}$ \\
			\hline
			DES + KiDS + HSC ($\ell < 8192$) & 998.0 & 1005.9 & 965 & 0.21 & 0.16 & $0.799^{+0.014}_{-0.018}$ & $0.213^{+0.016}_{-0.031}$ & $0.954^{+0.063}_{-0.048}$ & $12.6^{+2.0}_{-0.86}$ \\
			\hline
			DES + KiDS + HSC, no baryons ($\ell < 1000$) & 487.4 & 493.6 & 460 & 0.16 & 0.12 & $0.791\pm 0.012$ & $0.245^{+0.031}_{-0.041}$ & $0.883\pm 0.068$ & -- \\
			\hline
			DES + KiDS + HSC, no baryons ($\ell < 2048$) & 670.7 & 677.9 & 640 & 0.18 & 0.13 & $0.789^{+0.012}_{-0.011}$ & $0.226^{+0.022}_{-0.034}$ & $0.915\pm 0.059$ & -- \\
			\hline
			DES + KiDS + HSC, no baryons ($\ell < 4500$) & 841.4 & 850.9 & 820 & 0.28 & 0.21 & $0.787\pm 0.011$ & $0.221^{+0.020}_{-0.032}$ & $0.921^{+0.060}_{-0.052}$ & -- \\
			\hline
			DES + KiDS + HSC, no baryons ($\ell < 8192$) & 999.0 & 1002.6 & 965 & 0.20 & 0.18 & $0.788\pm 0.011$ & $0.220^{+0.020}_{-0.028}$ & $0.924\pm 0.051$ & -- \\
			\hline
			DES + KiDS + HSC + $H_0$ Planck 2018 prior  & 842.1 & 851.0 & 820 & 0.27 & 0.21 & $0.793^{+0.013}_{-0.015}$ & $0.236^{+0.012}_{-0.027}$ & $0.897^{+0.047}_{-0.032}$ & $12.2^{+2.2}_{-1.2}$ \\
			\hline
			DES + KiDS + HSC + $H_0$ SH0ES prior & 841.6 & 843.0 & 820 & 0.28 & 0.27 & $0.795\pm 0.014$ & $0.229^{+0.012}_{-0.034}$ & $0.913^{+0.065}_{-0.034}$ & $12.4^{+2.2}_{-1.1}$ \\
			\hline
			DES + KiDS + HSC + BAO prior & 841.4 & 850.0 & 820 & 0.28 & 0.21 & $0.795^{+0.012}_{-0.015}$ & $0.262^{+0.011}_{-0.017}$ & $0.853^{+0.029}_{-0.024}$ & $12.1^{+2.0}_{-1.2}$ \\
			\hline
			DES + KiDS + HSC + BAHAMAS prior & 841.5 & 849.3 & 820 & 0.28 & 0.22 & $0.789\pm 0.013$ & $0.218^{+0.020}_{-0.031}$ & $0.930\pm 0.054$ & $11.4^{+1.6}_{-1.4}$ \\
			\hline
			DES + KiDS + HSC + clusters prior & 841.9 & 847.5 & 820 & 0.27 & 0.23 & $0.794^{+0.014}_{-0.018}$ & $0.212^{+0.016}_{-0.032}$ & $0.951^{+0.063}_{-0.046}$ & $12.4^{+2.2}_{-0.98}$ \\
			\hline
			DES + KiDS + HSC + fixed cosmo (BF no-baryons) & 840.4 & 844.7 & 820 & 0.29 & 0.25 & -- & -- & -- & $12.5^{+1.7}_{-0.73}$ \\
			\hline
			DES + KiDS + HSC + fixed cosmo (Planck 2018) & 846.4 & 854.3 & 820 & 0.24 & 0.18 & -- & -- & -- & $14.25^{+0.59}_{-0.22}$ \\
			\hline
			DES + KiDS + HSC + Planck 2018 & 845.8 & 854.3 & 820 & 0.24 & 0.18 & $0.8126\pm 0.0095$ & $0.3108^{+0.0067}_{-0.0092}$ & $0.799^{+0.011}_{-0.0065}$ & $13.6^{+1.1}_{-0.30}$ \\
			\hline
			DES + KiDS + HSC (AE) & 841.6 & 848.7 & 820 & 0.28 & 0.22 & $0.777\pm 0.030$ & $0.224^{+0.021}_{-0.032}$ & $0.905\pm 0.069$ & -- \\
			\hline
			DES + KiDS + HSC (ST15) & 842.3 & 849.9 & 820 & 0.27 & 0.21 & $0.792^{+0.012}_{-0.015}$ & $0.222^{+0.019}_{-0.032}$ & $0.925^{+0.061}_{-0.053}$ & $12.71^{+0.29}_{-0.80}$ \\
        \hline
	\caption{Summary of results. The $\chi^2_{\rm WL}$ and $p$-value account only for the weak lensing data and not priors nor external likelihoods. In the case of BF, this correspond to the parameters that minimize the data $\chi^2$ (including BAO and CMB, when used) and MAP, to those that minimize that of the posterior distribution. If not stated otherwise, the fiducial $\lmax = 4500$ is used.}
	\label{tab:results}
    \end{longtable}
\end{landscape}

\begin{longtable}{|l|c|c|c|c|c|}
        \hline
        {\bf Combination} & $\Delta \chi^2$ & $\Delta S_8 [\sigma]$ & $\Delta \Omega_m [\sigma]$ & $\Delta \sigma_{S_8}[\%]$ & $\Delta \sigma_{\Omega_{\rm m}} [\%]$ \\
        \hline
        DES  & -0.5 & 0.3 & -0.1 & -17 & -11 \\
        \hline
        KiDS & 0.0 & 0.1 & 0.1 & -9 & -5 \\
        \hline
        HSC & -0.8 & 0.4 & 0.0 & -22 & 5 \\
        \hline
        DES + KiDS & -0.9 & 0.3 & -0.2 & -15 & -2 \\
        \hline
        DES + KiDS + HSC  ($\ell < 1000$) & -0.7 & 0.2 & -0.2 & -19 & 0 \\
        \hline
        DES + KiDS + HSC  ($\ell < 2048$) & 0.2 & 0.2 & 0.0 & -24 & -1 \\
        \hline
        DES + KiDS + HSC  ($\ell < 4500$) & -0.3 & 0.4 & -0.2 & -30 & 1 \\
        \hline
        DES + KiDS + HSC  ($\ell < 8192$) & -1.0 & 0.6 & -0.2 & -32 & -11 \\
        \hline
        DES + KiDS + HSC (AE) & 0.2 & -0.3 & 0.1 & -62 & 1 \\
        \hline
        DES + KiDS + HSC (ST15) & 0.9 & 0.3 & 0.0 & -24 & 0 \\
        \hline
\caption{Shifts introduced by not modelling baryonic effects; i.e. $\Delta X [\sigma] = \langle X_{\rm baryons} \rangle - \langle X_{\rm no\, baryons} \rangle / \sqrt{\sigma_{\rm baryons}^2 + \sigma_{\rm no\, baryons}^2}$ and $\Delta \sigma_X = (X_{\rm no\, baryons} / X_{\rm baryons} -1) \times 100$. The errors $\sigma^2_X$ have been estimated as the weighted covariance of the chains. If not stated otherwise, the fiducial $\lmax = 4500$ is used.}
\label{tab:diff}
\end{longtable}

\begin{table}
\centering
    \begin{tabular}{|l|c|c|}
			\hline
			{\bf Combination} & $\Delta S_8 [\sigma]$ & $\Delta \Omega_m [\sigma]$ \\
			\hline
			DES & -1.7 & -1.5 \\
			\hline
			DES, no baryons & -2.3 & -1.5 \\
			\hline
			KiDS & -2.6 & -0.8 \\
			\hline
			KiDS, no baryons & -3.0 & -0.9 \\
			\hline
			HSC & 0.0 & -1.2 \\
			\hline
			HSC, no Baryons  & -0.5 & -1.1 \\
			\hline
			DES + KiDS & -2.5 & -2.2 \\
			\hline
			DES + KiDS, no baryons & -3.2 & -2.1 \\
			\hline
			DES + KiDS + HSC ($\ell < 1000$) & -1.9 & -2.2 \\
			\hline
			DES + KiDS + HSC ($\ell < 2048$) & -2.0 & -2.9 \\
			\hline
			DES + KiDS + HSC ($\ell < 4500$) & -1.8 & -3.5 \\
			\hline
			DES + KiDS + HSC ($\ell < 8192$) & -1.6 & -3.5 \\
			\hline
			DES + KiDS + HSC, no baryons ($\ell < 1000$) & -2.3 & -1.9 \\
			\hline
			DES + KiDS + HSC, no baryons ($\ell < 2048$) & -2.5 & -2.9 \\
			\hline
			DES + KiDS + HSC, no baryons ($\ell < 4500$) & -2.6 & -3.1 \\
			\hline
			DES + KiDS + HSC, no baryons ($\ell < 8192$) & -2.6 & -3.5 \\
			\hline
			DES + KiDS + HSC + $H_0$ \planck 18 prior & -2.0 & -3.0 \\
			\hline
			DES + KiDS + HSC + $H_0$ SH0ES prior & -2.0 & -2.9 \\
			\hline
			DES + KiDS + HSC + BAO prior& -2.0 & -2.8 \\
			\hline
			DES + KiDS + HSC + BAHAMAS prior & -2.4 & -3.4 \\
			\hline
			DES + KiDS + HSC + clusters prior & -1.8 & -3.5 \\
			\hline
			DES + KiDS + HSC + \planck 2018 & -1.2 & -0.5 \\
			\hline
			DES + KiDS + HSC (AE) & -1.7 & -3.1 \\
			\hline
			DES + KiDS + HSC (ST15) & -2.0 & -3.1 \\
			\hline
    \end{tabular}
	\caption{Comparison with \planck 2018 TTTEEE+lowTT+lowEE+lensing+$m_\nu$. If not stated otherwise, the fiducial $\lmax = 4500$ is used.}
	\label{tab:P18}
\end{table}

\section{Fiducial full posterior distributions}
\begin{figure}
    \centering
    \includegraphics[width=\textwidth]{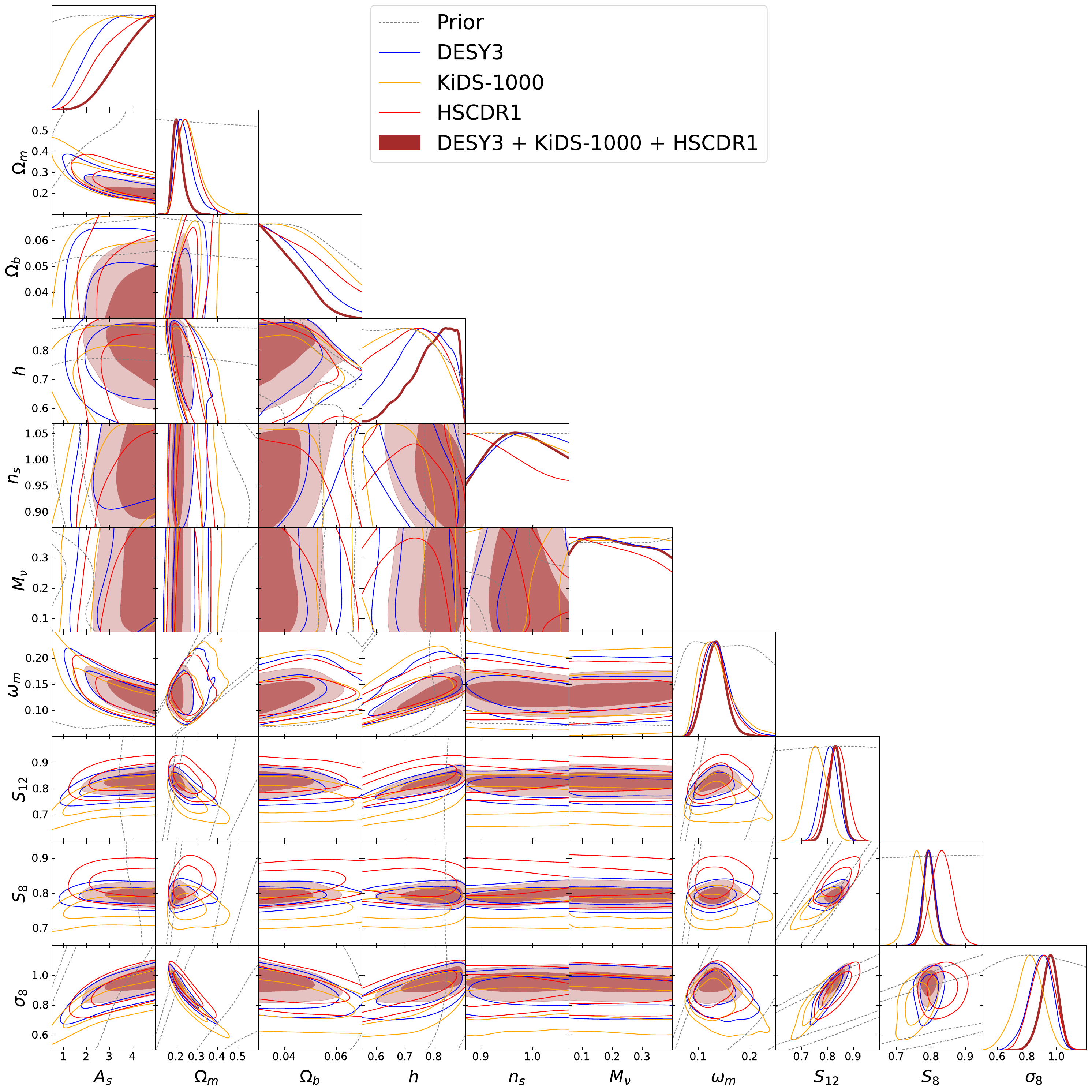}
    \caption{Full posterior distribution (68\% and 95\% C.L. regions) of the cosmological parameters for the fiducial $\lmax=4500$.}
    \label{fig:full_cosmo}
\end{figure}

\begin{figure}
    \centering
    \includegraphics[width=\textwidth]{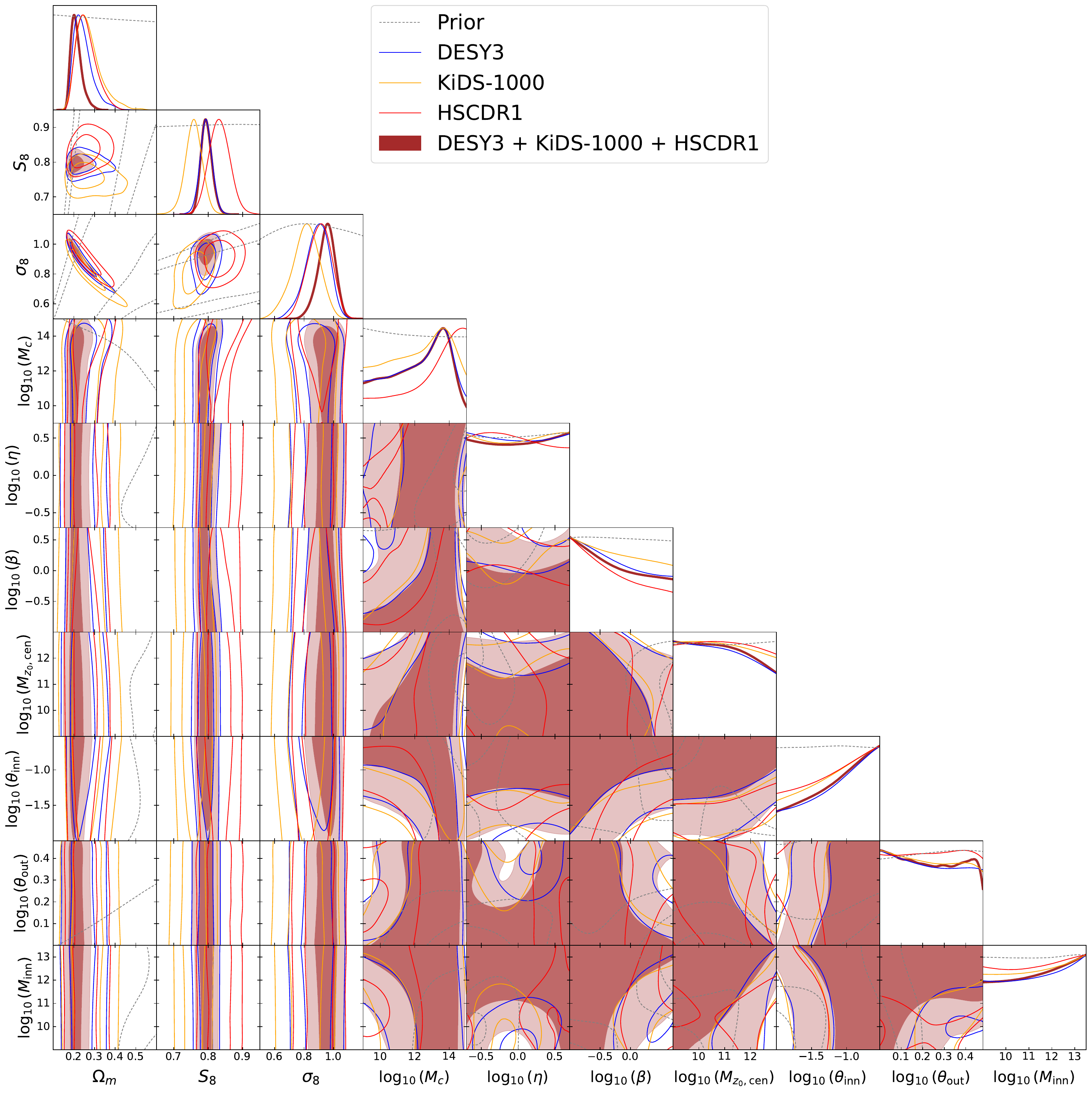}
    \caption{Full posterior distribution (68\% and 95\% C.L. regions) of the BCM paramters for the fiducial $\lmax=4500$}
    \label{fig:full_baryons}
\end{figure}

\begin{figure}
    \centering
    \includegraphics[width=\textwidth]{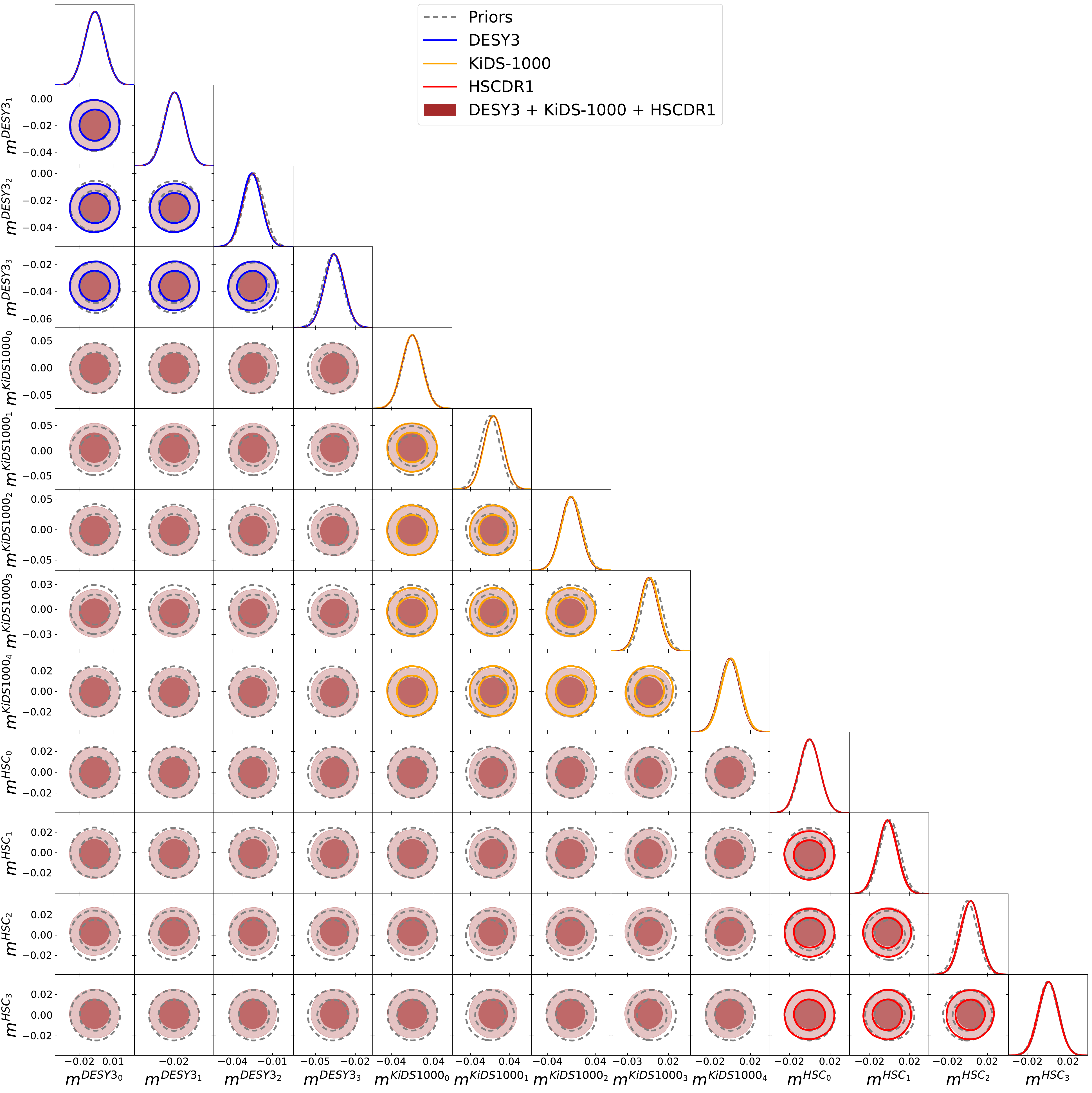}
    \caption{Full posterior distribution (68\% and 95\% C.L. regions) of the multiplicative biases for the fiducial $\lmax=4500$}
    \label{fig:full_mbias}
\end{figure}

\begin{figure}
    \centering
    \includegraphics[width=\textwidth]{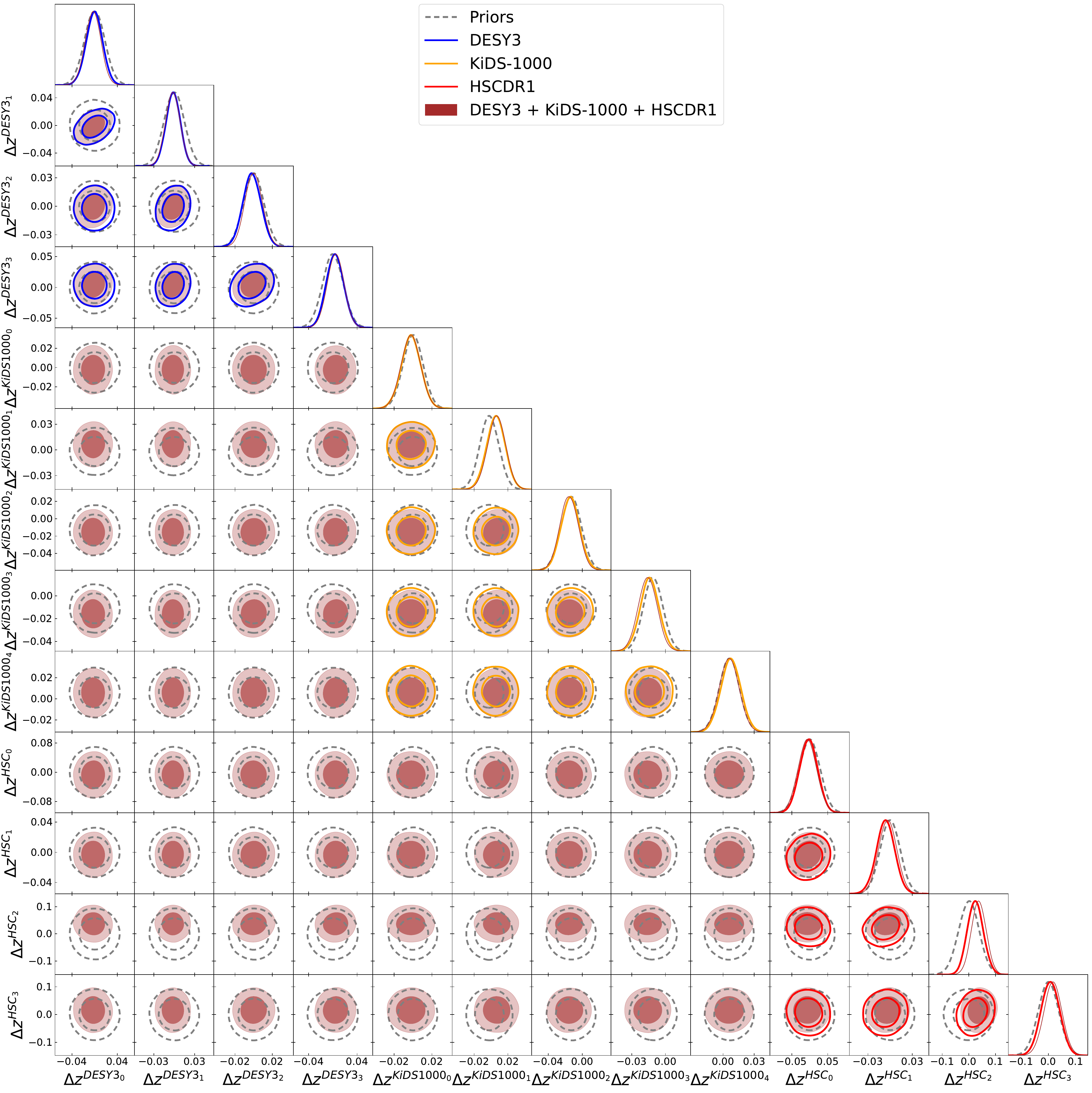}
    \caption{Full posterior distribution (68\% and 95\% C.L. regions) of the redshift shifts $\Delta z$ for the fiducial $\lmax=4500$}
    \label{fig:full_dz}
\end{figure}

\begin{figure}
    \centering
    \includegraphics[width=\textwidth]{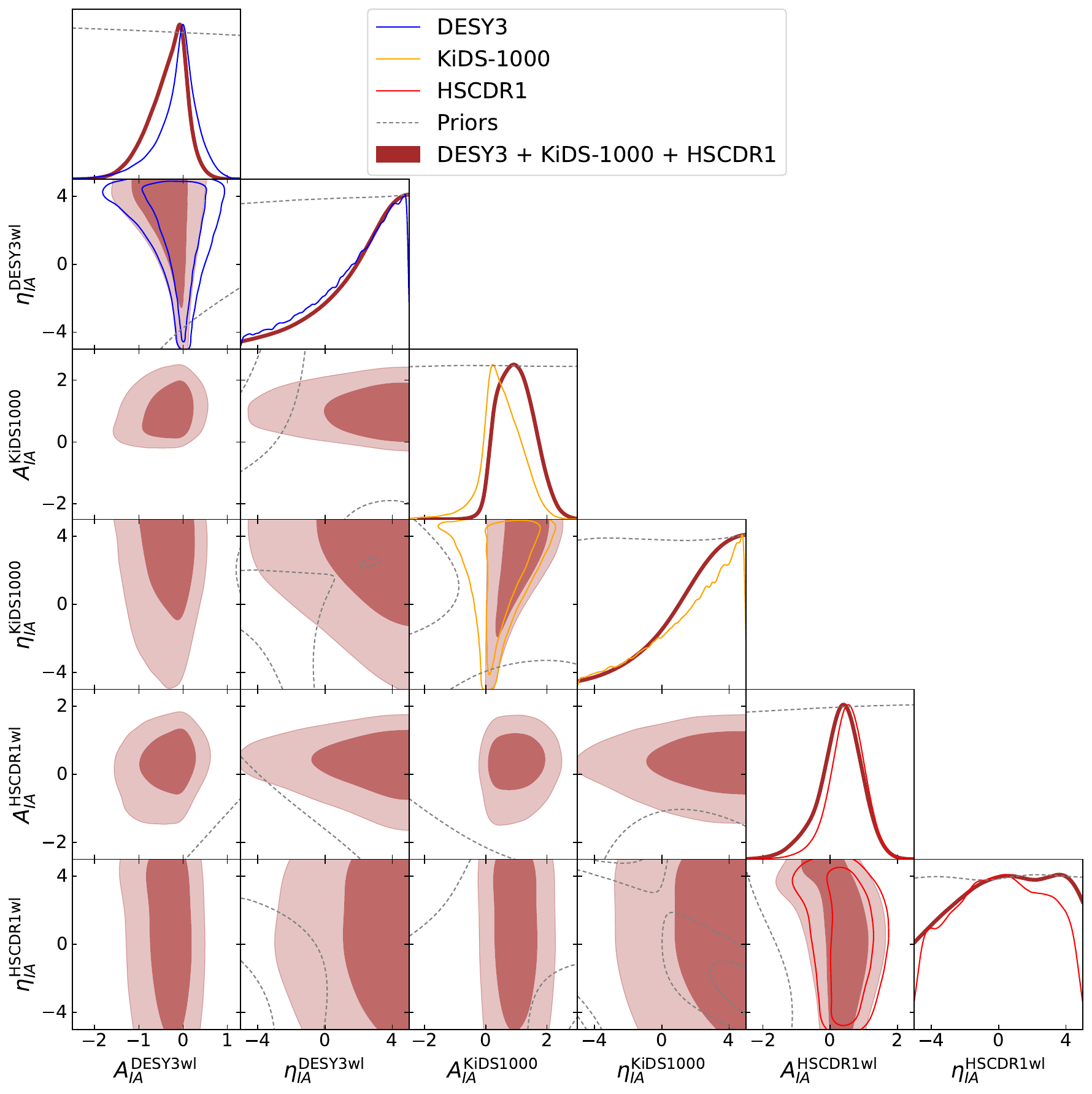}
    \caption{Full posterior distribution (68\% and 95\% C.L. regions) of the IA parameters for the fiducial $\lmax=4500$}
    \label{fig:full_IA}
\end{figure}

This appendix gathers all the full posterior distributions of the fiducial results for \desyt, \kidsot, \hscdro and their combination. Fig.~\ref{fig:full_cosmo} shows the cosmological parameter posterior distributions, Fig.~\ref{fig:full_baryons}, those of the BCM parameters, Fig.~\ref{fig:full_mbias} those of the multiplicative biases, Fig.~\ref{fig:full_dz} those of the redshift shifts and Fig.~\ref{fig:full_IA} those of the IA parameters.

\clearpage
\bibliography{main,non_ads}{}

\end{document}